\documentclass[twocolumn,twocolappendix,hypersetup]{openjournal}
\usepackage{amsmath}
\usepackage{amssymb} 
\usepackage{upgreek}
\usepackage{iondefs}

\usepackage[breaklinks,colorlinks,allcolors=blue,citecolor=blue,urlcolor=blue]{hyperref}
\usepackage{orcidlink}

\newcommand{\subE}{_{\hbox{\tiny E}}}
\newcommand{\supT}{^{\hbox{\tiny T}}}
\newcommand{\subT}{_{\hbox{\tiny T}}}
\newcommand{\supQ}{^{\hbox{\tiny Q}}}
\newcommand{\subLOS}{_{\hbox{\tiny LOS}}}

\newcommand{\supNFW}{^{\hbox{\tiny NFW}}}
\newcommand{\subcalC}{_{\cal C}}
\newcommand{\subCGM}{_{\hbox{\tiny CGM}}}
\newcommand{\sky}{_{\hbox{\tiny SKY}}}
\newcommand{\ihat}{\hat{\hbox{\bf e}}_x}
\newcommand{\jhat}{\hat{\hbox{\bf e}}_y}
\newcommand{\khat}{\hat{\hbox{\bf e}}_z}
\newcommand{\ihato}{\hat{\hbox{\bf e}}_{x_o}}
\newcommand{\jhato}{\hat{\hbox{\bf e}}_{y_o}}
\newcommand{\khato}{\hat{\hbox{\bf e}}_{z_o}}
\newcommand{\shat}{\hat{\hbox{\bf s}}}
\newcommand{\qhat}{\hat{\hbox{\bf e}}_k}
\newcommand{\rvec}{\hbox{\bf r}}

\newcommand{\rhat}{\hat{\hbox{\bf e}}_r}
\newcommand{\phihat}{\hat{\hbox{\bf e}}_\upphi}
\newcommand{\thetahat}{\hat{\hbox{\bf e}}_\uptheta}
\newcommand{\rhohat}{\hat{\hbox{\bf e}}_\uprho}
\newcommand{\Gsys}{$G$}
\newcommand{\Osys}{$O$}
\newcommand{\subG}{_{\hbox{\tiny G}}}
\newcommand{\subO}{_{\hbox{\tiny O}}}

\shorttitle{\sc SKAM I}
\shortauthors{\sc Churchill}

\defcitealias{churchill25-skamII}{Paper~II}

\begin{document}

\title{Spatial-Kinematic Absorption Models of the Circumgalactic Medium \\ I. Structures, Orientations, and Kinematics \vspace{-15mm}}





\author{Christopher W. Churchill\,\orcidlink{0000-0002-9125-8159}}

\affiliation{New Mexico State University\\
Department of Astronomy \\
1320 Frenger Mall \\
Las Cruces, NM 88011, USA}
\thanks{email: \href{mailto:cwc@nmsu.edu}{cwc@nmsu.edu}}

\begin{abstract}
In this two-paper series, we present a straightforward mathematical model for synthesizing quasar absorption line profiles from sight lines through idealized, spatial-kinematic models of the circumgalactic medium (CGM) and their host galaxies. Here, in Paper I, we develop the spatial geometries of multiple galaxy/CGM structures and populate these structures with 3D velocity fields. For arbitrary viewing angles and galaxy-quasar impact parameters, we derive observer coordinate-based expressions for the perceived azimuthal angle and galaxy inclination and a generalized scalar expression for the line-of-sight velocity as a function of position along the line of sight. We motivate and develop four idealized galaxy/CGM spatial-kinematic structures based on empirical data and theoretical predictions: (1) a rotating galactic disk/extra-planar gas, (2) a static or dynamic spherical halo, (3) an outflowing bi-polar galactic wind, and (4) an inward spiraling flared planar accretion. Using a small set of free parameters, the spatial geometries and velocity fields can be adjusted and explored, including velocity gradients, wind stalling, and accretion trajectories. These spatial-kinematic models are designed to be flexible and easily modified and can be tailored for studying individual galaxy-absorber pairs or galaxy group environments; they can be applied to real-world observations or hydrodynamic simulations of the baryon cycle as studied through quasar absorption line systems. These models also serve as tools for developing physical intuition. In \citetalias{churchill25-skamII}, we will present the formalism for populating the galaxy/CGM structures with multiphase photoionized and collisionally ionized gas and for generating absorption profiles for ions of interest.
\end{abstract}

\keywords{
Astronomical models (86) ---
Quasar-galaxy pairs (1316) --- 
Quasar absorption line spectroscopy (1317) ---
Circumgalactic medium (1879)}

\section{Introduction}
\label{sec:introduction}

The circumgalactic medium (CGM), a region of dynamically active multiphase gas surrounding individual galaxies, is a massive reservoir of baryons that plays a prominent role in the evolution of galaxies \citep[see][and references therein]{tpw-araa17, faucher-giguere23}. Dynamically, the CGM serves as a conduit through which infalling gas from the intergalactic medium (IGM) accretes into galaxies to form stars and through which stellar-driven winds are expelled from or recycled back into galaxies \citep[e.g.,][]{oppenheimer08, dave11, vandevoort11}. As such, this important medium channels and regulates the flow of baryons in and out of galaxies. Through this ``baryon cycle," metals and energy are transported between the galaxy's star forming interstellar medium (ISM), its CGM, and the IGM in which it is embedded.  The energetics include an array of dynamical, radiative, and collisional processes that induce photo-heating, shock heating, radiative cooling, turbulent mixing, advection and/or diffusion, and ionization balancing, all of which induce multiphase gas structure \citep[e.g.,][]{birnboim03, maller04, dekel06, ceverino09, faerman17, faerman20, oppenheimer18, fielding20, esmerian21, pandya22}.


Dark matter plus hydrodynamic simulations of galaxies in the cosmic setting show that these processes give rise to a variety of spatially and kinematically correlated regions surround galaxies on scales extending to hundreds of kiloparsecs, and that each is functionally engaged in the galactic baryon cycle in a fairly specific manner \citep[e.g.,][]{ford14, stewart17, nelson19, peroux20, hafen22, trapp22, faucher-giguere23, kocjan24, stern23}. Observations of the CGM using the technique of quasar absorption lines, which record the spatial distribution and the chemical, kinematic, and ionization conditions of CGM gas, have generally supported the growing body of theoretical predictions  \citep[e.g.,][]{bordoloi11, bouche12, kacprzak12, kacprzak15, rubin12, rubin14, nielsen15, ho17, martin19, roberts-borsani19, ho20, peroux20, zabl20, beckett23}.  As such, an idealized canonical composite picture of a spatially and kinematically coherent CGM has developed over the last decade.

In this picture, much of the gas that accretes onto galaxies originates from the IGM and cosmic web and is concentrated within infalling filaments \citep[e.g.,][]{ceverino09, vandevoort11, danovich15, nelson18}.  This infalling gas meets resistance from stellar-driven outflows \citep[e.g.,][]{habe82}, which are propelled outward perpendicular to the galaxy plane and concentrated mostly near the rotation axis \citep[e.g.,][]{nelson19, peroux20, nguyen22, trapp22, faucher-giguere23}.  The outflows tend to fan out as they rise above the disk, forming a bi-polar conical geometry \citep[see][]{rupke18, fielding22, nguyen22}. This overall dynamic, in tandem with angular momentum conservation, compels the infalling gas to spiral inward and to concentrate in an extended co-planar geometry \citep[e.g.,][]{keres05, stewart11, stewart17, stewart-proc17, hafen22, trapp22, kocjan24, stern23}. The relative flow rates and interactions between the  metal-enriched galactic outflows and presumably lower-metallicity accreting gas are influential in creating the observed universe of galaxies, including the stellar-mass to halo-mass function, the global galaxy mass-metallicity relationship, and the mean star formation rate density of the universe \citep[e.g.,][]{dalcanton07, dave11, behroozi13, madau14, chisolm18, wechsler18}.

\begin{figure}[tb]
\centering
\includegraphics[width=1.00\linewidth]{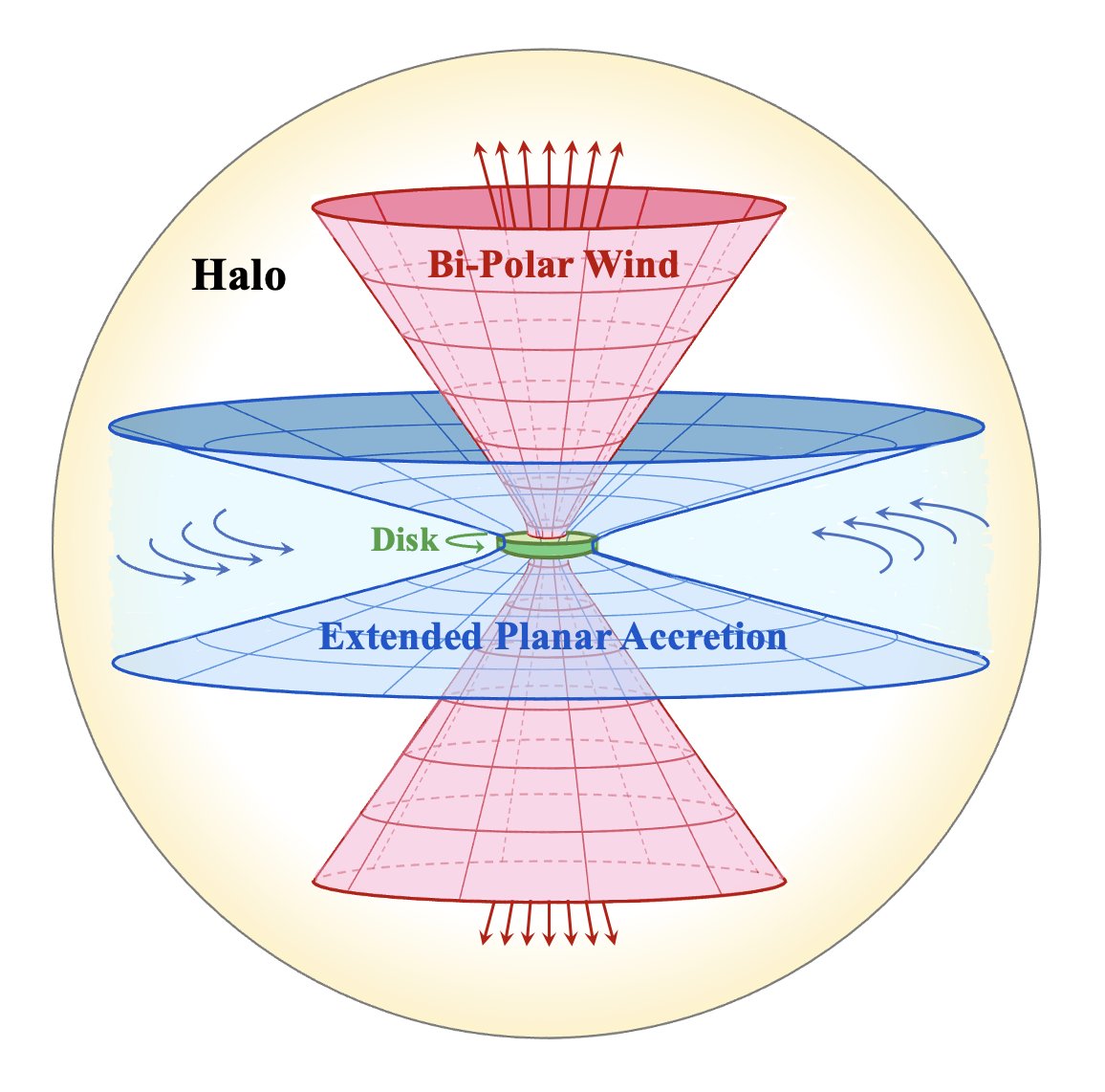}
\caption{\small Schematic of four idealized CGM spatial-kinematic components; the spherical halo (yellow), the bi-polar outflowing wind (red), the extended planar accretion (blue), and the galactic disk (green) with its ISM and extra-planar gas (EPG).  By populating the structures with density, temperature, and velocity fields and running sightlines through the structures, synthetic absorption line systems can be compared to observations.}
\label{fig:myskam}
\end{figure}

This simplified distillation of the CGM is schematically illustrated in Figure~\ref{fig:myskam}.  There are four main spatial-kinematic components: (1) A spherical halo (shaded yellow) encompasses the overall CGM. This region can be kinematically diverse, with outflowing, inflowing, polar streaming, recycling, and/or semi-static material. (2) Stellar-driven outflowing winds can be represented by a bi-polar structure (shaded red) confined within a conical geometry. (3) The ISM is confined within the galaxy disk (shaded green), above which extra-planar gas can form a ``lower halo." (4) The bulk of the accretion is confined to  a ``flared" extended planar structure (shaded blue) in which material spirals inward toward the disk.

To deepen and broaden our understanding, it is important that we continue to study the salient characteristics of the CGM with the goal of organizing our thoughts on its complex roles in the baryon cycle and galaxy evolution. One approach to testing and refining this canonical picture is to apply the observational technique of quasar absorption lines to hydrodynamical cosmological simulations. By running ``mock" quasar lines of sight through simulated galaxies over a range of impact parameters and synthesizing absorption line spectra, the simulated CGM can be studied in detail 
\citep[e.g.,][]{kacprzak10, kacprzak19, ford13, churchill15-direct, nelson18, peeples19, hassan20, appleby21, marra21-ovi, marra21-ionization, marra22, hafen23}. The synthetic spectra can be analyzed in similar fashion to the real-world spectra and the distribution of various properties, such as equivalent widths, column densities, metallicities, number of velocity components, and kinematic spreads, can be quantitative compared.  

A powerful benefit of using simulations is that the absorbing gas spatial location relative to both the galaxy and the line of sight (LOS) can be directly examined in relation to the LOS velocities measured from the spectra. However, with simulations it is challenging to establish clear spatial-kinematic relationship between the absorption component velocities and the absorbing gas; individual simulated halos do not necessarily reflect the distilled canonical picture as illustrated in Figure~\ref{fig:myskam} and LOS velocities measured in absorption do not always arise from a unique spatial location \citep[][]{churchill15-direct, kacprzak19, peeples19, ho-eagle20, marra21-ovi, marra22}.

Analytic models provide a useful bridge between simulations and observations. Simple models provide intuitive insight into the characteristic observables from the real-world universe.  They can be used to unravel the tangled complexity of simulations. Simple intuitive spatial-kinematic absorption models of the CGM can enhance our interpretation of LOS kinematics observed in high resolution quasar absorption line profiles.  

Such models are a rich part of the history of CGM studies.  \citet{weisheit78} developed a constant rotating spherical halo model with a declining power-law density profile and generated some of the first mock absorption profiles based on a model of the CGM.  \citet{lanzetta92} compared observed high-resolution absorption profiles to rotating disks and spherical infall spatial-kinematic models.  Since those early days, a array of idealized models of accreting disks, spherical halos, and bi-conical outflows have been developed and compared to observed absorption lines \citep[e.g.][]{charlton98, prochaska98, steidel02, kacprzak10, kacprzak11, bouche12, gauthier12, chen14, kacprzak19, bordoloi14, diamond16, ho17, lan18, martin19, schroetter19, french20, ho20, zabl20, nateghi21, beckett22, beckett23, carr22, casavecchia23}. With these idealized models, there is no ambiguity about which CGM component, what location relative to the central galaxy, or what underlying kinematics has given rise to absorption lines at a given LOS velocity.  Various spatial geometries and kinematics can be quickly adjusted and flexibly explored and tested without requiring time-intensive, expensive computations on competitive, high-powered computing facilities. 

The majority of previous studies that used idealized models to help explain observed quasar absorption lines have focused on only a single spatial-kinematic CGM component (such as simple disk rotation, or spherical infall/outflow, or bi-conical outflows with constant radial velocities) or compared and contrasted two components \citep[however, see][]{beckett22, beckett23}.  The growing maturity of our canonical picture of the CGM motivates a broader approach in which multiple idealized spatial-kinematic components are simultaneously included using spatial, kinematic, and gas phase properties that are able to reflect our continually evolving collective insights. 


Considering the potential benefits such a model would have for studying the CGM using quasar absorption lines, we have endeavored to develop a multi-component spatial-kinematic absorption model (SKAM). Ideally, SKAMs should be simple in their construction and straightforward to program. The hope is that SKAMs can serve as evolvable tools for guiding interpretations of observed  absorption line profiles through a wide range of astrophysically representative galaxy/CGM environments. We will present the development of a SKAM in two phases, each presented in a separate paper.  

The topic of this Paper~I is three fold. 
In Section~\ref{sec:definitions}, we develop the geometric and mathematical formalism of the model, including the relationship between the observer, background quasar, galaxy, and LOS. Idealized geometric spatial components of the CGM and galaxy are motivated and described in Section~\ref{sec:geometries}.  Also in Section~\ref{sec:geometries}, we show how to determine the precise LOS positions probing each galaxy/CGM structure. In Section~\ref{sec:kinematics}, we describe generalized 3D velocity fields centered on the galaxy and derive the scalar expression yielding the observed LOS velocity as a function of LOS position accounting for multiple arbitrary velocity fields. In Section~\ref{sec:skmodels}, velocity fields suitable to each galaxy/CGM component are proposed and described. This culminates the first installment of our SKAM modeling.  However, we present several examples of LOS velocities versus LOS position (so called $V\subLOS$--LOS plot) for a range of adjustable spatial-geometric and kinematic free parameters. Concluding remarks and a brief discussion are reserved for Section~\ref{sec:conclusion}. 

The topic of \citetalias{churchill25-skamII} is also three fold.  First, we motivate and develop adjustable models of the gas phases that we then populate into the individual galaxy/CGM structures. Second, we incorporate chemical and ionization models of the gas phases in the galaxy/CGM components in order to map out the 3D densities fields of absorbing ions. Third, we apply radiative transfer accounting for the kinematics of the absorbing ions intercepted by the LOS and generate synthetic quasar absorption line profiles. A rudimentary SKAM GUI, further described in \citetalias{churchill25-skamII}, is available as a Fortran 95 code at \href{https://github.com/CGM-World}{github.com/CGM-World}.



\section{Definitions}
\label{sec:definitions}

Here, we define the characteristics of the galaxy dark-matter halo of mass $M_{\rm vir}$ and radius $R_{\rm vir}$, which are specified by the circular velocity of the galaxy.  We then construct two coordinate systems with coincident origins, one in the frame of the galaxy and one in the observer frame. From these two coordinate systems, transformations are derived that yield the galaxy inclination and the LOS impact parameter and azimuthal angle.

\subsection{The Galaxy Dark Matter Halo}
\label{sec:defs-galaxy}

It is appropriate to characterize the galaxy by its virial mass,
\begin{equation}
    M_{\rm vir} = \frac{4\pi}{3}
    R_{\rm vir}^3 (\Delta_c \rho_c) \, ,
\label{eq:virialmass-rvir}
\end{equation}
where $R_{\rm vir}$ is the virial radius. The virial radius is defined such that the average mass density within this radius is $\Delta_c\rho_c$, where $\rho_c$ is the critical density of the universe.  Typically, an overdensity factor of $\Delta_c \sim 200$ is adopted. If we employ an ``NFW" dark matter halo \citep{NFW96}, we can also write
\begin{equation}
    M_{\rm vir} = 4\pi \rho_0 R_s^3 A\subcalC \, ,
\label{eq:virialmass-Rs}
\end{equation}
where $\rho_0$ is related to the mean density of the dark matter halo, $R_{\rm vir} = \mathcal{C} R_s$ where $R_s$ is the scale radius, and where
$A\subcalC = \ln(1+\mathcal{C}) - \mathcal{C}/(1+\mathcal{C})$, where $\mathcal{C}$ is  the ``concentration parameter."
Simulations predict that the median value of $\mathcal{C}$ follows the relation 
$\mathcal{C} \simeq 10.5 M_{12}^{-0.11}$, where $M_{12}= M_{\rm vir}/10^{12}$ in solar masses \citep{maccio07}, where the uncertainty is $\sigma_{\mathcal{C}} = \pm 2$ independent of $M_{\rm vir}$. This dispersion reflects the fact that, at a fixed mass, the halo concentration correlates to a galaxy's assembly time, and therefore redshift of formation and environment \citep[see][]{bullock01, wechsler02, zhao03}.

Given that we are developing spatial-kinematic models, it is desirable that we define the galaxy mass and virial radius in terms of its rotation kinematics.  For an NFW halo, the radius at which the circular velocity, $V_c$, is a maximum is given by $R_{c} = \xi R_s$, where $\xi=2.16258$ is a constant.\footnote{The value of $\xi$ is the positive root of the transcendental equation $ \ln(1+\xi) = \xi (1+ 2\xi)/(1+\xi)^2$.}  The relation between circular velocity and the mass enclosed within the radius $R_c$ is
\begin{equation}
    V^2_c(R_c) = \frac{GM(R_c)}{R_c} \, .
\label{eq:defcircvel}
\end{equation}
Applying the relation, 
\begin{equation}
    M(r) = M_{\rm vir}
    \frac{\ln (1\!+\!x_s)-x_s/(1\!+\!x_s)}
         {A\subcalC} \, ,
\label{eq:nfw-massinside}
\end{equation}
for the mass contained inside radius $r$ of an NFW halo, where $x_s = r/R_s$, we have 
\begin{equation}
    M(R_c) 
    = M_{\rm vir} \frac{A_\xi}{A\subcalC}
    \, ,
\label{eq:massinside}
\end{equation}
where $A_\xi = \ln(1\!+\!\xi) - \xi/(1\!+\!\xi) = 1.83519$.  

\begin{figure}[thb]
\centering
\includegraphics[width=0.80\linewidth]{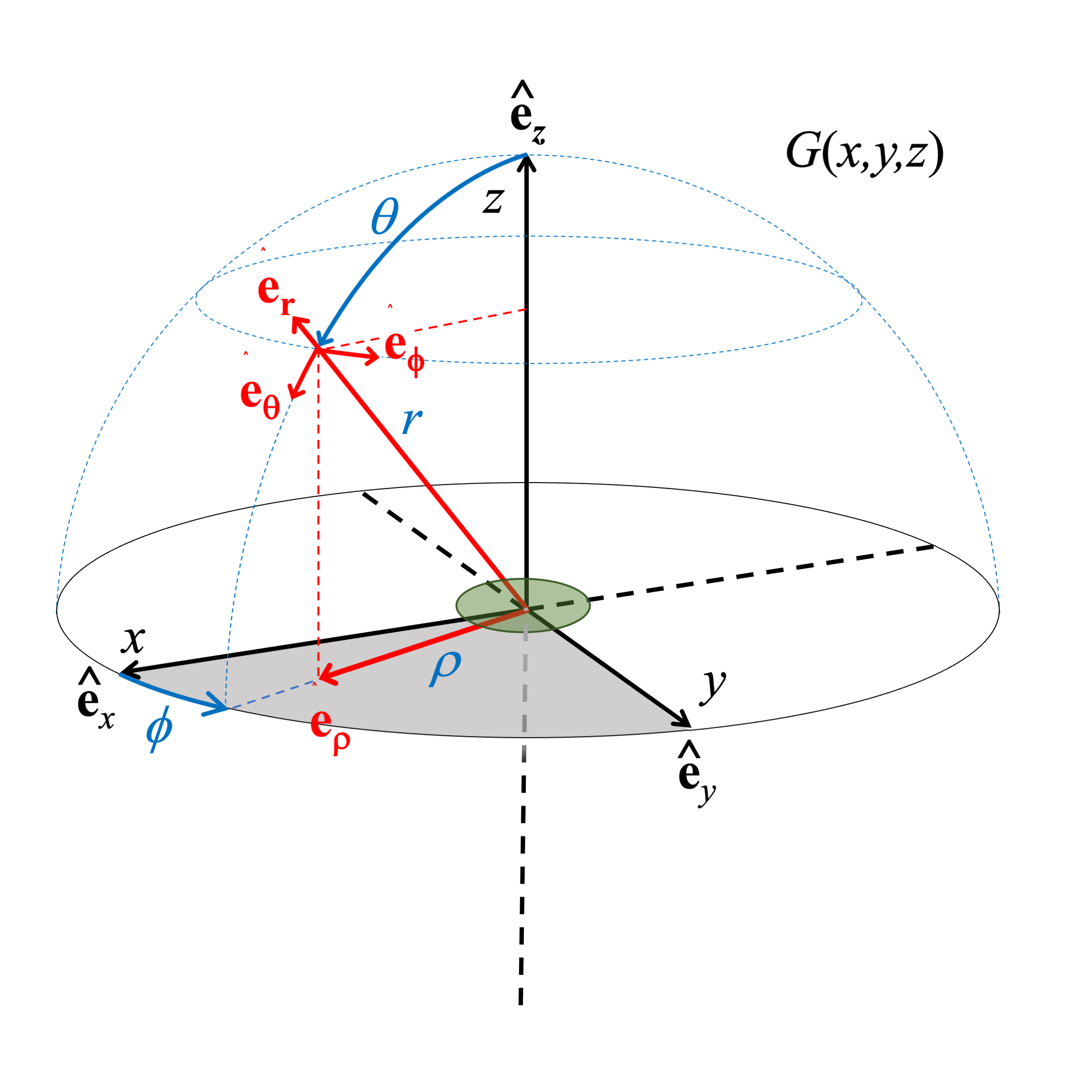}
\caption{\small The coordinate system {\Gsys}$(x,y,z)$ describing the geometry centered on the galaxy (green). Any point at location $P\subG(x,y,z)$ can equally be described by its spherical coordinate $P\subG(r,\theta,\phi)$. Similarly, any vector can be equally described by its Cartesian unit vectors, ${\ihat}$, ${\jhat}$, and ${\khat}$, or by its spherical unit vectors ${\rhat}$, ${\thetahat}$, and ${\phihat}$. The axial radius $\rho$, which has unit vector direction ${\rhohat}$, is the projection of ${\bf r}$ on planes of constant $z$. The galaxy and its associated structures are fixed in this coordinate system.}
\label{fig:thecoordinatesystem}        
\vglue 0.1in
\end{figure}

Combining Eqs.~\ref{eq:virialmass-rvir},  \ref{eq:defcircvel}, and \ref{eq:massinside}, and using the fact that  $R_c=(\xi/\mathcal{C}) R_{\rm vir}$, we obtain a direct relationship between $V_c$ and $R_{\rm vir}$,
\begin{equation}
 R_{\rm vir}^2 = 
  \frac{3}{4\pi G} 
  \frac{\xi/\mathcal{C}}{\Delta_c \rho_c}    \frac{A\subcalC}{A_\xi}  
\, V_c^2 \, .
\label{eq:virialradius}
\end{equation}
Using Eq.~\ref{eq:virialradius}, we can then compute the characteristic radius, $R_s=R_{\rm vir}/\mathcal{C}$, and the radius of maximum circular velocity, $R_c= \xi R_s$. The virial mass, $M_{\rm vir}$, can be computed from Eq.~\ref{eq:virialmass-rvir}. 

We thus have a formalism in which the galaxy virial mass and radius can be defined by only three free parameters, the overdensity factor, $\Delta_c$, the concentration parameter, $\mathcal{C}$, and the maximum circular velocity, $V_c$. For example, a galaxy with $V_c \simeq 220$~{\kms}, $\Delta_c = 200$, and $\mathcal{C}=10$, yields $\log (M_{\rm vir}/{\rm M}_\odot) \simeq 12$, with $R_{\rm vir} \simeq 200$~kpc, $R_s \simeq 20$~kpc, and
$R_c \simeq 45$~kpc.

\subsection{Galaxy-Quasar-Observer and Sky Plane}

\begin{figure}[thb]
\centering
\includegraphics[width=0.80\linewidth]{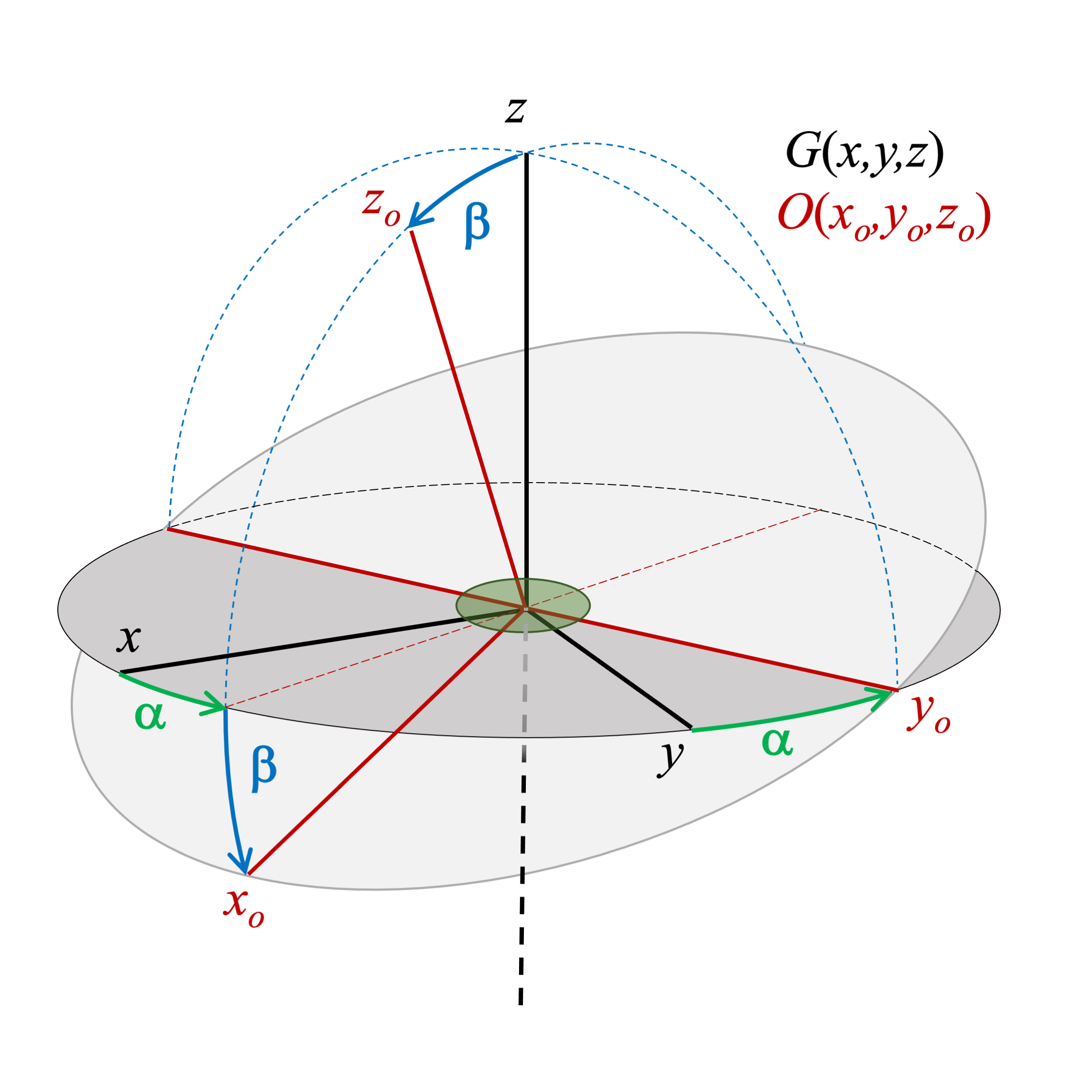}
\caption{\small The observer coordinate system {\Osys}$(x_o,y_o,z_o)$ is rotated about {\Gsys}. An observer in {\Osys} will then see the galaxy with a perspective and orientation dependent on the rotation. System {\Osys} is rotated twice, first by the angle $\alpha$ around the $z$-axis followed by the angle $\beta$ around the {\it rotated} $y_o$ axis.  The galaxy is centered at $P\subG (0,0,0) = P\subO(0,0,0)$.}
\label{fig:thecoordinaterotation}        
\vglue 0.1in
\end{figure}

Our aim is to develop a simple yet flexible formalism for allowing an observer to view a galaxy with an arbitrary spatial orientation.  Furthermore, the LOS and sky position of the background quasar has to be expressed in terms of the relationship between the observer and the galaxy. We first set up a fixed coordinate system (denoted {\Gsys}) for the frame of reference of the galaxy.  As shown in Figure~\ref{fig:thecoordinatesystem}, the principal axes are $x,y,z$ with respective unit vectors $\ihat$, $\jhat$, and $\khat$. The corresponding spherical coordinates are $r, \theta, \phi$ with respective unit vectors ${\rhat}$, ${\thetahat}$, and ${\phihat}$, where $\theta$ is the polar angle and $\phi$ is the azimuthal angle on planes of constant $z$. The galaxy is centered at the origin. The galaxy rotation (polar) axis is the $z$ axis and galactic plane is the $xy$ plane. Spatially, the galaxy and its CGM structures remain fixed in this system. 

As illustrated in Figure\ref{fig:thecoordinaterotation}, we also define a second coordinate system (denoted {\Osys}) in which the observer resides. In {\Osys}, the principle axes are $x_o,y_o,z_o$ with unit vectors are ${\ihato}$, ${\jhato}$, and ${\khato}$.  The origin of {\Osys} is coincident with that of {\Gsys}. To provide for arbitrary viewing angles of the galaxy, the observer coordinate system, {\Osys}, is rotated with respect to the galaxy coordinate system, {\Gsys}. 
Consider an observer placed some distance from the origin on the $x_o=x$ axes prior to coordinate rotation. A single rotation of {\Osys} about the $y$ axis of {\Gsys} through the angle $\beta$ would be sufficient for changing the observed galaxy inclination.  However, the polar axis of the galaxy ($z_o$) would always be confined to the coincident $xz$ and $x_oz_o$ planes; an oblique viewing angle could not be achieved. For a general orientation, or what we are calling an ``oblique inclination," a minimum of two rotations of {\Osys} are required. The first is a counterclockwise rotation of {\Osys} about the $z$ axis of {\Gsys} through an angle $\alpha$, with $\alpha \in (0^\circ,90^\circ]$.  The second is a counterclockwise rotation about the {\it rotated\/} $y_o$ axis by an angle $\beta$, with $\beta \in (-90^\circ,+90^\circ)$.  

These two rotations, carried out in this order, are illustrated in Figure~\ref{fig:thecoordinaterotation} and are described in detail in Appendix~\ref{app:A}.  They provide the means for the galaxy to be viewed in system {\Osys} from all possible orientations.  Assuming azimuthal symmetry in {\Gsys}, by limiting the first rotation to $0^\circ \leq \alpha < 90^\circ$, we avoid degeneracies. 

\begin{figure}[h!tb]
\centering
\includegraphics[width=.95\linewidth]{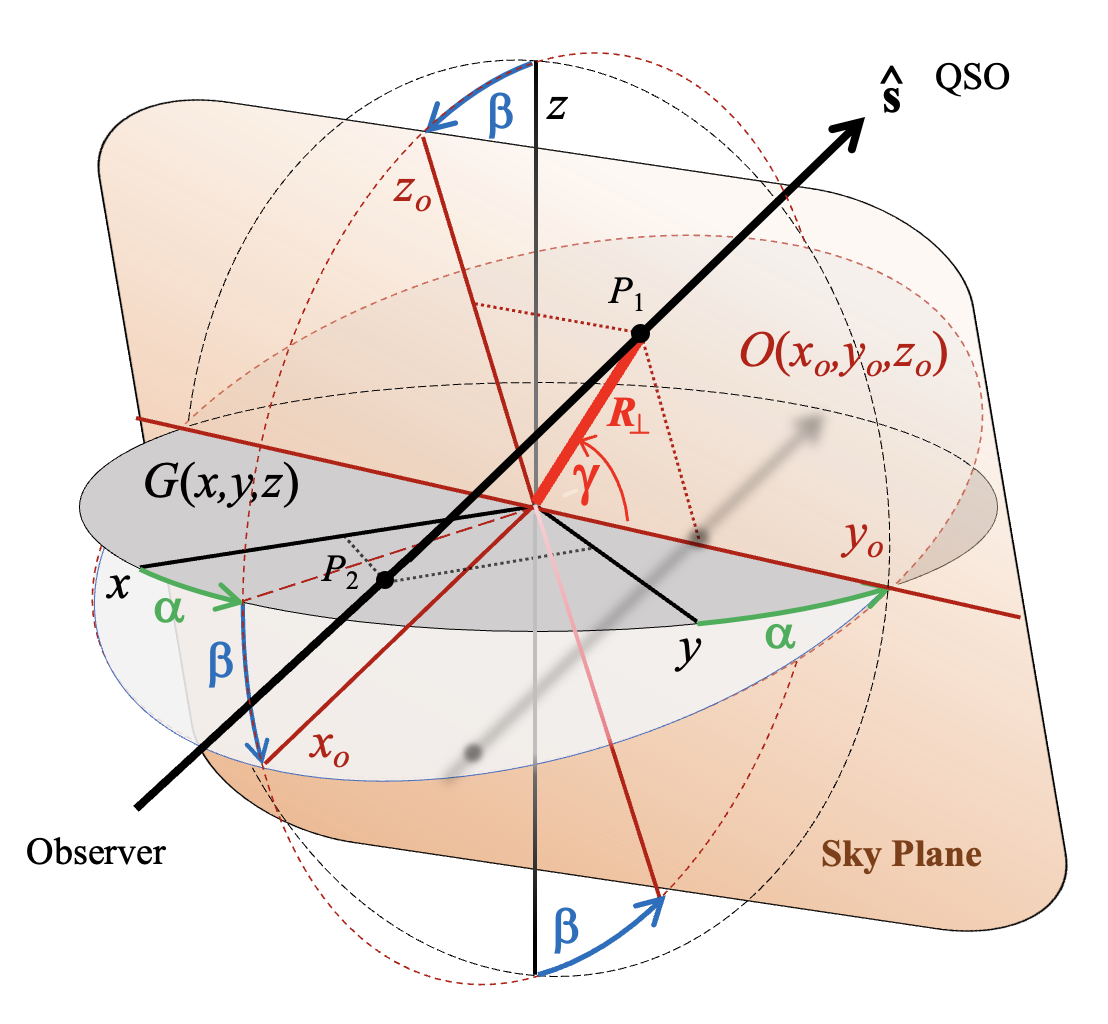}
\vglue 0.15in
\caption{\small The cosmic-eye view of the galaxy coordinate system {\Gsys}$(x,y,z)$ and the observer coordinate system {\Osys}$(x_o,y_o,z_o)$, which has been rotated twice, first by the angle $\alpha$ around the $z$-axis of {\Gsys} followed by the angle $\beta$ around the {\it rotated} $y_o$ axis.  The galaxy is centered at $P\subG (0,0,0) = P\subO(0,0,0)$. The observer is located at $x_o = +\infty$ and the background quasar is located at  $x_o = -\infty$.  The LOS (thick black arrow) intersects the sky plane at the coordinates $P_1 = P\subO(0,R_\perp \cos \gamma, R_\perp \sin \gamma)$, where $R_\perp$ is the impact parameter and $\gamma$ is the position angle of the quasar on the sky plane. In {\Gsys}, this point is denoted $P_1 = P\subG(X_0,Y_0,Z_0)$. The LOS is parallel to the $x_o$ axis and has unit vector  ${\shat}= - {\ihato}$. Note the LOS intersects the $z=0$ plane of the galaxy at point $P_2 = P\subG(x_g,y_g,0)$, which is given by Eq~\ref{eq:rhodisk}.}
\label{fig:skyplanedefs}
\end{figure}

As shown in Figure~\ref{fig:skyplanedefs}, we place the observer at $x_o=+\infty$, the quasar source at $x_o=-\infty$, and define the plane of the sky as the $z_o y_o$ plane ($x_o=0$), which we will call the ``sky plane." The LOS between the source and the observer is perpendicular to the sky plane and is thus parallel to the $x_o$ axis.  We define the LOS direction to be the unit vector $\shat$ and adopt the convention  ${\shat} = -{\ihato}$ such that it points from observer to source.
 
In {\Osys}, we defined the sky position of the background quasar with respect to the galaxy in terms of the impact parameter, $R_\perp$, which is the sky-projected separation between the background quasar and the galaxy center, and its position angle, $\gamma$.  Whereas observationally, position angles are measured relative to the north celestial pole and rotate positive into the direction of increasing right ascension, we defined the quasar position angle to be measured positive counterclockwise from the $y_o$ axis in the sky plane.

Thus, in {\Osys}, the point at which the LOS intersects the sky plane (labeled $P_1$ in Figure~\ref{fig:skyplanedefs}) is given by 
\begin{equation}
\begin{array}{rcl}
 x_o  \!\!&=&\!  X\supQ\sky   = 0 \, , \\[3pt] 
 y_o  \!\!&=&\! Y\supQ\sky  = R_\perp \cos \gamma \, , \\[3pt]
 z_o  \!\!&=&\! Z\supQ\sky    = R_\perp \sin \gamma \, .
\end{array}
\label{eq:SKYimpact}
\end{equation} 
Since the equation of line in a 3D space requires its direction, in this case ${\shat} = -{\ihato}$, and a single point, in this case  $P\subO(0,R_\perp \cos \gamma,R_\perp \sin \gamma)$, we have the minimal information to fully describe the LOS.

\subsection{The Line of Sight (LOS)}

We define the location, or position, along the LOS using the parametrized coordinate $t$, measured in physical length units (such as parsecs), over the domain $t\in [-\infty, +\infty]$.  We adopt the convention that $t$ increases from the observer to the quasar, with the observer placed at $t=-\infty$ and the quasar placed at $t=+\infty$. We define $t=0$ as the point where the LOS intersects the sky plane. Since the LOS is perpendicular to the sky plane, $t$ directly provides the LOS distance from the sky plane in physical length units, where $t>0$ lies on the quasar side and $t<0$ lies on the observer side. 

In the observer frame of reference, {\Osys}, the position along the LOS is very simply described by the relations 
\begin{equation}
\begin{array}{rcl}
x_o(t) \!\! &=& - t \, , \\[3pt]
y_o(t) \!\! &=& \! Y\supQ\sky \, , \\[3pt]
z_o(t) \!\! &=& \! Z\supQ\sky \, ,
\end{array}
\end{equation}
In the galaxy frame of reference, {\Gsys}, where the spatial-kinematic models of the CGM are defined, the LOS location is given by the parametric equation 
\begin{equation}
\begin{array}{rcl}
x(t) \!\! &=& \! X_0 + \sigma_x t \, , \\[3pt]
y(t) \!\! &=& \! Y_0 + \sigma_y t \, , \\[3pt]
z(t) \!\! &=& \! Z_0 + \sigma_z t \, ,
\end{array}
\label{eq:defineLOSPOS}
\end{equation}
where $P\subG(X_0,Y_0,Z_0)$ is the point on the LOS in {\Gsys} when $t=0$, and $\sigma_x$, $\sigma_y$, and $\sigma_z$ are the LOS direction cosines.
To fully describe the LOS position as a function $t$ in {\Gsys}, we need to obtain the point $P\subG(X_0,Y_0,Z_0)$ and the three direction cosines $\sigma_x$, $\sigma_y$, and $\sigma_z$.  The location along the LOS where the sky plane is intersected ($t=0$) is a function of the rotation angles, $\alpha$ and $\beta$, and the sky position of the quasar $(R_\perp,\gamma)$, whereas the direction of the LOS through {\Gsys} is a function of $\alpha$ and $\beta$ only.  At $t=0$, the LOS position is at the spatially coincident points $P\subG(X_0,Y_0,Z_0) = P\subO(X\supQ\sky,Y\supQ\sky,Z\supQ\sky)$.  In other words, point $P\subG(X_0,Y_0,Z_0)$ gives the location in {\Gsys} where the LOS intersects the sky plane in {\Osys}, which is given by Eq.~\ref{eq:SKYimpact}.  The LOS direction is  ${\shat} =  \sigma_x {\ihat} + \sigma_y {\jhat} + \sigma_z {\khat}$ in {\Gsys} and (by definition) is ${\shat} = -{\ihato}$ in {\Osys}.  Thus we have  $\sigma_x  = -({\ihato} \!\!\cdot {\ihat})$,  $\sigma_y  = -({\ihato} \!\!\cdot {\jhat})$, and $\sigma_z  = -({\ihato} \!\!\cdot {\khat})$.

In Appendix~\ref{app:A}, we derive the full expression for the LOS position. We obtain 
\begin{equation}
\begin{array}{rcl}
x(t) \!\! & = & \! 
R_\perp ( \cos\gamma \cos\beta \sin\alpha - \sin\gamma \sin\beta ) \\ & & - \, t \cos\beta \cos\alpha \, , \\[4pt]
y(t) \!\! & = & \! 
R_\perp \cos\gamma \cos\alpha + t \sin\alpha \, , \\[5pt] 
z(t) \!\! & = & \! 
R_\perp (\cos \gamma \sin\beta \sin \alpha  + \sin\gamma \cos\beta )
\\ & & -\, t \sin\beta \cos\alpha \, ,
\end{array}
\label{eq:lospos}
\end{equation}
where the point $P\subG(X_0,Y_0,Z_0)$ is given by Eq.~\ref{eq:skypos-derived} and the three direction cosines, $\sigma_x$, $\sigma_y$, and $\sigma_z$, are given by Eq.~\ref{eq:shat-derived}. The galactocentric distance of the LOS position $t$ is
\begin{equation}
\begin{array}{rcl}
r(t) \!\! &=&\! \left[ x^2 (t) + y^2 (t) + z^2 (t) \right] ^{1/2} \\[5pt]
\!\ &=&\! \left[ R^2_\perp + t^2 \right] ^{1/2} \, ,
\label{eq:rlosoft}
\end{array}
\end{equation}
which is an invariant quantity between {\Gsys} and {\Osys}. 

\subsection{The Observed Galaxy Orientation}

The two observable quantities that define the orientation of the galaxy with respect to the LOS are the inclination and the sky-projected azimuthal angle. It would be desirable that these two observed quantities could be defined to determine the coordinate rotations, $\alpha$ and $\beta$.  However, degeneracies in the projection of the galaxy on the sky plane render this impossible to geometrically map with a unique one-to-one relation.  Instead, we express the observed inclination in terms of the coordinate rotation angles, i.e., $i=i(\alpha,\beta)$, and express the azimuthal angle in terms of both the rotation angles and the quasar position angle, i.e., $\Phi = \Phi(\alpha,\beta,\gamma)$.

The inclination angle is defined as the angle between the $z$-axis of the galaxy and the LOS.  When this angle is a right angle, the galaxy is viewed edge on. When the two vectors are parallel, the galaxy is viewed face on.  As such, the inclination is given by the dot product between these two vectors, $\cos i = {\shat} \cdot {\khat}$.  In Appendix~\ref{app:B}, we show that the observed inclination can be written, 
\begin{equation}
 \cos i = \sin|\beta| \cos\alpha \, .
\label{eq:inclination}
\end{equation}

The observed sky-projected azimuthal angle, $\Phi$, is defined as the angle between the projected major axis of the galaxy and a line connecting the galaxy center to the LOS position on the sky plane. 
In Appendix~\ref{app:C}, we show that the sky-projected azimuthal angle can be written 
\begin{equation}
\Phi =  \gamma + \cot ^{-1}
\left( \frac{\cos\beta}{\sin\beta \sin\alpha}  \right) \, .
\label{eq:phi}
\end{equation}
We define $\Phi \in [0^{\circ},90^{\circ}]$ such that it is always the primary angle.  In the special case where $\alpha = 0$, we have ${\cos i = \sin \beta}$, or $i=90^\circ - \beta$, and $\Phi = \gamma$. Also note that when $\beta = 0^\circ$ the galaxy is viewed edge-on so that any rotation by $\alpha$ will always yield $i=90^\circ$, and $\Phi = \gamma$. When $\beta \rightarrow \pm 90^\circ$, we have the less intuitive result that $i \rightarrow \alpha$ and $\Phi \rightarrow \gamma - 90^\circ$. However, for $\beta = \pm 90^\circ$, $\alpha$ is a degenerate quantity from the point of view of the observer, giving $i=0$, and $\Phi$ cannot be measured.

\subsection{Intersecting the Galaxy Mid-Plane}

The LOS intersects the mid-plane of the galaxy at $z_g = 0$.  Setting $z(t)=0$ in Eq.~\ref{eq:defineLOSPOS}, we find that this intersection occurs at LOS location $t = -Z_0/\sigma_z$. From Eq.~\ref{eq:shat-derived} and Eq.~\ref{eq:inclination},  $t = Z_0 / (\sin \beta \cos \alpha) = Z_0\sec i$, where $Z_0 \in [-R_\perp,+R_\perp]$ and $i \in [-90^\circ,90^\circ]$. Note that as  $i \rightarrow \pm 90^\circ$ or $\beta \rightarrow \pm 0^\circ$, the LOS intersection position tends toward $t \rightarrow \pm \infty$ because the galaxy is viewed edge on. Recall that $\alpha \in (0^\circ , 90^\circ ]$, so $\cos \alpha > 0$ always.  Therefore, for $|i| < 90^\circ$, the galaxy disk mid-plane LOS intersection point is
\begin{equation}
\begin{array}{rcl}
x_g \!\!&=&\! X_0 + \sigma_x Z_0 \sec i \, , \\[3pt]
y_g \!\!&=&\! Y_0 + \sigma_y Z_0 \sec i \, ,\\[4pt]
\rho_g \!\!&=&\! \sqrt{x^2_g + y^2_g} \, ,
\label{eq:rhodisk}
\end{array}
\end{equation}
where $\rho_g$ is the axial radial distance from the galaxy center, $X_0$, $Y_0$, and $Z_0$ are given by Eq.~\ref{eq:skypos-derived}, and $\sigma_x$ and $\sigma_y$ are given by Eq.~\ref{eq:shat-derived}.
This intersection is illustrated as point $P_2 = P\subG(x_g,y_g,0)$ in Figure~\ref{fig:skyplanedefs}.

\section{Idealized CGM Structures}
\label{sec:geometries}

As described in the Section~\ref{sec:introduction}, observations and theory suggest that the CGM can be segregated into a few somewhat unique spatial-kinematic components each playing a dominant role in the baryon cycle. These components were schematically illustrated in Figure~\ref{fig:myskam}.

In summary, the CGM itself can be loosely visualized as a spherical region surrounding the central galaxy. Distributed throughout the CGM are galactic winds, accreting filaments, and recycling gas flows. Tidal streams from past mergers and/or from orbiting satellite galaxies may also be present. The inner halo (within $\sim\! 10$~kpc above and below the galaxy disk) is populated by extra-planar structures such as galactic fountains and high velocity clouds. 

\begin{figure}[htb]
\centering
\includegraphics[width=0.85\linewidth]{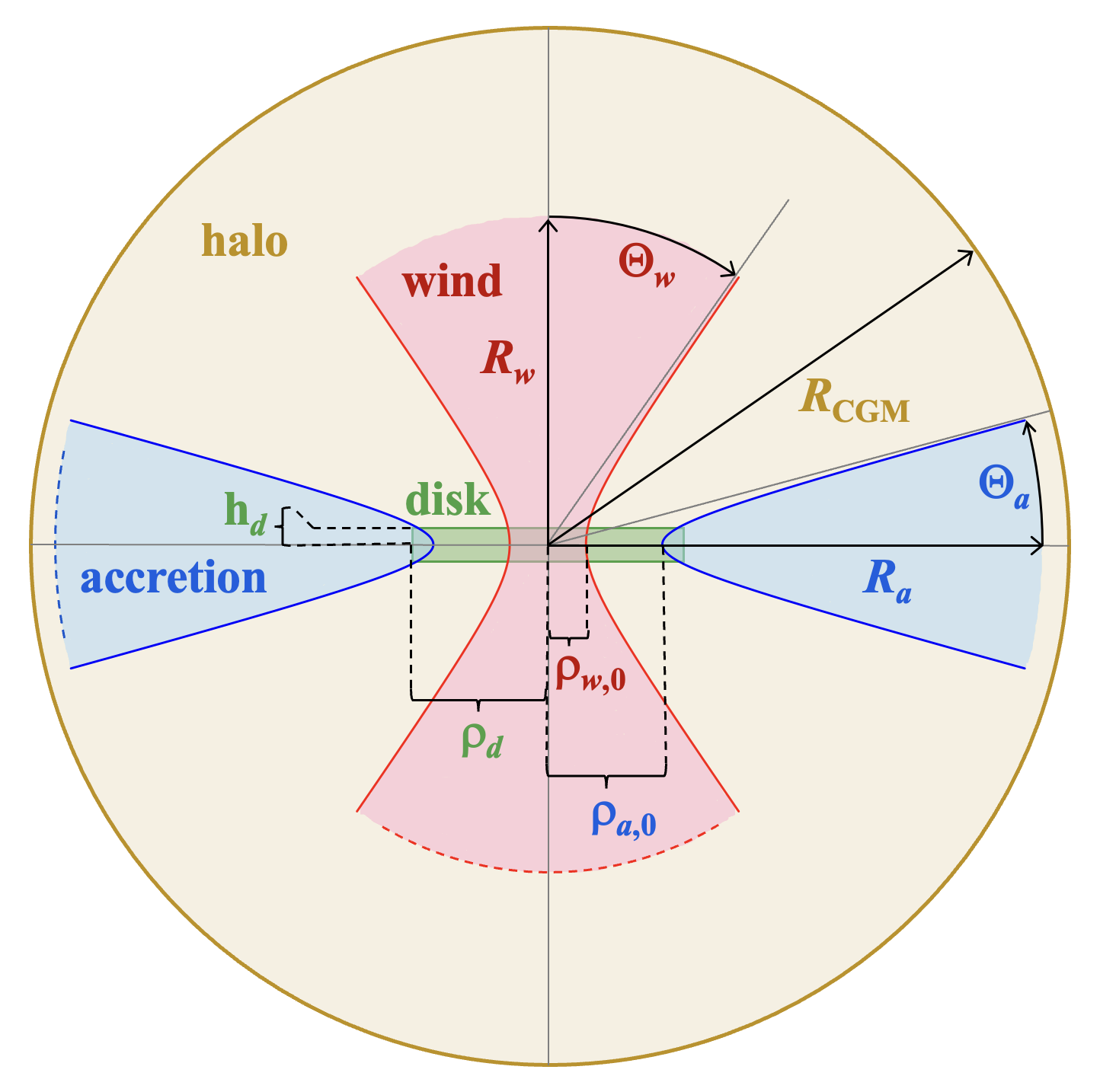}
\vglue 0.1in
\caption{\small A cross-sectional schematic (not to scale) of the idealized spatial structures of the CGM as seen in the galaxy frame, {\Gsys}. The ``halo" is modeled as a sphere of radius $R\subCGM$. The  ``bi-polar wind" (red) is modeled as a hyperboloid of one sheet described by its opening angle $\Theta_w$, base radius $\rho_{w,0}$, and maximum extent $R_w$. The ``disk" (green) is modeled as a finite cylinder with axial radius $\rho_d$ and height $h_d$.  The ``extended planar accretion" (blue) is also modeled as a hyperboloid of one sheet described by its inner (accretion) radius $\rho_{a,0}$ and maximum extent $R_a$, but the accretion material is confined to void volume outside the solid geometry of the hyperboloid so that the complimentary angle to the opening angle is used; we call it the flare angle, $\Theta_a$.  }
\label{fig:structure}
\end{figure}

In Figure~\ref{fig:structure}, we represent the four idealized spatial components of the CGM as common geometric solids.  A spherical ``halo''  readily represents the CGM region (shaded beige), whereas a cylinder of finite height represents a disk-like central galaxy and its extra-planar gas (shaded green). The bi-polar wind is represented by a hyperboloid (shaded red), as is the planar accretion (shaded blue).  The free parameters that describe the structures are geometric quantities. The spherical halo has a single parameter, the CGM radius, $R\subCGM$.  The cylindrical disk has two, its axial radius and height, $\rho_d$ and $h_d$, respectively. The wind hyperboloid is described by its ``skirt" radius, opening angle, and radial extend, $\rho_{w,0}$, $\Theta_w$, and $R_w$, whereas the accretion hyperboloid is similarly described by $\rho_{a,0}$, $\Theta_a$, and $R_a$.  All these geometric quantities are labeled in the schematic illustrated in Figure~\ref{fig:structure}.

As a LOS passes through the CGM, it likely probes or partially probes multiple CGM components.  As each spatial structure is associated with its own predominant systematic kinematics pertaining to its role in the baryon cycle, we expect to see characteristic signatures in the LOS velocities of quasar absorption line profiles. Therefore, for modeling purposes, it is important we determine the locations on the LOS that intersect and probe each geometric structure, as this allows us to determine LOS velocities.  For any geometric structure, the point(s) of LOS intersection are derived from the solution  $x(t)\!=\! X$, $y(t) \!=\! Y$, and $z(t)  \!=\! Z$, where $x(t)$, $y(t)$, and $z(t)$ are the LOS position at location $t$ as given by Eq.~\ref{eq:lospos}, and $X$, $Y$, and $Z$ are the parametric equations defining the surface of the CGM structure (geometric solids) in the galaxy frame, {\Gsys}.

\subsection{The Spherical Halo}
\label{subsec:thesphere}

The historically motivated and simplest ``halo'' model for interpreting quasar absorption line data is an extended spherical geometry \citep[e.g.,][]{bahcall69, weisheit78, lanzetta90, bergeron91, steidel95, charlton96}.  Efforts to develop increasingly sophisticated analytic models of extended gaseous halos has remained an active field of study \citep[e.g.,][]{li92, mo96, maller04, fang13, sharma14, stern16, stern18, stern19, stern23, faerman17, faerman20, mathews17, pezzulli17, qu18, sormani18, faerman23}. These models are  computationally inexpensive, and based on a few simple parameters, often provide density, $n({\bf r})$, column density, $N({\bf r})$, temperature, $T({\bf r})$, and/or rotational velocity, $V_\upphi({\bf r})$, distribution functions.

All points on the surface of a sphere of radius $R_h$ are given by 
\begin{equation}
    X_h^2 + Y_h^2 + Z_h^2 = R^2_h \, .
    \label{eq:sphere}
\end{equation}
where the subscript ``$h$'' denotes ``halo.'' The sphere can be representative of the physical extent, or ``boundary,'' of the CGM surrounding a galaxy of virial mass, $M_{\rm vir}$, which would relate to the virial radius, $R_{\rm vir}$.  
We would then define $R\subCGM \!=\! R_h$, and substitute $R\subCGM \!=\! \eta\subCGM R_{\rm vir}$, into Eq.~\ref{eq:sphere} where $\eta\subCGM \in (0,\infty]$ is a free parameter.  The choice of $\eta\subCGM$ should be physically motivated based on theory or observations. Following \citet{shull14}, we point out the vagaries in the nomenclature ``extended gas in galactic halos," ``CGM," ``filaments," and ``IGM,'' as adopted by observational astronomers; if one aims to enclose all gravitationally bound gas within their adopted CGM boundary, then the CGM can extend beyond the virial radius ($\eta\subCGM > 1$), but then one would likely enclose some gas at the CGM/IGM interface that is unbound.

To obtain the spatial location where the LOS probes the sphere, we equate Eq.~\ref{eq:sphere} and Eq.~\ref{eq:defineLOSPOS} with $R_h = R\subCGM$, and write,
\begin{equation}
    (X_0 \!+\! \sigma_x t)^2 + (Y_0 \!+\! \sigma_y t) ^2 + (Z_0 \!+\! \sigma_z t) ^2 = R^2\subCGM \, ,
\end{equation}
which yields a quadratic equation of the form 
\begin{equation}
  f(t) =   At^2 + Bt + C = 0 \, ,
\label{eq:quadeq}
\end{equation}
with solution
\begin{equation}
   t = \frac{-B \pm \sqrt{\Delta}}{2A} \, ,
\label{eq:quadsol}
\end{equation}
where $\Delta = B^2 - 4AC$ is the determinant, and where the geometric coefficients are
\begin{equation}
\begin{array}{rcl}
A \!\!&=&\! 1 \, ,\\[3pt] 
B \!\!&=&\! 2(\sigma_x X_0 + \sigma_y Y_0 + \sigma_z Z_0) \, ,\\[3pt] 
C \!\!&=&\! R^2_\perp - R^2\subCGM \, ,
\end{array}
\label{eq:ABC-halo}
\end{equation}
employing 
$\sigma_x^2 + \sigma_y^2 + \sigma_z^2= 1$ and $X_0^2 + Y_0^2 + Z_0^2 = R^2_\perp$. 

For $R_\perp < R\subCGM$, we have $\Delta > 0$ and there are two points of intersection, one on each side the sky plane.  If $R_\perp \!=\! R\subCGM$, there is one intersection point tangent to the surface at $t_0 = 0$ (on the sky plane).  For $R_\perp > R\subCGM$, we have $\Delta <0$ and there is no intersection. When there are two points of intersection, we adopt the convention that $t_1 < t_2$. The gas structure is being probed by the LOS for all $t \in (t_1,t_2)$. This domain of $t$ always satisfies the trivial condition
\begin{equation}
    r(t) \leq R\subCGM \, .
\label{eq:insideshpere}
\end{equation}
where $r(t)$ is given by Eq.~\ref{eq:rlosoft}.  Note that Eq.~\ref{eq:insideshpere} is equivalent to the condition $f(t) \geq 0$.

\subsection{The Bi-Polar Wind}
\label{subsec:thehyperboloid}

Observations indicate clear trends between outflow and inflow velocities and galaxy inclination, star formation rate, and stellar mass that are consistent with bi-polar outflow and planar accretion spatial kinematic distributions
\citep[e.g.,][]{bordoloi11, bouche12, kacprzak12, rubin12, rubin14, chisolm15, nielsen15, rupke18, zhang18, roberts-borsani19, veilleux20, bizyaev22-outflows}.  Simulations are generally consistent with bi-polar outflows, as they predict that outflowing gas driven by stellar winds and/or supernovae is preferentially directed perpendicular to the disk \citep[e.g.,][]{nelson19, peroux20, trapp22, faucher-giguere23}. 

\begin{figure*}[htb]
\centering
\vglue -0.05in
\includegraphics[width=0.85\linewidth]{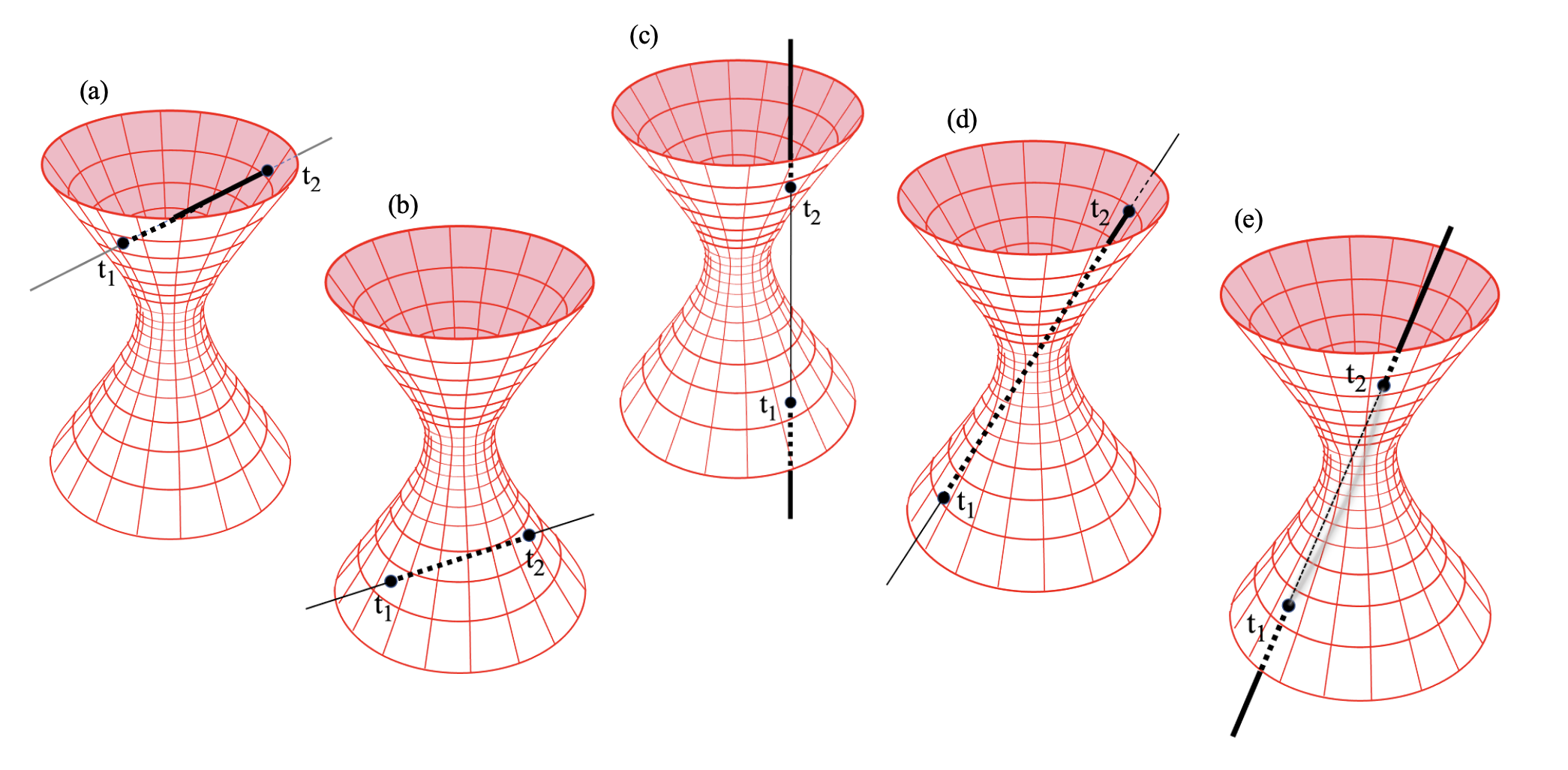}
\caption{\small Five scenarios for LOS intersections with the wind hyperboloid. Intersection points are shown as solid dots and are labeled $t_1$ and $t_2$. Thick lines represent LOS locations that reside inside the hyperboloid and thus probe the wind structure.  (a) The upper surface of the hyperboloid is intersected at two points. (b) The lower surface is intersected at two points. (c) Both lower and upper surfaces are intersected with the LOS exiting and then reentering the structure; this can occur for nearly face-on galaxies. (d) Both surfaces are intersected such that the LOS passes within the skirt radius, a scenario possible with small impact parameters. (e) A scenario similar to (c) for an oblique inclination. Not illustrated is that the hyperboloid is capped by $R_w = \eta_w R_{\rm vir}$. Multiple other scenarios can occur, such as the highly special case when the LOS intersects tangent to the structure surface at a single point, $t_1$.} 
\label{fig:intersections}
\end{figure*}

A cone or hyperboloid of one sheet serves as an appropriate geometry for modeling bi-polar outflowing winds \citep[e.g.,][]{bustard16, rupke18, zhang18, fielding22, nguyen22}. As concluded by \citet{rupke18} in their review of stellar driven galactic winds, ``The simple model of a modest-velocity, biconical flow of multiphase gas and dust perpendicular to galaxy disks continues to be a robust descriptor of these flows."  Indeed, an idealized cone model of the spatial distribution of bi-polar winds for interpreting quasar absorption line data has been explored and supported by several observational studies \citep[e.g.,][]{bouche12, chen14, schroetter19, zabl20, beckett23}.  

For a conical surface, assuming the $z$ axis is
the axis of symmetry, we have 
\begin{equation}
    \frac{X_w^2}{a^2} +
     \frac{Y_w^2}{b^2} -
      \frac{Z_w^2}{c^2} 
      = d \, ,
\end{equation}
where $d=0$ maps the upper and lower nappes of a cone and $d=1$ maps the continuous surface  of a hyperboloid of one sheet. For conic sections with circular cross sections, $b = a$. The inverse tangent of the ratio $a/c$ defines the opening angle, which we will hereafter call the wind opening angle, $\Theta_w$, where $\tan \Theta_w = a/c$. We have adopted the subscript ``$w$'' to denote ``wind.''

A shortcoming of a cone geometry is that there is a vertex at the origin;  the wind would thus originate from a single point at the galaxy center.  Studies of winds indicate that they are launched and entrain material over extended region of the galaxy disk \citep[see][]{nguyen22}.  An extended base can be modeled by adopting a hyperboloid of one sheet. The key difference between the hyperboloid ($d=1$) and the cone ($d=0$) is that the geometric length $a$ now defines the hyperboloid ``skirt radius,'' which is the radius of the circular cross section in the $z=0$ plane.  We will adopt the notation $\rho_{w,0}$ for the skirt radius and hereafter denote it as the base radius of the wind. 

The parametric equations for the surface of a hyperboloid of one sheet are written
\begin{equation}
\begin{array}{rcl}
X_w \!\!&=&\! \left(\rho_{w,0}^2 + z^2 \tan^2 \Theta_w \right) ^{1/2} \cos \phi \, , \\[3pt]
Y_w \!\!&=&\! \left(\rho_{w,0}^2 + z^2 \tan^2 \Theta_w \right) ^{1/2} \sin \phi \, ,\\[4pt]
Z_w \!\!&=&\! z \, .
\end{array}
\label{eq:hypersurface}
\end{equation}
In Eq.~\ref{eq:hypersurface}, a cone is trivially recovered from the hyperboloid when the base radius of the wind is nulled, i.e., $\rho_{w,0}=0$.  As written, Eq.~\ref{eq:hypersurface}, represents a hyperboloid of infinite height above and below the $z=0$ (galactic) plane. If we wish to give the ``wind" a finite extent we can apply a ``cap,'' or maximum wind radius, $R_w = \eta_w R_{\rm vir}$, where $\eta_w \in (0,1)$ would confine the wind structure to extend no further than the boundary of the CGM.  However, values of $\eta_w >1$ can be explored.

As with the spherical structure, the point(s) of intersection of the LOS with the surfaces of the cone or hyperboloid wind model are determined by equating the LOS position at $t$ with the surface points.  For the following, we first assume a structure of infinite extent and later apply constraints for a structure of finite extent. Equating Eq.~\ref{eq:hypersurface} and Eq.~\ref{eq:defineLOSPOS} yields, 
\begin{equation}
\begin{array}{rcl}
\left(\rho_{w,0}^2 + z^2 \tan^2 \Theta_w \right) ^{1/2}  \cos \phi \!\! & = & \! X_0 + \sigma_x t \, , \\[3pt]
\left(\rho_{w,0}^2 + z^2 \tan^2 \Theta_w \right) ^{1/2}  \sin \phi \!\! & = & \! Y_0 + \sigma_y t \, , \\[5pt]
z \!\! & = & \! Z_0 + \sigma_z t \, .
\end{array}
\end{equation}
We substitute $z$ to obtain
\begin{equation}
\begin{array}{rcl}
\left[\rho_{w,0}^2 + T^2_\Theta (Z_0\!+\! \sigma_z t) ^2 \right] ^{1/2}  \cos \phi \!\! & = & \! X_0 + \sigma_x t  \, , \\[4pt]
\left[\rho_{w,0}^2 + T^2_\Theta (Z_0\!+\! \sigma_z t) ^2 \right] ^{1/2} \sin \phi \!\! & = & \! Y_0 + \sigma_y t \, ,
\end{array}
\end{equation}
where we have introduced the notation $T_\Theta =\tan \Theta_w$.  We square both equations and add them.  After rearranging and collecting terms of $t^2$ and $t$, we obtain the quadratic equation given by Eq.~\ref{eq:quadeq} with geometric coefficients
\begin{equation}
\begin{array}{rcl}
A \!\!&=&\!  \sigma_z^2 T^2_\Theta - \sigma_x^2 - \sigma_y^2 \, ,  \\[3pt]
B \!\!&=&\! 2(\sigma_z Z_0 T^2_\Theta - \sigma_x X_0 - \sigma_y Y_0) \, , \\[3pt]
C \!\!&=&\!  \rho^2_{w,0}  + Z^2_0 T^2_\Theta  - (X^2_0 \!+\! Y^2_0) \, ,
\end{array}
\label{eq:ABC-wind}
\end{equation}
and solution given by Eq.~\ref{eq:quadsol}.  In Eq.~\ref{eq:quadsol}, if $\Delta < 0$ there is no intersection. If $\Delta = 0$ there is a single intersection point at $t_1=-B/2A$.  If $\Delta > 0$, then there are two points of intersection, $t_1$ and $t_2$.  As with the spherical model, we adopt $t_1 < t_2$. In Figure~\ref{fig:intersections}, we present multiple examples of possible LOS intersections with the wind structure. 

Alternatively, the LOS probes the wind structure wherever it resides within the wind surface and this can be established by testing for the condition that the axial distance of the LOS position, $\rho(t)= [x^2(t)+y^2(t)]^{1/2}$, is smaller than the axial distance of the wind surface, $\rho_w(t)$. Applying this condition yields 
\begin{equation}
\rho^2 (t) \leq  \rho^2_{w,0} + T^2_\Theta z^2(t)  \, .
\label{eq:equal-x2y2}
\end{equation}
If we substitute Eq.~\ref{eq:defineLOSPOS} into Eq.~\ref{eq:equal-x2y2}, we obtain a quadratic inequality of the form $f(t) \geq 0$, where $f(t)$ is given by Eq.~\ref{eq:quadeq} invoking Eq.~\ref{eq:ABC-wind}.  To account for wind structures truncated at a finite radial extent, if we cap the hyperboloid at $R_w = \eta_w R_{\rm vir}$, then the LOS is probing inside the wind structure when 
\begin{equation}
   f(t) \geq 0, \quad \hbox{and} \quad r(t) \leq R_w \, ,
\label{eq:hypercap}
\end{equation}
are simultaneously satisfied.

\subsection{The Galaxy Disk}
\label{subsec:thegalaxydisk}

For the galaxy disk, we adopt a cylindrical structure with axial radius $\rho_d$ and height $h_d$. The parametric equations for the surface of this cylinder are
\begin{equation}
    X_d = \rho_d \cos \phi \, , \quad
    Y_d = \rho_d \sin \phi \, , \quad
    Z_d = z \, ,
\label{eq:diskstructure}
\end{equation}
where we have adopted the subscript ``$d$'' to denote ``disk,'' and where $z \in [-h_d,+h_d]$. However, this would be a model component more appropriate for the ISM, rather than the CGM. 

To obtain the LOS locations probing the disk, we equate the $x$ and $y$ components of Eq.~\ref{eq:diskstructure} and Eq.~\ref{eq:defineLOSPOS}, and square and sum them to obtain 
\begin{equation}
    (X_0 + \sigma_x t)^2 + (Y_0 + \sigma_y t)^2  = \rho_d^2  \, ,
\label{eq:diskmodel}
\end{equation}
which yields the quadratic equation $f(t)$ given by Eq.~\ref{eq:quadeq}, with geometric coefficients
\begin{equation}
\begin{array}{rcl}
A \!\!&=&\!  \sigma_x^2 + \sigma_y^2 \, , \\[3pt]
B \!\!&=&\! 2(\sigma_x X_0  + \sigma_y Y_0 ) \, , \\[3pt]
C \!\!&=&\!  X^2_0 + Y^2_0 - \rho_d^2 \, ,
\end{array}
\label{eq:ABC-disk}
\end{equation}
and solution given by Eq.~\ref{eq:quadsol}. Since the disk is given a finite height, $h_d$, we find that  the LOS is probing the disk when the conditions
\begin{equation}
f(t) \leq 0 \quad \hbox{and} \quad |z(t)| \leq h_d \, ,
 \label{eq:diskinside}
\end{equation}
are simultaneously satisfied.

Disk warping can be approximated using a simple flare model, such as the one adopted by \citet{kalberla08} for the Milky Way,  
\begin{equation}
h_d(\rho) = h_d(0) \exp \{ (\rho\!-\! \rho_{f,0})/\rho_f \}  \, , 
\label{eq:diskflare}
\end{equation}
valid for $\rho \in (\rho_{f,0}, \rho_d)$, where $\rho_f$ is the exponential flare scale length, and $\rho_{f,0}$ is the axial radius at which the flare begins.  For the Milky Way, \citet{kalberla08} adopted  $h_d(0) = 0.15$~kpc, $\rho_f = 9.8$~kpc, and $\rho_{f,0} = 5$~kpc. For this augmentation to the disk model, the region where the LOS probes the disk is still given by Eq~\ref{eq:diskinside} but with $h_d(t)$ replacing the constant $h_d$.  In this work, we will not treat the flaring of the disk using Eq.~\ref{eq:diskflare}, but will adopt a separate CGM structure.

\subsection{Extended Planar Accretion}
\label{subsec:theflareddisk}

Theoretically, planar accretion is predicted to be a significant component to the building of galaxy disks \citep[e.g.,][]{keres05, stewart11, stewart13, stewart17, stewart-proc17, vandevoort11, hafen22, trapp22, gurvich23, kocjan24, stern23}.  In simulations, \citet{stewart11, stewart13} reported the formation of a transient ``cold flow disk" structure around their galaxies. These massive extended planar structures were often warped, not rotationally supported, and comprise inflowing cold halo gas.  Most interestingly, these structures are kinematically aligned with the angular momentum of the central galaxy rather than with the angular momentum of the inflowing cold filamentary gas.  

Subsequent theoretical work using hydrodynamic cosmological simulations has confirmed that the infalling material spirals inward, and as doing so, tends to align with the plane of the galactic disk (see Fig.~2 of \citealp{hafen22} and Fig.~1 of \citealp{stern23}).  \citet[][see their Fig.~7]{trapp22} shows that this accretion fans inward over a flare angle ranging from $15^\circ$--$30^\circ$.  According to \citet[][]{hafen22}, \citet[][]{trapp22}, and \citet[][]{stern23} rotational support occurs within a narrow annulus at the galaxy disk edge. 

As such, in terms of an idealized spatial structure, an extended planar accretion CGM component can be modeled as a flared extension to the galactic disk. Geometrically, this simple spatial structure should exhibit azimuthal symmetry and have a vertical height that increases with axial distance from the galaxy center. A modified hyperboloid of one sheet is well suited. In addition to having an azimuthal symmetry about the galaxy rotation axis, the hyperboloid skirt radius can be interpreted as the axial distance at which the accretion interfaces to the galactic disk. 

To model the expected spatial location of the extended planar accreting gas, we modify Eq.~\ref{eq:hypersurface} by replacing the wind opening angle $\Theta_w$ with the complementary angle $\Theta_a$, which we call the accretion opening angle, or ``flare" angle.  
As illustrated schematically in Figure~\ref{fig:structure}, this angle defines the flare of the extended accretion disk.  The surface of the flared disk is then  written, 
\begin{equation}
\begin{array}{rcl}
X_a \!\!&=&\!  \left(\rho_{a,0}^2 + z^2 \cot^2 \Theta_a \right) ^{1/2} \cos \phi \, , \\[3pt]
Y_a \!\!&=&\! \left(\rho_{a,0}^2 + z^2 \cot^2 \Theta_a \right) ^{1/2}  \sin \phi \, , \\[5pt]
Z_a \!\!&=&\! z \, .
\end{array}
\label{eq:planarsurface}
\end{equation}
where we have adopted the subscript ``$a$'' to denote ``accretion,'' and where we have written the skirt radius as $\rho_{a,0}$, which we call the ``accretion radius." 

To obtain the LOS intersection points for the extended planar accretion structure, we following the steps for the wind hyperboloid; however, note that the ``inside" of the accretion structure is the technically the void of the hyperboloid solid geometry.  We equate Eq.~\ref{eq:planarsurface} and Eq.~\ref{eq:defineLOSPOS} and derive the quadratic equation given by Eq.~\ref{eq:quadeq} with solution given by Eq.~\ref{eq:quadsol} with geometric coefficients
\begin{equation}
\begin{array}{rcl}
A \!\!&=&\!  \sigma_z^2 C^2_\Theta - \sigma_x^2 - \sigma_y^2  \, , \\[3pt]
B \!\!&=&\! 2(\sigma_z Z_0 C^2_\Theta - \sigma_x X_0 - \sigma_y Y_0) \, , \\[3pt]
C \!\!&=&\!  \rho^2_{a,0}  + Z^2_0 C^2_\Theta  - (X^2_0 \!+\! Y^2_0) \, ,
\end{array}
\label{eq:ABC-planar}
\end{equation}
where we have introduced the notation $C_\Theta =\cot \Theta_a$. As with the wind hyperboloid, the value of $\Delta$ in Eq.~\ref{eq:quadsol} dictates the number of LOS intersection points.

The LOS probes the planar material wherever the axial distance of the LOS position, $\rho(t)$, is {\it greater\/} than the axial distance of the accretion hyperboloid surface, giving
\begin{equation}
\rho^2(t) \geq \rho^2_{a,0} + C^2_\Theta z^2(t) \, .
\label{eq:equal-x2y2-planar}
\end{equation}
If we substitute Eq.~\ref{eq:defineLOSPOS} into Eq.~\ref{eq:equal-x2y2-planar}, we obtain a quadratic inequality of the form $f(t) \leq 0$, where $f(t)$ is given by Eq.~\ref{eq:quadeq} using Eq.~\ref{eq:ABC-planar}. To account for the maximal extent of the planar accretion structure, we adopt $R_a = \eta_a R_{\rm vir}$. Therefore, the LOS is probing the accretion when 
\begin{equation}
   f(t) \leq 0, \quad \hbox{and} \quad r(t) \leq R_a \, ,
\label{eq:hyper2cap}
\end{equation}
are simultaneously satisfied.

\begin{figure}[h!tb]
\centering
\vglue -0.05in
\includegraphics[width=0.75\linewidth]{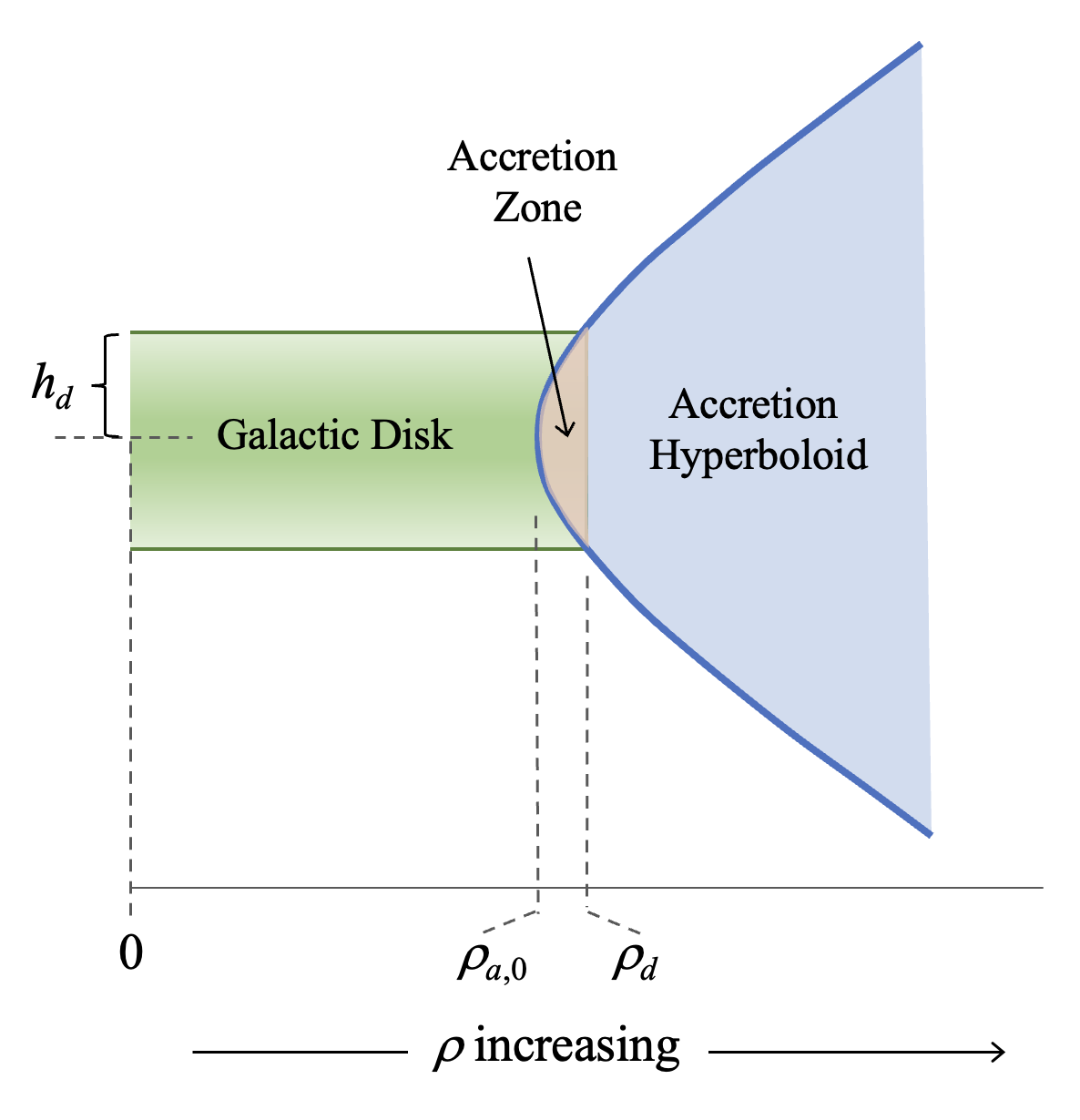}
\caption{\small A schematic cross section of the spatial relationship between the galaxy disk (shaded green; a cylinder of axial radius $\rho_d$ and height $h_d$ above the galactic plane) and the extended planar accretion structure (shaded blue; a hyperboloid of one sheet with skirt radius $\rho_{a,0}$ and accretion opening angle $\Theta_a$).  Theory suggests that accretion transpires at the edge of the disk over a very narrow axial radius \citep[e.g.,][]{hafen22,trapp22}.  Application of Eq.~\ref{eq:makeAzone}, in which the hyperboloid surface intersects the disk at height $z = \pm h_d$ creates such a region of intersection, which we call the ``accretion zone" (shaded pink).}
\label{fig:Azone}
\end{figure}

One might adopt $\rho_{a,0} \simeq \rho_d$ so that the accretion radius exactly coincides with the disk ``edge."   However, theory suggests that planar accretion occurs over a finite annular region overlapping the disk edge \citep[e.g.,][]{hafen22, trapp22, gurvich23, stern23}. A smooth transition between the cylindrical disk and accretion hyperboloid  structures that naturally yields an annular finite region of overlap can be obtained by applying the condition that the height of the hyperboloid structure matches the disk height, $h_d$ at the disk edge, i.e.,
\begin{equation}
    \rho_{a,0} = \sqrt{ \rho_d^2 - 
    h_d^2 C^2_\Theta }\, .
\label{eq:makeAzone}
\end{equation}

In Figure~\ref{fig:Azone}, we show a schematic of the cross section of the galactic disk and the accretion hyperboloid when Eq.~\ref{eq:makeAzone} is satisfied.  Note that this configuration is constrained by the geometric criterion  $\rho_d > h_d C_\Theta$.  We call the overlap region the ``accretion zone" and interpret it as the range of axial radial distances over which the kinematics of the planar accreting gas transitions to the galactic disk. We will describe these kinematics further in Section~\ref{sec:XPAkinematics}.

\begin{figure*}[h!tb]
\centering
\includegraphics[width=0.85\linewidth]{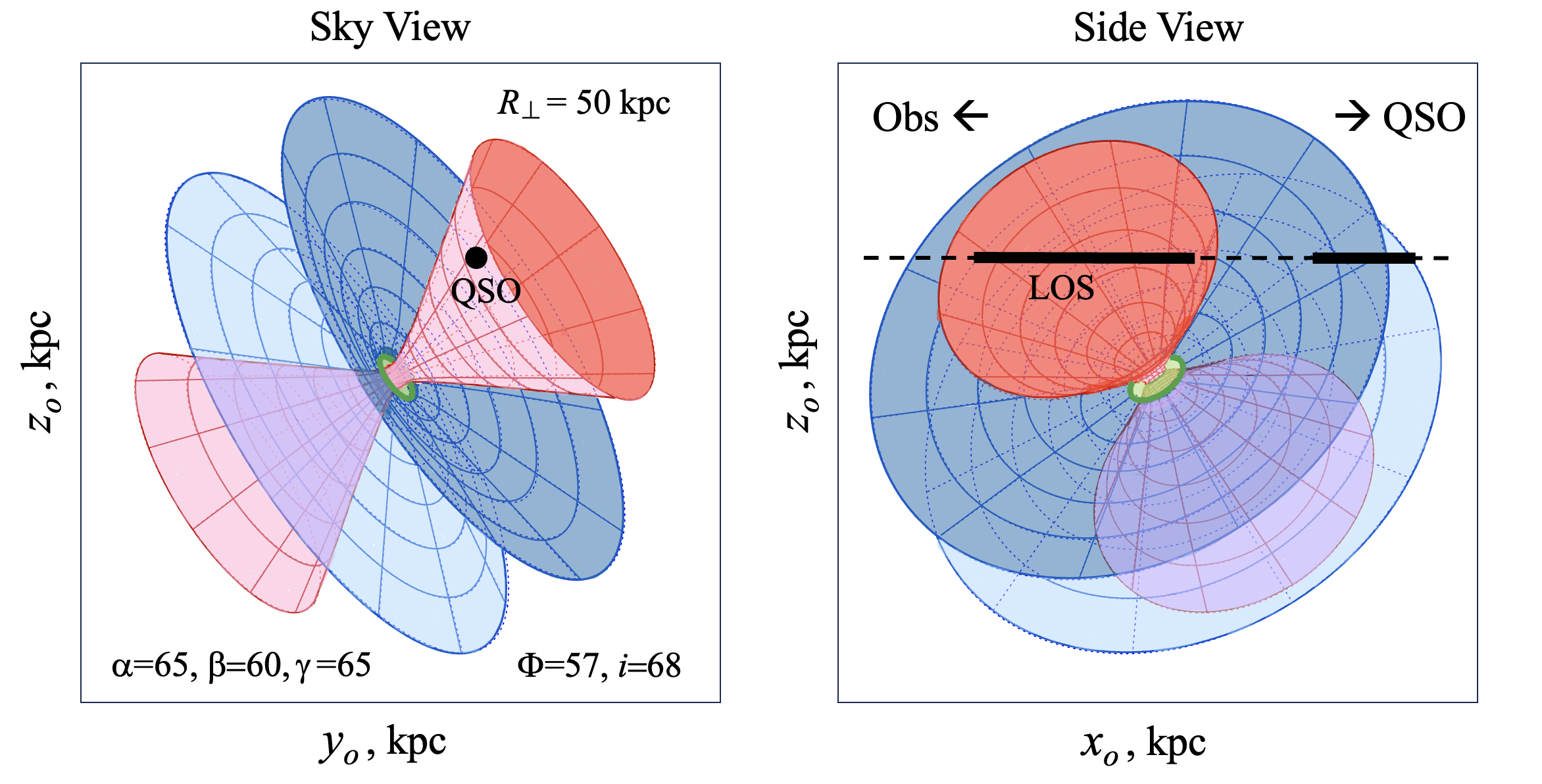}
\caption{\small A schematic of a spatial-kinematic model and LOS in the observer frame, {\Osys}.  The galaxy is obliquely inclined with $\alpha,\beta = 65^\circ, 60^\circ$, which yields inclination $i=68^\circ$. The quasar is placed at $R_\perp = 50$~kpc  with position angle $\gamma = 65^\circ$, which yields $\Phi = 57^\circ$.  The wind parameters are $\rho_{w,0}=10$~kpc, $\Theta_w=40^\circ$, and $R_w=R_{\rm vir}$, where $R_{\rm vir}= 200$~kpc. The accretion parameters are $\rho_{a,0}= 23$~kpc, $\Theta_w=20^\circ$, and $R_a=R_{\rm vir}$. The disk parameters are $\rho_d=25$~kpc, and $h_d=5$~kpc. (left) The sky plane, showing the observer perspective, where the quasar LOS is eclipsed by the upper surface of the wind hyperboloid on the near side of the galaxy and by the accretion hyperboloid on the far side of the galaxy. (right) The side view showing where LOS is probing the wind and accretion structure, as shown by the thicker portion of the LOS. Recall that the observer is located at $x_0=\infty$, whereas the quasar is located at $x_0=-\infty$ and the LOS vector direction is ${\shat}=-{\ihato}$. 
}
\vglue -0.1in
\label{fig:LOSskyview}
\end{figure*}

\begin{table*}[bth]
\centering
\caption{Summary of Spatial Components and LOS Probing Conditions}
\begin{tabular}{lllll}
\hline\\[-8pt]
Spatial & Geometric & Structural & Geometric          & LOS Probing  \\
Component     &  Structure   &  Parameters          &  Coefficients    & Conditions$^{\rm a}$ \\[2pt]
\hline\\[-8pt]
halo & sphere          &  $R\subCGM$  & Eq.~\ref{eq:ABC-halo} & $f(t) \geq 0$; $r(t) \leq R\subCGM $\\
disk & cylinder & $\rho_d; h_d$ & Eq.~\ref{eq:ABC-disk} & $f(t) \leq 0$; $|z(t)| \leq h_d$ \\
wind & hyperboloid &   $\Theta_w; \rho_{w,0}; R_w$ & Eq.~\ref{eq:ABC-wind} & $f(t) \geq 0$; $r(t) \leq R_w $  \\
accretion & hyperboloid &   $\Theta_a; \rho_{a,0}; R_a$  & Eq.~\ref{eq:ABC-planar} & $f(t) \leq 0$; $r(t) \leq R_a $\\[2pt]
\hline \\[-8pt]
\multicolumn{5}{l}{(a) The function $f(t)=At^2+Bt+C$ (given by Eq.~\ref{eq:quadeq}) is evaluated using the corresponding} \\
\multicolumn{5}{l}{\phantom{(a)} geometric coefficients.} \\ \\
\end{tabular}
\label{tab:downtocases}
\end{table*}

\subsection{Where the LOS Goes to Probe}
\label{sec:downtocases}

In Figure~\ref{fig:LOSskyview}, we present an example of a LOS through an obliquely inclined galaxy ($\alpha,\beta = 65^\circ, 60^\circ$) in the observer frame of reference.  This galaxy has inclination $i=68^\circ$ and the quasar is located at $R_\perp=50$~kpc with azimuthal angle $\Phi = 57^\circ$.  The upper surface of the wind hyperboloid is located on the observer side of the plane of the sky, and is probed by the LOS, whereas the lower surface is on the far side of the galaxy. Also probed is the accretion hyperboloid (in the void between the upper and lower surfaces) on the far side of the galaxy.  

In Table~\ref{tab:downtocases}, we list the four idealized CGM components and their geometric representations.  We also list the free parameters that define these geometric structures.  The equations for solving the LOS intersection positions are tabulated, as are the conditions for determining for what values of $t$ the LOS is probing the structures. Finally, if a maximum extent is being applied to cap the structures, this condition for determining when the LOS is probing within the structure is listed.  

Given the complexity of the multiple permutations by which the LOS can pass through the wind and planar accretion structures, the most straight-forward strategy is to scan along the LOS from some $t_{\rm min}$ to some $t_{\rm max}$ and track the conditions within the structures as outlined in Table~\ref{tab:downtocases}. To ensure that all structures are bracketed by $t \in (t_{\rm min},t_{\rm max})$, we can apply Eq.~\ref{eq:quadeq} with Eq.~\ref{eq:ABC-halo} while replacing $R\subCGM$ with the maximal radial extent of the structures being modeled. 

Note that, with this scanning approach, it is not required that the points of intersection are computed in order to determine where the LOS enters and exits the galaxy/CGM structures. However, the intersection points can provide insight into the overall geometric relationship between the galaxy and the LOS ingress and egress locations for the various structures.  This scanning approach is appropriate.  Once the velocity fields are populated in the structures, the scanning allows LOS velocities to be computed simultaneously at each LOS position. 

\section{Velocity Fields and LOS Velocities}
\label{sec:kinematics}

The motions of the gas associated with given dynamical structures, such as galactic winds, extended galactic disks, intergalactic infall and accretion, recycling gas, and galactic fountains, etc., can be modeled by simple velocity fields in which the velocity vector of the material is defined over a range of spatial regions.  We define a velocity field as the vector field ${\hbox{\bf V}}({\rvec})$ at spatial location ${\rvec}$ in coordinate system {\Gsys}.  At LOS location $t$, the LOS velocity is the dot product of the velocity vector  at ${\rvec}={\rvec}(t)$ and the LOS unit vector,
\begin{equation}
 V\subLOS(t) = 
 {\hbox{\bf V}}(t) \cdot {\shat}  \, ,
\label{eq:vlosdotprod}
\end{equation}
where the LOS unit vector is ${\shat} = \sigma_x {\ihat} + \sigma_y {\jhat} + \sigma_z {\khat}$ (see Eq.~\ref{eq:shat-derived}) and ${\rvec}(t) = x(t) {\ihat} + y(t) {\jhat} + z(t) {\khat}$, where $x(t)$, $y(t)$, and $z(t)$ are given by Eq.~\ref{eq:lospos}. Note that the LOS velocity is negative when it is directed toward the observer and is positive when directed away from the observer.

Any velocity field can be defined in terms of its principle velocity components. The principle components are those parallel to the principle unit vectors in the coordinate system. Once defined, the principle components can be flexibly combined via vector addition to describe general velocity fields applicable to the kinematics of the CGM. Here, we derive the LOS velocities for the principle components of an arbitrary velocity field.  We consider radially directed velocities, ${\hbox{\bf V}}_r({\rvec}) = V_r({\rvec})\,{\rhat}({\rvec})$, azimuthal velocities, ${\hbox{\bf V}}_\upphi({\rvec}) = V_\upphi({\rvec})\,{\phihat}({\rvec})$, polar orbital velocities, ${\hbox{\bf V}}_\uptheta({\rvec}) = V_\uptheta ({\rvec})\,{\thetahat}({\rvec})$,  vertical velocities, ${\hbox{\bf V}}_z({\rvec}) = V_z({\rvec}) \,{\khat}$, and axial flow, ${\hbox{\bf V}}_\uprho({\rvec}) = V_\uprho({\rvec})\, {\rhohat}({\rvec})$.  We present convenient multiplicative ``LOS projection functions" for each principle component.

\subsection{Radial Velocities}

If we assume a purely radial directed velocity, $V_r({\rvec})$, we have ${\hbox{\bf V}}_r(t) = V_r(t){\,\rhat}(t)$, where the unit radial vector at LOS position $t$ is 
\begin{equation}
{\rhat}(t) = \frac{{\bf r}(t)}{r(t)} = 
\frac{x(t)}{r(t)} \,{\ihat} + 
\frac{y(t)}{r(t)} \, {\jhat} + 
\frac{z(t)}{r(t)} \, {\khat} \, ,
\label{eq:rhat}
\end{equation}
where $r(t) = [x^2(t) + y^2(t) + z^2(t)]^{1/2}$. We define the LOS projection function as $\mathcal{P}_r (t) ={\rhat}(t)\cdot {\shat}$. Recalling that ${\shat} = \sigma_x {\ihat} + \sigma_y {\jhat} + \sigma_z {\khat}$, we have
\begin{equation}
\mathcal{P}_r (t) 
= \frac{x(t)}{r(t)} \sigma_x
+ \frac{y(t)}{r(t)} \sigma_y 
+ \frac{z(t)}{r(t)} \sigma_z
 \, .
\label{eq:rdots}
\end{equation}
The observed LOS velocity of radially directed velocity at LOS position $t$ is then
\begin{equation}
 V^{(r)}\subLOS(t) = V_r(t) \mathcal{P}_r (t) \, .
\label{eqLVlosrad}
\end{equation}

Eq.~\ref{eqLVlosrad} can be employed to model constant velocity spherical outflow or infall.  If we wish to model outflow, we adopt $V_r({\rvec}) >0$. If we wish to model inflow, we adopt $V_r({\rvec}) <0$.  

\subsection{Azimuthal/Circular Velocities}

For an azimuthal/circular velocity, $V_\upphi({\rvec})$, we have ${\hbox{\bf V}}_\upphi(t) = V_\upphi(t)\,{\phihat}(t)$, where the unit azimuthal rotation vector at LOS position $t$ is 
\begin{equation}
{\phihat}(t) = 
\frac{x(t)}{\rho(t)} \, {\jhat} 
- \frac{y(t)}{\rho(t)} \, {\ihat} \, ,
\label{eq:phihat}
\end{equation}
and where the axial radius is $\rho(t) = [x^2(t) + y^2(t)]^{1/2}$.  The LOS projection function of the unit azimuthal rotation vector is $\mathcal{P}_{\upphi} (t) = {\phihat}(t)\cdot {\shat}$, which is written
\begin{equation}
\mathcal{P}_{\upphi} (t)  
= \frac{x(t)}{\rho(t)} \sigma_y 
- \frac{y(t)}{\rho(t)} \sigma_x 
  \, .
\label{eq:phidots}
\end{equation}
The observed LOS velocity of a circular velocity at LOS position $t$ is then 
\begin{equation}
 V^{(\upphi)}\subLOS(t) = V_\upphi(t) \mathcal{P}_\upphi (t) \, .
\label{eq:Vloscirc}
\end{equation}
A simple example would be a flat rotation curve, which can be obtained for ``solid body'' rotation for which we would adopt a constant $V_\upphi({\rvec})=V_\upphi$.

\begin{table*}[htb]
\centering
\caption{Summary of LOS Velocity Projection Functions}
\begin{tabular}{llll}
\hline\\[-8pt]
Velocity & Vector & Reference & Projection Function, ${\cal P}_k(t)= {\qhat} (t) \! \cdot {\shat}$ \\
Component & Direction & Equation \\[2pt]
\hline\\[-8pt]
vertical & ${\khat}$ &Eq.~\ref{eq:zdots}& ${\cal P}_z = \sigma_z$\\[-2pt]
radial & ${\rhat}({\rvec})$ &Eq.~\ref{eq:rdots}& ${\cal P}_r(t) = \sigma_x[x(t)/r(t)]+\sigma_y[y(t)/r(t)]+\sigma_z[z(t)/r(t)]$ \\
azimuthal & ${\phihat}({\rvec})$ &Eq.~\ref{eq:phidots}& ${\cal P}_\upphi(t) = \sigma_y[x(t)/\rho(t)]-\sigma_x[y(t)/\rho(t)]$ \\
axial & ${\rhohat}({\rvec})$  &Eq.~\ref{eq:rhodots}& ${\cal P}_\uprho(t) = \sigma_x[x(t)/\rho(t)]+\sigma_y[y(t)/\rho(t)]$ \\
polar & ${\thetahat}({\rvec})$   &Eq.~\ref{eq:thetadots}&  ${\cal P}_\uptheta(t) = [z(t)/r(t)] {\cal P}_\uprho(t) - \sigma_z [\rho(t)/r(t)]$ \\[2pt]\hline \\
\end{tabular}
\label{tab:vlosprojections}
\end{table*}

\subsection{Vertical and Axial Flow Velocities}

For a purely vertical velocities, $V_z({\rvec})$, for example perpendicular to the galaxy disk, we have ${\hbox{\bf V}}_z (t) = V_z(t)\,{\khat}$. The LOS projection function is 
$\mathcal{P}_x (t) = {\khat}\!\cdot {\shat}$, which yields
\begin{equation}
\mathcal{P}_z  = \sigma_z  \, ,
\quad 
 V^{(z)}\subLOS(t) = V_z (t) \mathcal{P}_z 
 \, .
\label{eq:zdots}
\end{equation}
Note that the vector projection function is independent of LOS position.  We also trivially find that
\begin{equation}
\begin{array}{rcl}
\mathcal{P}_x  &=& \sigma_x  \, ,
\quad 
 V^{(x)}\subLOS(t) = V_x (t) \mathcal{P}_x 
 \, , \\[4pt]
\mathcal{P}_y  &=& \sigma_y  \, ,
\quad 
 V^{(y)}\subLOS(t) = V_y (t) \mathcal{P}_y 
 \, .
\end{array}
\label{eq:xdots-ydots}
\end{equation}

For a purely axial velocity flow, $V_\uprho({\rvec})$, we have ${\hbox{\bf V}}_\uprho (t) = V_\uprho(t){\,\rhohat}(t)$, where
the unit axial radius vector at LOS position $t$ is 
\begin{equation}
{\rhohat}(t) = 
\frac{x(t)}{\rho(t)} \, {\ihat} + 
\frac{y(t)}{\rho(t)} \, {\jhat} \, .
\label{eq:rhohat}
\end{equation}
This velocity is pointing directly outward from the polar axis (is divergent) if $V_\uprho ({\rvec})>0$ and directly toward the polar axis (is convergent) if $V_\uprho ({\rvec}) <0$. The LOS projection function is 
$\mathcal{P}_\uprho (t) = {\rhohat}(t)\cdot {\shat}$, is written
\begin{equation}
\mathcal{P}_\uprho (t)  
= \frac{x(t)}{\rho(t)} \sigma_x 
+ \frac{y(t)}{\rho(t)} \sigma_y
 \, ,
\label{eq:rhodots}
\end{equation}
from which we obtain 
\begin{equation}
 V^{(\uprho)}\subLOS(t) = V_\uprho (t) \mathcal{P}_\uprho (t) \, .
\label{eq:Vlosacc}
\end{equation}

\subsection{Polar Orbital Velocities}

If we assume a purely polar velocity, $V_\uptheta ({\rvec})$, we have ${\hbox{\bf V}}_\uptheta (t) = V_\uptheta(t)\,{\thetahat}(t)$, where the unit polar orbit vector at LOS position $t$ is 
\begin{equation}
{\thetahat}(t) =
 \frac{z(t)}{r(t)} \, {\rhohat}(t) - \frac{\rho(t)}{r(t)} \, \, {\khat} 
\label{eq:thetahat}
\end{equation}
The LOS projection function is $\mathcal{P}_\uptheta (t) = {\thetahat}(t)\cdot {\shat}$, which is written
\begin{equation}
\mathcal{P}_\uptheta (t) = 
\frac{z(t)}{r(t)} \mathcal{P}_\uprho (t)  
- \sigma_z \frac{\rho(t)}{r(t)} \, .
\label{eq:thetadots}
\end{equation}
The observed LOS velocity of a polar orbital velocity at LOS position $t$ is then 
\begin{equation}
 V^{(\uptheta)}\subLOS(t) = V_\uptheta (t) \mathcal{P}_\uptheta (t) \, .
\label{eq:Vlostheta}
\end{equation}

\subsection{The Full Velocity Field}

Having expressed the principle velocity vectors and their LOS projection functions, we can construct an arbitrary velocity vector field and determine its LOS velocity projection at any location along the LOS. Consider a velocity field at location ${\rvec}$ with principle velocity components $V_k({\rvec}) {\qhat} ({\rvec})$, where $k$ represents a principle component and ${\qhat}({\rvec})$ represents its unit vector direction at ${\rvec}$.
The velocity vector field is written
\begin{equation}
 {\bf V}({\rvec}) =  \sum_k V_k({\rvec}) {\qhat} ({\rvec})\, .
 \label{eq:V-general}
\end{equation}
At LOS location ${\rvec}={\rvec}(t)$, the LOS projected velocity is
$V\subLOS(t) =  {\bf V}(t) \! \cdot {\shat}$, which is
\begin{equation}
\begin{array}{rcl}
 V\subLOS(t) = \displaystyle  \sum _k V_k(t) [{\qhat} (t) \! \cdot {\shat}] = \sum _k V_k(t) \mathcal{P}_k(t) \, ,
\end{array}
  \label{eq:Vlos-general}
\end{equation}
where $\mathcal{P}_k(t)$ is the LOS projection function for principle component $k$. In Table~\ref{tab:vlosprojections}, we list the principle component projection functions. As we will describe below, unique velocity fields can be spatially ``confined'' to the various CGM spatial structures using the constraints listed in Table~\ref{tab:downtocases}.

\section{Idealized CGM Kinematics}
\label{sec:skmodels}

In this section, we describe idealized kinematics for the four spatial galaxy/CGM components.  

\subsection{The Spherical Halo}

Though the halo is modeled as a sphere, it can be argued there is no single velocity field that applies throughout the spherical space.  By design, portions of this volume are occupied by the wind and the accretion structures and these two spatial regions are expected to be dominated by outflow and inflow, respectively. As such, we could adopt a working definition that halo kinematics, per se, applies within the volume {\it not\/} occupied by the disk, wind, and accretion structures.  In other words, the LOS probing conditions in Table~\ref{tab:downtocases} are simultaneously false for the disk, wind, and accretion while the condition for the halo is true (most simply, where $r(t) \leq R\subCGM$).

This halo region can be kinematically tangled due to added complexities such as high-velocity clouds \citep[][]{richter15, marasco22}, tidal streams, such as the Galactic Magellanic Stream \citep[e.g.,][]{d-onghia116}, and polar rings \citep[e.g.,][]{maccio06, brook08, spavone10, tepper-garcia15}, all of which  are likely to have a prominent polar component to their kinematics. 
Other models characterize the halos as ``cloudy with a chance of rain" and invoke accretion braking of cold clouds that results in terminal infall velocities \citep[e.g.,][]{benjamin97, tan23}. 

A spherically symmetric spatial-kinematic model would have the form  ${\bf V}_h({\rvec}) = V_r({\rvec})\,{\rhat}({\rvec})$, with $V_r({\rvec})$ being positive everywhere for outflow or negative everywhere for infall \citep[e.g.,][]{lanzetta92}. For pure radial kinematics, Eq.~\ref{eq:Vlos-general} simplifies to Eq.~\ref{eqLVlosrad} and we obtain $V\subLOS ^{(h)}(t) = V_r(t) \mathcal{P}_r(t)$ for the LOS velocity. Alternately, the radial velocity $V_r({\rvec})$ could have a functional form such as $V_r(r) = V_0 (r/r_0)^\epsilon$ for a simple radial dependency, where $\epsilon$ is a free parameter.  

Varieties of radial dependent infall and outflow velocities can be explored.  For inflows, \citet{stern23}, for example, developed an analytical model (see their Eq.~11) to estimate the density, temperature, pressure, sound speed, velocity, and Mach number for a radiative cooling adiabatic ideal gas in a gravitational potential.  For outflows, \citet{chisolm15}, for example, observationally determined that outflow velocities scale as $V_r = V_c^\epsilon$ with $\epsilon = 0.44$--$0.87$.  We also note that outflow or infall may decelerate.  To approximate deceleration, one could invoke a ``stall" or ``breaking function,'' $\mathcal{S}(r)$, which we introduce in Section~\ref{subsubsec:deceleration-function}. 

Another possibility is that the spherical CGM has a rotational component \citep[e.g.,][]{weisheit78}, in which case the velocity field is ${\bf V}_h({\rvec}) = {\bf V}_r({\rvec})+{\bf V}_\upphi({\rvec})$, which yields LOS velocity $V\subLOS ^{(h)} (t) = V_r(t) \mathcal{P}_r(t) + V_\upphi(t) \mathcal{P}_\upphi(t)$, where we note that the rotational velocity vector, ${\bf V}_\upphi({\rvec})$, can also be a function of radius or height above the disk, etc.

Polar orbits would be modeled as,  ${\bf V}_h({\rvec}) = {\bf V}_\uptheta({\rvec})$. For a predominantly polar orbit with a rotational component, we would write ${\bf V}_h({\rvec}) = {\bf V}_\uptheta({\rvec}) +{\bf V}_\upphi({\rvec})$ with $|V_\uptheta| > |V_\upphi|$, which would represent a precessing polar orbit.  One could invoke a predominantly polar component with a radial component, ${\bf V}_h({\rvec}) = {\bf V}_\uptheta({\rvec}) +{\bf V}_r({\rvec})$ with $|V_\uptheta| > |V_r|$,, which could represent slowly infalling polar orbit. Finally, we can write a general case for which  ${\bf V}_h({\rvec}) = {\bf V}_\uptheta({\rvec}) + {\bf V}_\upphi({\rvec}) + {\bf V}_r({\rvec})$.  

Tidal streams are typically not distributed spherically about a galaxy.  A modification of the full spherical spatial structure to form a segment of a spherical shell to model such streams would be straight forward using the formalism of Section~\ref{sec:geometries}, which is designed for flexibility. For example, the surface of a torus is easily computed and has well-behaved closed form solution for the LOS intersection points  \citep[see][]{skala23}.

\subsubsection{Wind Stalling and Accretion Breaking}
\label{subsubsec:deceleration-function}

Whether we are modeling spherical radial outflow or bi-polar outflows confined within cones or hyperboloids, we might expect the gas to decelerate. The wind models of \citet{fielding22}, for example, predict that hot galactic winds quickly become supersonic and then, under certain conditions, can stall.  Their models predict that in galaxies with $V_c \simeq 150$~{\kms} and mass loading factors of $M_w/M_\ast \simeq 1$--3, the wind velocity is on the order of a 200--300~{\kms}, and when subject to cooling and gravity, the winds can stall out. The stall heights can vary and the stall can occur over distance scales as small as few percent of the height of the wind material.  

\begin{figure}[htb]
\centering
\includegraphics[width=0.95\linewidth]{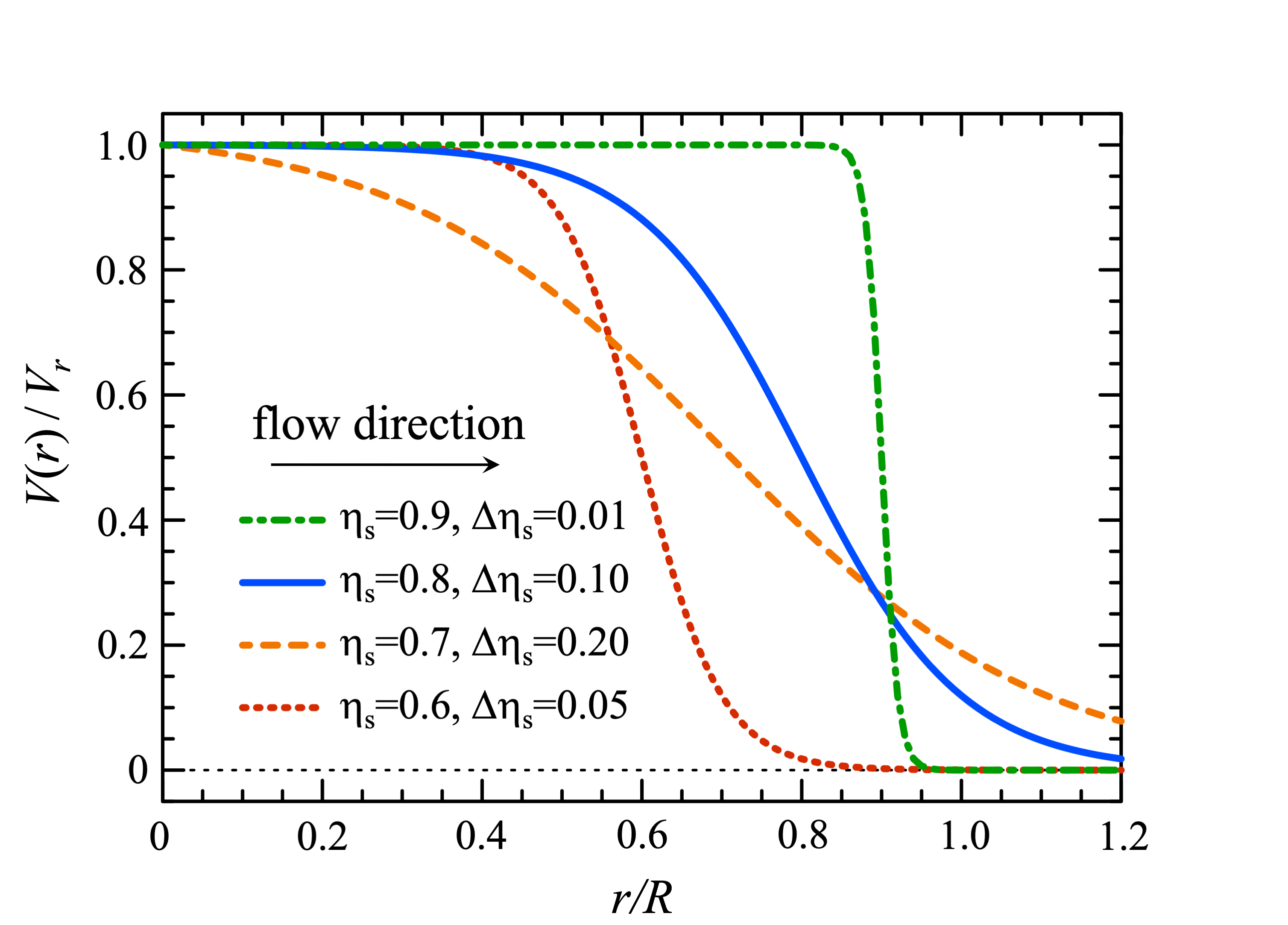}
\caption{\small The outflow stall function given by Eq.~\ref{eq:fermi-dirac} for four parameter combinations of the stall radius, $\eta_s$ and stall ratio, $\Delta \eta_s/\eta_s$. Application of the scenario $\eta_s=0.8$ and $\Delta\eta_s=0.1$ for the wind is shown in the cross sectional illustration in Figure~\ref{fig:vwind}(b).  The curves shown here can be compared to the ``cooling + gravity" wind velocity curves in Fig.~2 of \citet[][]{fielding22}.  }
\label{fig:stallfunc}
\end{figure}

A physics-based expression characterizing decelerating wind gas is beyond the scope of this work.  What we desire is a flexible function that can model a range of deceleration, either a smooth continuous deceleration as the wind height increases or a stall (rapid deceleration) at arbitrary height above the galactic plane. A normalized modified Fermi-Dirac function provides such flexibility. We introduce the stall function,
\begin{equation}
 {\cal S}(r; \eta_s,\Delta \eta_s) = \frac
 {1+ \exp\left\{ -\eta_s R / \Delta\eta_s\right\}}
 {1+ \exp\left\{ (r-\eta_s R)/\Delta\eta_s \right\}}  \, ,
\label{eq:fermi-dirac}
\end{equation}
where $R$ is the maximum extent of the spatial structure and $\eta_s$ defines the ``stall height,'' $\eta_s R$, at which the outflow velocity has decreased to $\sim\! 50$\% of its launch velocity.  The parameter $\Delta\eta_s$ defines the ``stall ratio,'' $\Delta \eta_s/\eta_s$.   The smaller the stall ratio, the more rapid the ``stall.'' As long as the ratio $\Delta\eta_s/\eta_s \leq 0.2$, then at height $(\eta_s \!-\! \Delta\eta_s) R $ the wind velocity has decreased to $\sim\! 75$\% of its launch value, and for $(\eta_s \!+\! \Delta\eta_s) R $ it has decreased to $\sim\! 25$\%.  

In Figure~\ref{fig:stallfunc}, we show four scenarios of the stall function for outflowing gas with constant radial velocity $V_r$. As the distance of the outflowing material increases, the velocity decreases.  For example, consider $\eta_s=0.8$ and $\Delta\eta_s = 0.1$ (blue curve). The outflow velocity has decreased to $\sim\! 75$\% of its launch value at a height $0.7R$, to $\sim\!50$\% at $0.8R$, and to $\sim\!25$\% at a height $0.9R$ (this scenario is illustrated in Figure~\ref{fig:vwind}(b) for a constant velocity radial outflow). For $\eta_s=0.9$ and $\Delta\eta_s = 0.05$ (magenta curve), the launch velocity has decreased to $\sim\! 75$\% at a height $0.85R$, to $\sim\! 50$\% at $0.9R$, and $\sim\! 25$\% at a height $0.95R$.  Ratios $\Delta \eta_s/\eta_s \simeq 0.1$--0.2 with $\eta_s \leq 0.7$ will effectively result in stalled velocities at some height within the radius $R$.  The behavior of Eq.~\ref{eq:fermi-dirac} can be highly flexible; mildly to moderately decelerated winds can be modeled if one considers stall ratios in the regime $\Delta \eta_s/\eta_s > 0.2$ with $\eta_s \simeq 0.5$.

Radially infalling cool clouds in a hot halos will experience ``accretion braking." \citep[e.g.,][]{tan23}. Such clouds may begin their journey through the outer halo following a ballistic trajectory, but they can grow through accretion as they develop turbulent radiative mixing layers. This growth and turbulence contributes an additional drag compared to simple ram pressure drag models. Consequently, the infall can rapidly reach sub-virial terminal velocities.   

\begin{figure}[htb]
\centering
\includegraphics[width=0.95\linewidth]{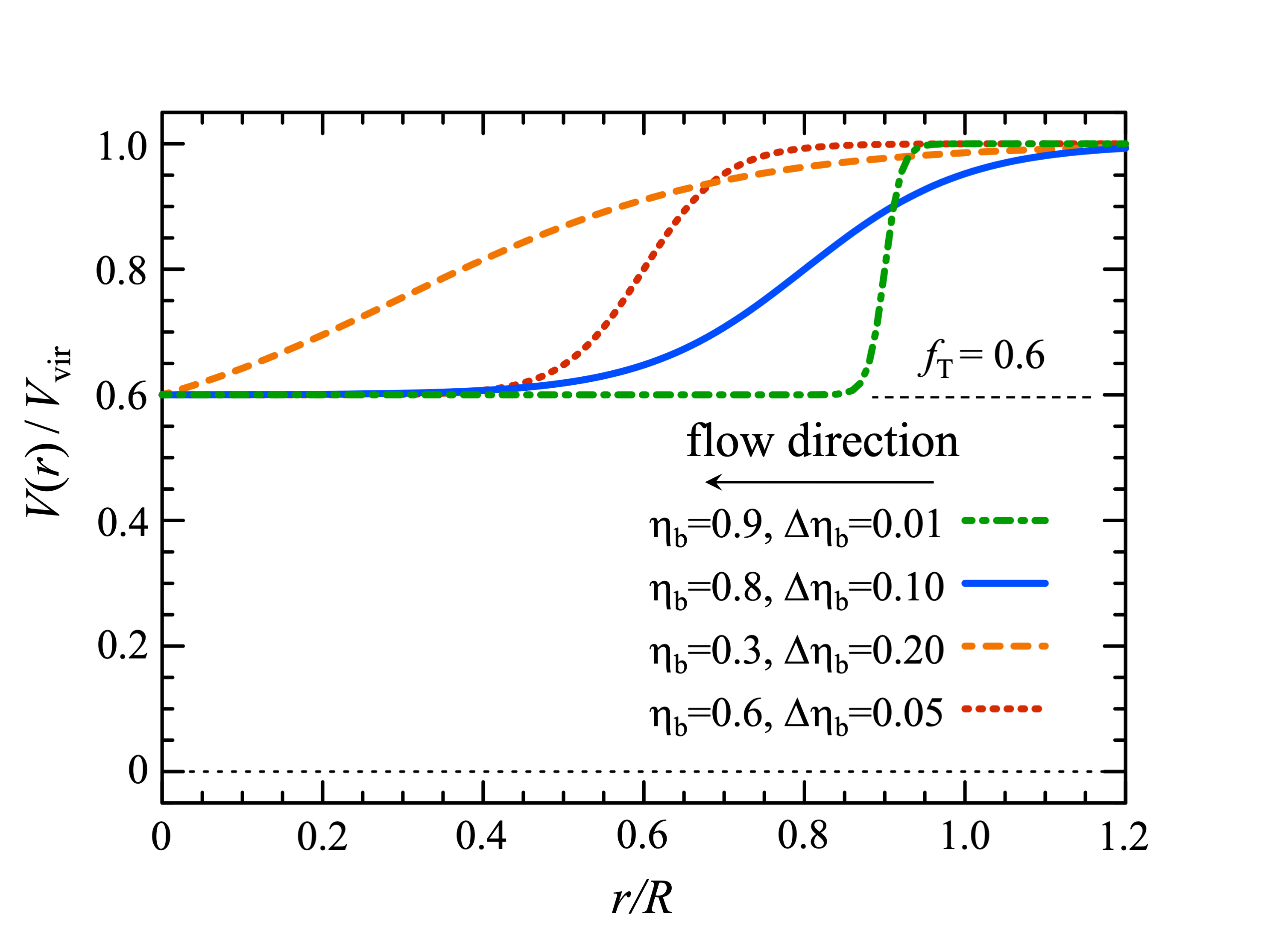}
\caption{\small The accretion breaking function given by Eq.~\ref{eq:breaking-function} for terminal velocity $f\subT = V\subT/V_{\rm vir} = 0.6$ and four parameter combinations of the breaking radius, $\eta_b$ and breaking ratio, $\Delta \eta_b/\eta_b$. The curves shown here are inspired by the models of \citet[][]{tan23} for their predicted value of $f\subT \simeq 0.6$. }
\label{fig:breakfunc}
\vglue 0.1in
\end{figure}

To describe accretion breaking with a terminal velocity, $V\subT$, we adopt a free parameter to describe the sub-virial terminal velocity as the fraction $f\subT = V\subT/V_{\rm vir}$ \citep{tan23}, where the virial velocity is effectively the circular velocity, $V_{\rm vir} = V_c$. For their models, values of $f\subT \simeq 0.6$ are predicted. As with the stall function for outflows, we aim to model accretion breaking with flexibility and simplicity.  Conveniently, we can define an accretion breaking function. As defined, the breaking function is the complementary stall function, 
\begin{equation}
\mathcal{B}(r;f\subT,\eta_b,\Delta \eta_b) = 1 - (1\!-\! f\subT ) \mathcal{S}(r;\eta_b,\Delta \eta_b) \, , 
\label{eq:breaking-function}
\end{equation}
where $\mathcal{S}(r;\eta_b,\Delta \eta_b)$ is given by Eq.~\ref{eq:fermi-dirac} with $\eta_b$ and $\Delta \eta_b$ replacing $\eta_s$ and $\Delta \eta_s$.

In Figure~\ref{fig:breakfunc}, we show four scenarios of the accretion breaking function for infalling gas starting with velocity $V_{\rm vir} = V_c$ at galactocentric distance $\sim\! R$. As the distance of the infalling material decreases, the velocity approaches the terminal velocity.  
Eq.~\ref{eq:breaking-function} can model slowly decelerated infall or rapid accretion breaking.  

For radially directed outflow with a constant positive velocity, $V_r>0$, the stall function would be applied as a multiplicative function,
\begin{equation}
 {\bf V}_h({\rvec}) = V_r\, {\cal S}(r) \,{\rhat}({\rvec}) \, ,
\label{eq:cgm-stalling}
\end{equation}
For radially directed infall onto the galaxy disk with a constant negative velocity, $V_r = - V_c$, the accretion breaking function would be also applied as a multiplicative function,
\begin{equation}
 {\bf V}_h({\rvec}) = -{\cal B}(r;f\subT) \, V_c\,  {\rhat}({\rvec}) \, ,
\label{eq:cgm-breaking}
\end{equation}
Carrying out Eq.~\ref{eq:Vlos-general}, we the LOS velocities for stalling outflow and breaking infall through the halo are respectively written
\begin{equation}
\begin{array}{rcl}
 V^{(h)}\subLOS(t) &=&  \mathcal{S}(t) \, V_{r}\, \mathcal{P}_r (t) \, , \\[5pt]
 V^{(h)}\subLOS(t) &=&  -\mathcal{B}(t;f\subT) \,V_{c}\, \mathcal{P}_r (t) \, ,
 \end{array}
\label{eq:vlos-cgmstalling}
\end{equation}
for a constant velocities $V_r$ and $V_c$.


\subsection{Bi-Polar Outflows/Winds}
\label{subsec:bipolar-kinematics}

To ensure that the velocity field associated with the outflows is confined within the geometry of the hyperboloid wind structure, we apply the conditions $f(t) > 0$ and $r(t) \leq R_w$ as given by Eq.~\ref{eq:hypercap} to identify the LOS locations that probe the wind. Outside the wind structure, we assume there is no wind material and therefore disregard the wind velocity field for all LOS locations external to the wind structure.  Wind models can be explored for a range of opening angles, $\Theta_w$, wind base radii, $\rho_{w,0}$, and wind heights, $R_w$.

A constant radial velocity field, ${\bf V}_w({\rvec}) = V_w\,{\rhat}({\rvec})$, where $V_w$ is a free parameter, provides the simplest kinematic model for the winds.  This radial velocity field works appropriately enough for a wind confined to a cone for which the radial wind trajectories diverge from a single point at the cone vertex. However, for a hyperboloid wind structure with a finite base radius, a radial wind velocity in the plane of the galaxy disk, or at small heights ($|z| \sim h_d$) above the disk plane, would have velocity primarily directed horizontally to the galactic plane, when in fact material in this region should be directed vertically and perpendicular to the plane.  Thus, we adopt a velocity field of the form
\begin{equation}
   {\bf V}_w({\rvec}) = V_\uprho({\rvec})\, {\rhohat}({\rvec}) + V_z ({\rvec}){\, \khat} \, ,
\label{eq:thehyperveltotal}
\end{equation}
and determine $V_\uprho({\rvec})$ and $V_z ({\rvec})$ such that ${\bf V}_w({\rvec})$ is perpendicular to the plane for small $|z|$, i.e., we achieve $|V_\uprho({\rvec})| \ll |V_z({\rvec})|$ and asymptotically approaches radial for large $|z|$ , i.e., $|V_\uprho({\rvec})| \simeq |V_z({\rvec})|$.

\begin{figure}[bht]
\centering
\vglue -0.025in
\includegraphics[width=0.8\linewidth]{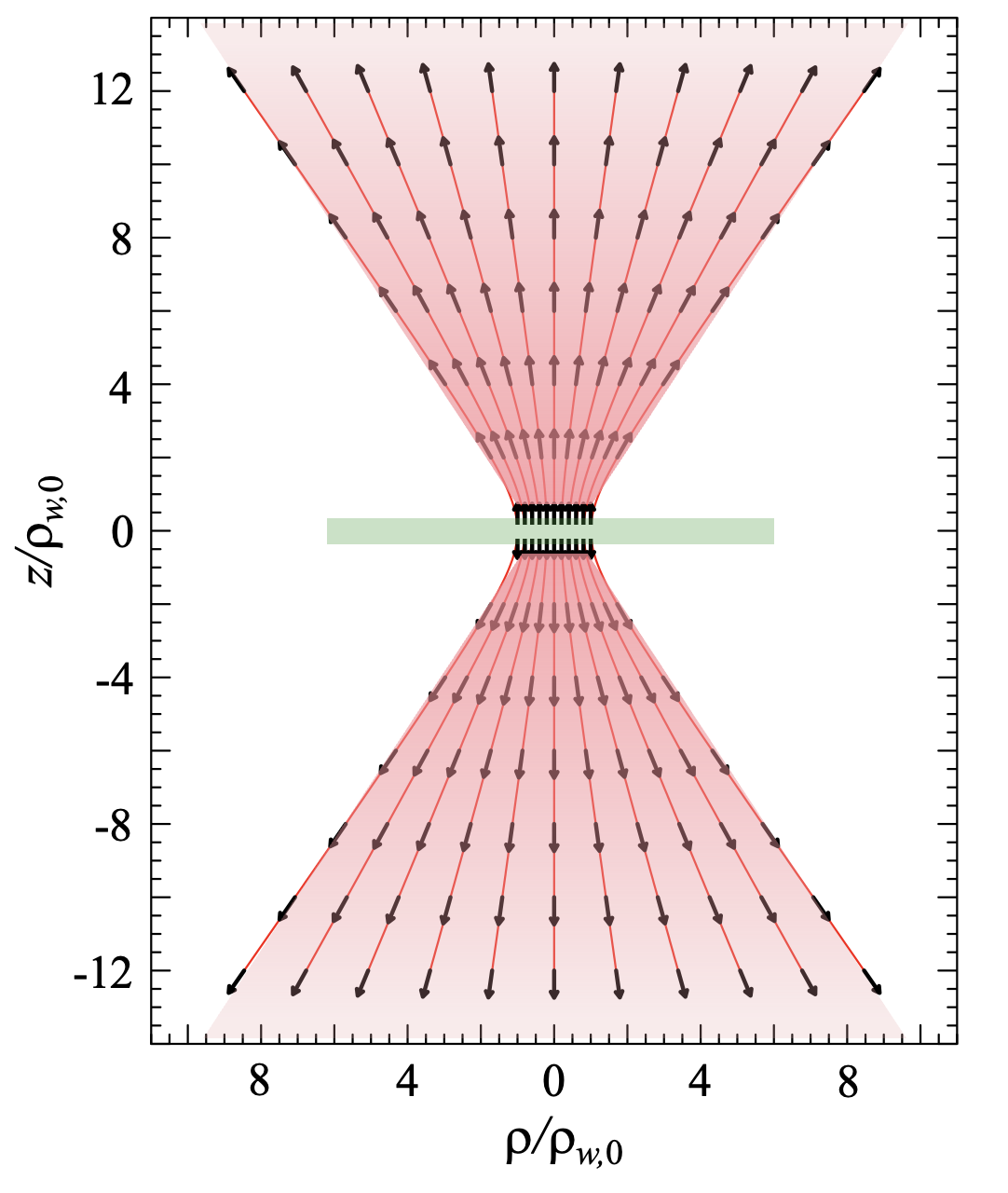}
\caption{\small A schematic cross section of the bi-polar wind kinematic model as given by Eqs.~\ref{eq:thehyperveltotal}, \ref{eq:upsilon4wind}, and \ref{eq:mywindmodel}. The red curves represent the trajectories of the wind material and the black arrows represent the velocity field for a hyperboloid wind structure with opening angle $\Theta_w=35^\circ$. The green stripe represents the galaxy disk, where all coordinates are normalized to the wind base radius, $\rho_{w,0}$. This model assumes a constant $V_w$ for the wind.}
\label{fig:hyperwind}
\end{figure}

For opening angle $\tan \Theta_w = T_\Theta = \rho_{w,0}/c$, the streamline trajectories bound within a hyperboloid would follow paths constrained by $(\rho/\tilde{a})^2+(z/c)^2=1$, where $c=\rho_{w,0}/T_\Theta$. Rearranging and substituting $c$, we write
\begin{equation}
    \rho^2 = \tilde{a}^2 + z^2  
    \left( \frac{\tilde{a}}{c} \right)^2 
    = \tilde{a}^2 \left[ 1 +  T^2_\Theta
    \left( \frac{z}{\rho_{w,0}} \right)^2
\right] \, ,
\label{eq:tildea}
\end{equation}
where $\tilde{a}$ is the azimuthal distance in the disk plane at which the trajectory is launched.  Adopting $V_\uprho = d\rho/dt$ and $V_z=dz/dt$ and  differentiating both sides yields,
\begin{equation}
    V_\uprho({\rvec}) = \frac{z}{\rho} 
    \left( \frac{\tilde{a}}{\rho_{w,0}} \right)^2 \!\! T^2_\Theta \, V_z({\rvec}) \, .
\label{eq:windVrho}
\end{equation}
Solving for $\tilde{a}$ in Eq.~\ref{eq:tildea} and substituting into Eq.~\ref{eq:windVrho}, we obtain $V_\uprho ({\rvec})= \Upsilon({\rvec}) |V_z({\rvec})|$, where 
\begin{equation}
    \Upsilon({\rvec}) =  \Upsilon(\rho,z) =\frac{|z|\rho T^2_\Theta}
    {\rho^2_{w,0} + z^2 T^2_\Theta} \, .
\label{eq:upsilon4wind}
\end{equation}
From $V_w^2({\rvec}) = V^2_\uprho({\rvec}) + V_z^2({\rvec})$, we derive
\begin{equation}
V_\uprho({\rvec}) = \displaystyle \frac{\Upsilon({\rvec}) V_w}{\sqrt{1+ \Upsilon^2({\rvec})}} \, , \quad
V_z({\rvec}) = \displaystyle \frac{{\rm sgn}(z)V_w}{\sqrt{1+ \Upsilon^2({\rvec})}}  \, , 
\label{eq:mywindmodel}
\end{equation}
where $V_w$ is a free parameter.  The signum function ensures that the vertical component of the wind velocity, $V_z({\rvec})$, is always outflowing. Note that $\Upsilon({\rvec}) =0$ for $z=0$ (in the galactic plane) and/or for $\rho=0$ (along the wind axis), yielding $V_\uprho(\rho,0)=V_\uprho(0,z)= 0$ and also $V_z(\rho,{z\!\rightarrow\!0}) = V_w$.  A schematic of this wind kinematic model is illustrated in Figure~\ref{fig:hyperwind}.  

The LOS wind velocity is then
\begin{equation}
    V\subLOS^{(w)}(t) = 
    \frac{V_w\left[ \Upsilon(t) P_\uprho(t) + 
    {\rm sgn}(z(t)) P_z \right]}
    {\sqrt{1+ \Upsilon^2(t)}} 
     \, .
\end{equation}

For galaxies that do not classify as starburst or ultra-luminous galaxies, the maximum outflow velocities are $V_w \sim 200$~{\kms} and have a mild dependence on the central star formation surface density \citep{chisolm15, bizyaev22-outflows}. \citet{chisolm15} reports outflow velocities relationships with star formation rate,  $V_w \propto SFR^{\, 0.08-0.2}$, stellar mass, $V_w \propto M_\ast^{0.12-0.20}$, and circular velocity, $V_w = V_c^{0.44-0.87}$. Further, guidance on the magnitude of $V_w$ can be gleaned from theoretical works \citep[][]{zhang18, fielding22, nguyen22} and observational scaling relations between galactic outflows and their host galaxies \citep[e.g.,][]{rupke05, weiner09}. In addition, observationally and/or theoretically inspired variations can easily be introduced to the wind velocity field, for example deceleration or stalling of outflows and/or axial variations in the velocity flows. 

\subsubsection{Wind Stalling}

Wind stalling has been described in some detail by \citet{fielding22}, and its modeling was discussed in Section~\ref{subsubsec:deceleration-function} using a simple stall function (see Eq.~\ref{eq:fermi-dirac}).  Incorporating the stall function, the wind velocity is written,
\begin{equation}
 {\bf V}_w({\rvec}) =  {\cal S}(r) \left[ V_\uprho({\rvec})\, {\rhohat}({\rvec}) + V_z({\rvec}) \, {\khat} \right]  \, .
\label{eq:wind-stalling}
\end{equation}

An example velocity field with wind stalling is illustrated in Figure~\ref{fig:vwind}(a) for a constant outflow velocity, $V_w$, for a hyperboloid wind structure with an opening angle of $\Theta_w = 35^\circ$ and maximum extent $R_w = 12\rho_{w,0}$.  We show the scenario for  $\eta_s=0.8$ and $\Delta\eta_s = 0.1$. For these stall parameters, the wind velocity has decreased to $\sim\! 75$\% of its launch value at a height $0.7R_w$, to $\sim\!50$\% at $0.8R_{w}$, and to $\sim\!25$\% at a height $0.9R_{w}$. 
Carrying out Eq.~\ref{eq:Vlos-general} , we obtain the LOS velocity
\begin{equation}
 V^{(w)}\subLOS(t) =  
 \frac{\mathcal{S}(t) V_{w}\left[ \Upsilon(t) P_\uprho(t) + {\rm sgn}(z)P_z \right]}{\sqrt{1+ \Upsilon^2(t)}} \, ,
\label{eq:vlos-windstalling}
\end{equation}
for a constant wind velocity, $V_w$.

\subsubsection{Enhanced Wind Core Velocity}
\label{sec:ehanced-vwind}

We might expect that velocities of the outflowing wind material launched nearest the galactic polar axis are higher than those launched nearest the ``walls'' of the wind structure (along the circumference of the wind base radius, $\rho_{w,0}$).  As a first-order approximation, this expectation may serve to account for a higher kinetic energy density injected by supernovae in the galactic nucleus that drives the core of the wind, while kinetic energy is lowered due ISM entrained along the wall of the wind structure \citep[e.g.,][]{bustard16, rupke18, zhang18, gronke20, fielding22}.   

\begin{figure*}[tbh]
\centering
\vglue -0.05in
\includegraphics[width=0.9\linewidth]{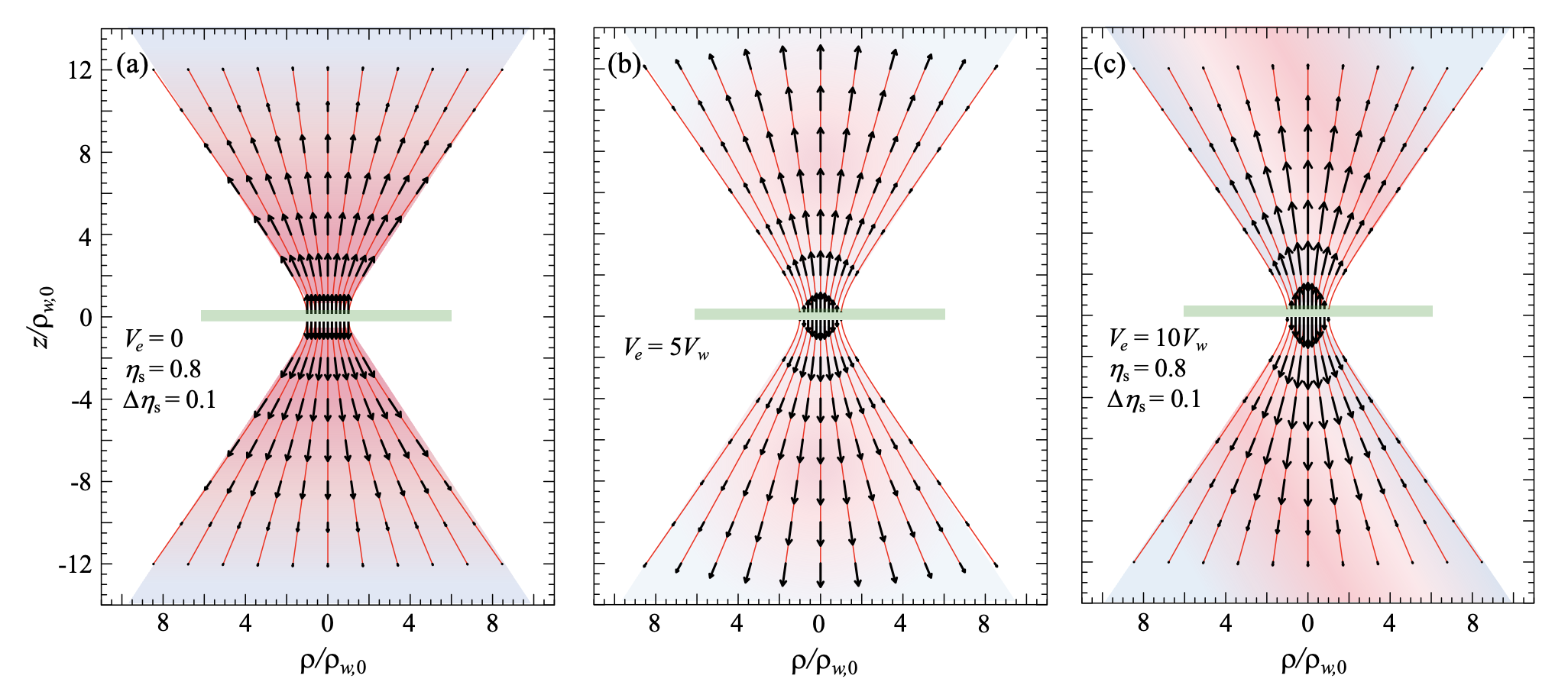}
\caption{\small Idealized bi-polar wind velocity fields viewed in cross section. Spatial coordinates are in units of the wind base radius, $\rho_{w,0}$. The green region represents the galactic disk. The red curves show the kinematic trajectories and the back arrows show the velocities in units of $V_w$. (a) A constant velocity velocity field (see Figure~\ref{fig:hyperwind}) with wind stalling applied via Eq.~\ref{eq:wind-stalling} with $\eta_s = 0.8$ and $\Delta \eta_s = 0.1$. (b) An enhanced wind model with $\tilde{V}\subE = 5V_w$, showing higher velocities in the wind core. (d) A model that combines the kinematics in panels (a) and (b) using an enhanced velocity of $\tilde{V}\subE=10V_w$ with stall function parameters $\eta_s = 0.8$ and $\Delta \eta_s = 0.1$.} 
\label{fig:vwind}
\end{figure*}

One approach to modeling this enhanced outflow velocity in the wind core is to employ a functional form that exhibits axial symmetry with peak amplitude on the wind axis.  We write the ``enhanced wind'' velocity as 
\begin{equation}
V'_w({\rvec}) =  V_w (r) + V\subE (\rho,z)  \, ,
\label{eq:vw-enhancement}
\end{equation}
where $V_w (r)$ can be the characteristic wind speed, i.e., $V_w(r) = V_w$. We define $V\subE(\rho,z)= \tilde{V}\subE f\subE (\rho,z)$ as the ``enhancement velocity," where $\tilde{V}\subE$ is amplitude in velocity units, which can be positive or negative in magnitude, and where $f\subE (\rho,z)$ is an arbitrary axial-symmetric ``enhancement function."  The enhancement function is required to have the properties that, at every height $z$, $f\subE(0,z)=1$  (on the wind axis) and $f\subE(\rho_{w}(z),z)$ vanishes at $\rho_w(z) = [\rho_{w,0}^2 + z^2 T^2_\Theta ]^{1/2}$ (on the surface of the wind.  Thus, we see that the enhancement is evaluated over the axial range $\rho(z) \in [0,\rho_w(z)]$ for the domain $z\in [0,\pm R_w]$. With this formalism, the velocity along the wind surface will be $V'_w(\rho_w(z),z) = V_w$ and in the wind core will be $V'_w(0,z) = V_w + \tilde{V}\subE$.

In Appendix~\ref{app:D}, we further discuss the properties of enhancement functions and provide selected examples. One simple enhancement function is a truncated cosine function, $f\subE(\rho,z) = \cos \{(\pi/2)(\rho/\rho_w(z))\}$, which is illustrated in Figure~\ref{fig:fE-efuncs}(c).  Wind models using this truncated cosine enhancement function are illustrated in Figure~\ref{fig:vwind}(b,c).  In panel~\ref{fig:vwind}(b), we show enhancement in the core with $\tilde{V}\subE =5V_w$ with no wind stalling. For example, if $V_w = 50$~km~s$^{-1}$, then the speed of the wind along the surface of the structure is $50$~km~s$^{-1}$ and the speed along the central axis of the wind is $250$~km~s$^{-1}$. We can include stalling of enhanced wind models by employing the stalling function.  This latter scenario is illustrated in Figure~\ref{fig:vwind}(c) with $\tilde{V}\subE=10V_w$ and stalling function parameters $\eta_s = 0.8$ and $\Delta \eta_s = 0.1$, which are the same as those employed for the illustration in panel~\ref{fig:vwind}(a). 

Since arbitrary enhancement functions can be applied, there is a great deal of versatility for approximating theoretical expectations.  As we will explore in \citetalias{churchill25-skamII}, similar ``enhancement" functions can serve to model the density and temperature profiles to approximate spatially varying changes in the gas phases.  

Carrying out Eq.~\ref{eq:Vlos-general}, we obtain the LOS velocity
\begin{equation}
 V^{(w)}\subLOS(t) =  
\frac{\mathcal{S}(t) V'_{w}(t)
\left[ \Upsilon(t) P_\uprho(t) + {\rm sgn}(z)P_z \right] } {\sqrt{1+ \Upsilon^2(t)}}\, ,
\label{eq:vlos-cosine+stalling}
\end{equation}
where we have incorporated enhanced velocity field $V'_w({\rvec})$ and wind stalling.

\subsection{Galactic Disks}

The morphology and kinematics of gaseous galactic disks have exhibit turbulence, asymmetry, and warps \citep[e.g.,][]{sancisi08, marasco11}.  Nevertheless, suitable idealistic models can provide insight into the absorption signatures of the galaxy disk itself. A simple approach would be to confine the disk material to a cylinder with axial radius $\rho_d$ and height $h_d$, as given by Eq.~\ref{eq:diskstructure}.  This might be appropriate for an ISM component to the model.  However, we will adopt a more general approach in which the velocity field and (as will be discussed in \citetalias{churchill25-skamII}) the density field will define the ``the disk'' extent.  Here, we discuss the kinematics.


\subsubsection{Disk and Extra-planar Gas}

For ISM gas, the simplest kinematic model is solid-disk rotation, ${\bf V}({\rvec}) = V_c\, {\phihat} ({\rvec})$ confined within the disk spatial structure.  Here, $V_c$ could be the maximum circular velocity employed to defined the galaxy virial mass and radius (see Eq.~\ref{eq:defcircvel}). However, the galactic rotation of extra-planar gas (EPG) is seen to slow with increasing height above the disk \citep[e.g.,][]{sancisi08, zschaechner15ApJ799, zschaechner15ApJ808, bizyaev17, bizyaev22-lags, levy19}. 

The origin of this rotational lag gradient
is not fully understood. Accretion from galactic fountains \citep{bregman80} is expected to occur because of cooling/condensation of the hot CGM (corona) triggered by the fountain predicts a disc growth rate compatible with the observations \citep[e.g.,][]{fall80}. Fountain-driven corona condensation is also a likely mechanism to sustain star formation, as well as the disc inside-out growth in local disc galaxies \citep[][]{marasco11, tan23}.  Based on absorption lines from both the Galaxy and from external galaxies, the observed gas kinematics in the ``lower halo" is inferred to trace galactic fountain flows that persist to $R_\perp > 5$ kpc \citep{marasco11, rubin22}. Simulations indicate that galactic fountains have complex mixing histories in the lower halo in that their angular momentum is a result of interactions with lower angular momentum wind-recycled gas and with higher angular momentum accreting IGM gas such that prograde accretion enhances their angular momentum whereas retrograde accretion reduce their angular momentum \citep[e.g.,][]{grand19} .

\citet{bizyaev17, bizyaev22-lags} reported that the lag correlates with galactic stellar mass, rotation curve amplitude, and central stellar velocity dispersion; they suggested larger lags are preferentially observed to be associated with higher-mass early-type disk galaxies as a consequence of higher proportions of infalling IGM gas relative to the local extra-planar gas in these galaxies as compared to lower-mass (slower rotating) galaxies \citep[also see][]{fraternali07}.  An alternative hypothesis is that more massive galaxies host hotter more massive gaseous halos that interact with the EPG and induce lagging velocities \citep[e.g.,][]{marinacci11, pezzulli17, sormani18}.  Theoretical predictions from galactic fountains are that larger extra-planar velocity lags are expected in the central regions of galaxies as compared to their peripheries due to momentum conservation 
\citep[e.g.,][]{bregman80, fraternali07, marasco11, grand19, li23}.

In view of EPG rotational lags, an alternative approach to modeling galaxy disk kinematics is to eliminate a physical geometric disk boundary and specify a velocity field reflecting EPG halo lag.  This approach was adopted by \citet{steidel02}, who allowed for an inclined disk of infinite radius and height, but introduced an exponential scale height, $H_v$, to the circular velocity,  
\begin{equation}
{\bf V} ({\rvec}) = V_\upphi(\rho,0) \exp \left\{ - |z|/H_v \right\} {\phihat} ({\rvec})\, ,
\label{eq:vsteidel02}
\end{equation}
where $V_\upphi(\rho,0)$ is the rotation velocity in the galaxy plane ($z=0)$ at axial radius $\rho$ on the galaxy disk. 
\citet{steidel02} adopted $V_\upphi(\rho,0) = V_c$.  Many studies have investigated rotational kinematics in quasar absorption line systems using this ``Steidel model" \citep[e.g.,][]{kacprzak10, kacprzak11, kacprzak19,
ho17, ho20, martin19, french20, beckett23}. Alternatively, \citet{bizyaev17} characterized rotation lags using the velocity model 
\begin{equation}
{\bf V} ({\rvec}) = \left[ V_\upphi(\rho,0) + |z| \frac{dV_\upphi}{dz} \right] {\phihat} ({\rvec}) \, , 
\label{eq:vrbizaev17}
\end{equation}
where
\begin{displaymath}
V_\upphi(\rho,0) = \left\{
\begin{array}{lcl}
\displaystyle V_c (\rho/\rho_{d,0}) && (\rho \leq \rho_{d,0}) \\[4pt]
\displaystyle V_c + (\rho - \rho_{d,0}) \frac{dV_\upphi}{d\rho}  && (\rho> \rho_{d,0}) \, ,
\end{array}
\right.
\end{displaymath}
where $\rho_{d,0}$ is the axial radius on the disk out to which the rotation curve increases linearly. Typically ${\rho_{d,0} \simeq 0.1 \rho_d}$. In Eq.~\ref{eq:vrbizaev17}, the rotation cut-off height is $|z| = V_c/|dV_\upphi / dz|$.  Typical lag gradients, $dV_\upphi / dz$, are on the order of $-10$~{\kms} kpc$^{-1}$, though they can range up to $-60$~{\kms} kpc$^{-1}$ \citep[e.g.,][]{zschaechner15ApJ799, zschaechner15ApJ808, levy19, bizyaev17, bizyaev22-lags, li23}. \citet{marasco11} modeled the observed properties of the {\HI} halo of the Milky Way and reported a vertical velocity gradient of $dV_\upphi/dz = -15 \pm 4$ {\kms} kpc$^{-1}$.  

Interestingly, \citet{bizyaev22-lags} found that the mean axial radius lag gradient, $dV_\upphi / d\rho$, is consistent with zero (90\% of galaxies show no significant axial radius lag gradient). This suggests $V_\upphi(\rho,0)$ is a constant in both Eq.~\ref{eq:vsteidel02} and in Eq.~\ref{eq:vrbizaev17} for $\rho \geq\rho_{d,0}$. Theoretical disk and extra-planar velocity fields based on equilibrium dynamical models \citep[e.g.,][]{sormani18} and magnetized galactic halo models \citep[e.g.,][]{henriksen16} also provide viable halo lag kinematic expressions. 

Finally, we note that the theoretical velocity profile for an NFW halo is \citep{NFW96},
\begin{equation}
V_\upphi\supNFW(\rho,z) = \frac{V_c}{x_s}
\frac{\ln(1\!+\!\mathcal{C}x_s) - x_s/(1\!+\!\mathcal{C}x_s)}
{A\subcalC}   \, ,
\label{eq:NFWvel}
\end{equation}
where $x_s = r/R_s = [\rho^2 + z^2]^{1/2}/R_s$ and 
${A\subcalC}$ was introduced in Section~\ref{sec:defs-galaxy}. 
This profile also exhibits a circular velocity lag with height $z$ above the plane.

\begin{figure*}[htb]
\centering
\includegraphics[width=0.82\linewidth]{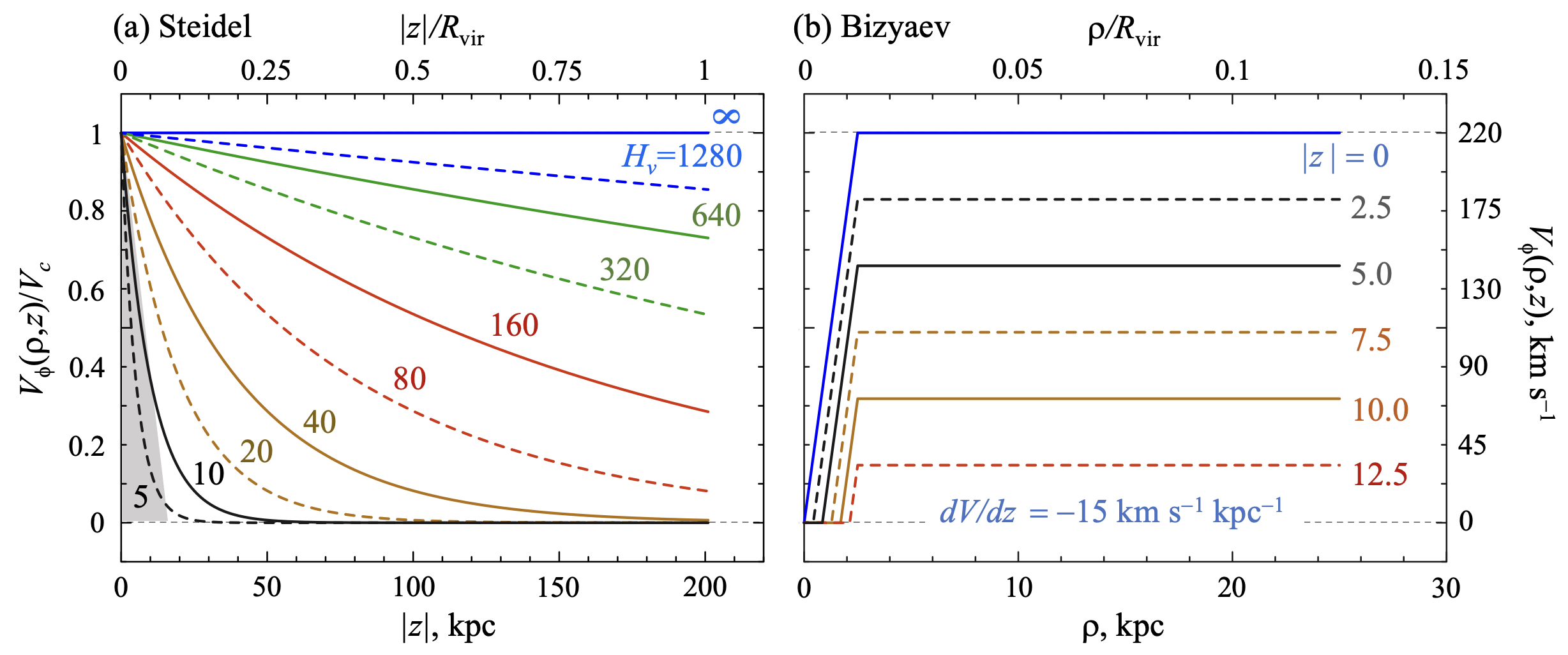}
\caption{\small The rotation velocity lag models, $V_\upphi (\rho,z)$, for the \citet{steidel02} model via Eq.~\ref{eq:vsteidel02} and the \citet{bizyaev17} model via Eq.~\ref{eq:vrbizaev17}. We adopt $V_c=220$~{\kms} for a galaxy with $R_{\rm vir}= 200$~kpc and disk axial radius $\rho_d=25$~kpc. (a) The rotation velocity as a function of height above the disk for the Steidel model for various velocity scale heights, $H_v$ [kpc], labeling each curve (there is no $\rho$ dependence). Smaller $H_v$ yield steeper velocity lag gradients. The gray shaded region is for a lag gradient $dV_\upphi/dz \leq -15$~{\kms} kpc$^{-1}$, corresponding to the mean value from \citet{marasco11} for the Milky Way. Note this corresponds to $H_v$ ranging from 5 to 10 kpc. (b) The lag velocities for the Bizyaev model for increasing height above the disk (the height of each curve is labeled) as a function of axial distance along the disk for a gradient $dV_\upphi/dz = -15$~{\kms} kpc$^{-1}$. We have assumed $\rho_{d,0} = 0.1\rho_d$ for the linear increase in the inner disk and $dV_\upphi/d\rho=0$ for the outer disk. }
\vglue 0.1in
\label{fig:vdisk-lags}
\end{figure*}

In Figure~\ref{fig:vdisk-lags}, we illustrate selected EPG halo lag curves for the \citet{steidel02} and \citet{bizyaev17} models.  We adopt a Milky Way-like galaxy with $R_{\rm vir}=200$~kpc having a disk axial radius of $\rho_d=25$~kpc. As observational works discussed above report that the halo lag gradients are on the order 10--60~{\kms}~kpc$^{-1}$, we find that a range of velocity scale heights for the \citet{steidel02} model that are consistent with this gradient are in the range of $5 \leq H_v \leq 10$~kpc.  Several of the aforementioned previous studies employing the \citet{steidel02} model have tended to adopt much larger values on the order of $H_v > 1000$~kpc.  As can be seen in  Figure~\ref{fig:vdisk-lags}(a), this effectively removes the halo lag to disk heights well beyond the viral radius, which can artificially elevate the portion of ``galactic disk" LOS velocities that are consistent with an observed range of absorption line velocities. As can be seen in Figure~\ref{fig:vdisk-lags}(a), we note that the \citet{bizyaev17} model has a discontinuous derivative at $\rho_{d,0}$.  

Adopting Eq.~\ref{eq:vsteidel02} and Eq.~\ref{eq:vrbizaev17}, and carrying through Eq.~\ref{eq:Vlos-general}, we write the LOS velocities as,
\begin{equation}
V^{(d)}\subLOS (t) = V_c \exp \left\{ - |z(t)|/H_v \right\} {\cal P}_\upphi(t) \, ,
\label{eq:VLOS-diskexp}
\end{equation}
for the \citet{steidel02} model, 
\begin{equation}
V^{(d)}\subLOS (t) = \left[ V_\upphi(\rho(t),0) - |z(t)| \frac{dV_\upphi}{dz} \right]  {\cal P}_\upphi(t) \, ,
\label{eq:VLOS-bizaev}
\end{equation}
for the \citet{bizyaev17} model.
Similarly, adopting Eq.~\ref{eq:NFWvel}, we have
\begin{equation}
V^{(d)}\subLOS (t) = V_\upphi\supNFW (t) {\cal P}_\upphi(t) \, ,
\label{eq:VLOS-NFW}
\end{equation}
for an \citet{NFW96} NFW model.  However, we note that NFW velocity field applies to the dark matter, and it is well-known that the velocity and angular momentum of the baryons do not strictly follow the dark matter, as evident from the spin parameters of galactic disks \citep[][]{m098, stewart17, stewart-proc17, yang23}.

\subsubsection{Axial and Vertical Flows}

Models of the Milky Way gaseous disk have required not only lag gradients above the plane, but also a vertical infall component and axial ``migration" of gas in the plane.  For example, the best fit model by \citet{marasco11} included a lag component of $dV_\upphi /dz  \simeq -15$ {\kms}~kpc$^{-1}$, which is consistent with the findings for external galaxies \citep[][]{bizyaev17, bizyaev22-lags}.  They also found 
vertical infall onto the galactic plane with $V_z \simeq -20$~{\kms} and an inward radial axial migration of $V_\uprho \simeq -30$~{\kms}.  Similarly, \citet{qu20} found $dV_\upphi /dz  \simeq -8$ {\kms}~kpc$^{-1}$ and a larger radial axial migration of $V_\uprho \simeq -70$~{\kms} toward the upper (northern) half of the disk and $V_\uprho \simeq -40$~{\kms} for the lower half (southern), though the latter is not statistically significant.  They also report a ``random component," which perhaps may be analogous to a vertical component with $|V_z| = 18$~{\kms}.  

These observationally constrained velocities are, for the most part, consistent with theoretical expectations; for example, the vertical infall velocity component may be related to the condensation of baryonic gas above the galactic plane that then ``rains" on to the disk \citep[e,g,][]{grand19,tan23}. The simulations focused on planar accretion show that an inward axial migration of material continues to small axial disk radii and that the magnitude of this velocity is on the order of 10\% of the circular velocity \citep[e.g.,][]{hafen22}. However, observational evidence that an axial inward velocity component is a {\it common\/} feature of disk galaxies is not compelling \citep{diteodoro21}. 

The inclusion of a vertical infall and/or axial inward migration into the kinematic model for the disk would be straight forward. We could assume a constant vertical velocity of the form ${\bf V}({\rvec}) = V_z{\khat}({\rvec})$ and/or axial velocity component, ${\bf V}({\rvec}) = V_\uprho{\rhohat}({\rvec})$.  For vertical infall we adopt $V_z < 0$ for $z>0$ and $V_z > 0$ for $z<0$, and for axial migration we adopt $V_\uprho < 0$ for axial inward migration \citep[see][]{ho17, ho20} or $V_\uprho > 0$ to for axial outward migration.  We would then simply add the LOS velocity components $V_z \mathcal{P}_z$ and $V_\uprho \mathcal{P}_\uprho(t)$ to the disk LOS velocities given by either Eq.~\ref{eq:VLOS-diskexp} or Eq.~\ref{eq:VLOS-bizaev}.

\subsection{Extended Planar Accretion}
\label{sec:XPAkinematics}

The spatial scale heights of the ISM and EPG from galactic fountains and accretion are on the order of 1--10 kpc \citep[e.g.][]{sancisi08, gaensler08, li23}. As such, spatial-kinematic models of the ISM and EPG are most relevant for LOS with impact parameters on the order of $R_\perp \leq \rho_d$ \citep[e.g.,][]{bregman13, rubin22}.

As we mentioned in Section~\ref{subsec:theflareddisk}, extended planar accretion in the form of inward spiraling infall kinematics is suggested by theory \citep[e.g.,][]{keres05, stewart11, stewart13, stewart17, danovich15, defelippis21, hafen22, trapp22, gurvich23, stern23, kocjan24}.  
Absorption line observations of low-ionization gas also corroborate a coupling between the angular momentum of the rotating galaxy disk and extended planar gas \citep[e.g.,][]{steidel02, kacprzak10, kacprzak25, ho17, martin19, zabl19}.  
Such signatures of extended planar accretion are expected if the formation of galactic disks is predominantly outside-in \citep[see][]{fall80}. 

In simulations of $z\simeq 0$ Milky-Way like galaxies, \citet{trapp22} found that planar infalling gas approaches the galaxy disk such that it nearly matches the angular momentum at the outer edge of the gaseous disk.  On approach, the inward radial velocities are suppressed as the gas piles-up on the disk edge and enters a regime of rotational support.  From there the accreting gas slowly moves axially inward parallel to the disk plane and, joining the galaxy in the disk outskirts. Similar findings were reported by \citet{hafen22} for the formation of thin, star-forming disk galaxies (but not for warped and amorphous disks).  The simulations of \citet{stern23} and \citet{kocjan24} corroborated these findings. \citet{stern23}  also derived an azimuthally symmetric analytical model for the rotating inflowing gas in the slow rotating limit that provides analytical expressions for $V_r(r,\theta)$, $V_\upphi(r,\theta)$, and $V_\uptheta(r,\theta)$.  

We will show that a simple Keplerian model of the infall gas trajectories can be configured to capture many of the essential properties just described.  In particular, the infall trajectories described by these models are inspired by the trajectories shown in Fig.~11 of \citet{kocjan24}.


\subsubsection{Keplerian Velocity Fields}
\label{sec:kepler-vfields}

A viable kinematic model for infalling spiral paths of the accreting material is a Keplerian trajectory using conic sections with the center of the galaxy at a focus point.\footnote{We also examined the kinematics along logarithmic spiral trajectories $\rho = \rho_0 e^{k\phi}$  \citep[known as {\it ewige Linie} or the eternal line,][]{durer1525}. This curve has the dual benefit of providing both a velocity field expressible in closed form that conserves angular momentum and is asymptotically consistent with an NFW gravitational potential for $r \gg R_s$ \citep[e.g.,][]{bassetto22}. It also yields a spiral structure compatible with density wave theory \citep[e.g.,][]{bertin96, pringle19}. This {\it spira mirabilis\/} kinematic model is described in Appendix~\ref{app:E}.} The material falls inward along a conic section orbital path from a galactocentric axial distance $\rho_2$ to its periapsis at $\rho_1$.  If we adopt $\rho_1$ such that the periapsis is the axial distance at which the infalling material ``accretes" on to the outer host galaxy on the order of the cooling time, \citep[see][]{stewart-proc17, hafen22, trapp22}.  The Keplerian model comprises a highly simplified picture in which no friction or viscosity is accounted in the dynamics of the infalling material on its trajectory to periapsis.  However, it does allow for conservation of specific angular momentum, ${\bf h}={\bf L}/m = \dot{\boldsymbol \rho} \times {\boldsymbol \rho}$, and first-order approximations of the energies involved.

\begin{figure*}[htb]
\centering
\includegraphics[width=0.96\linewidth]{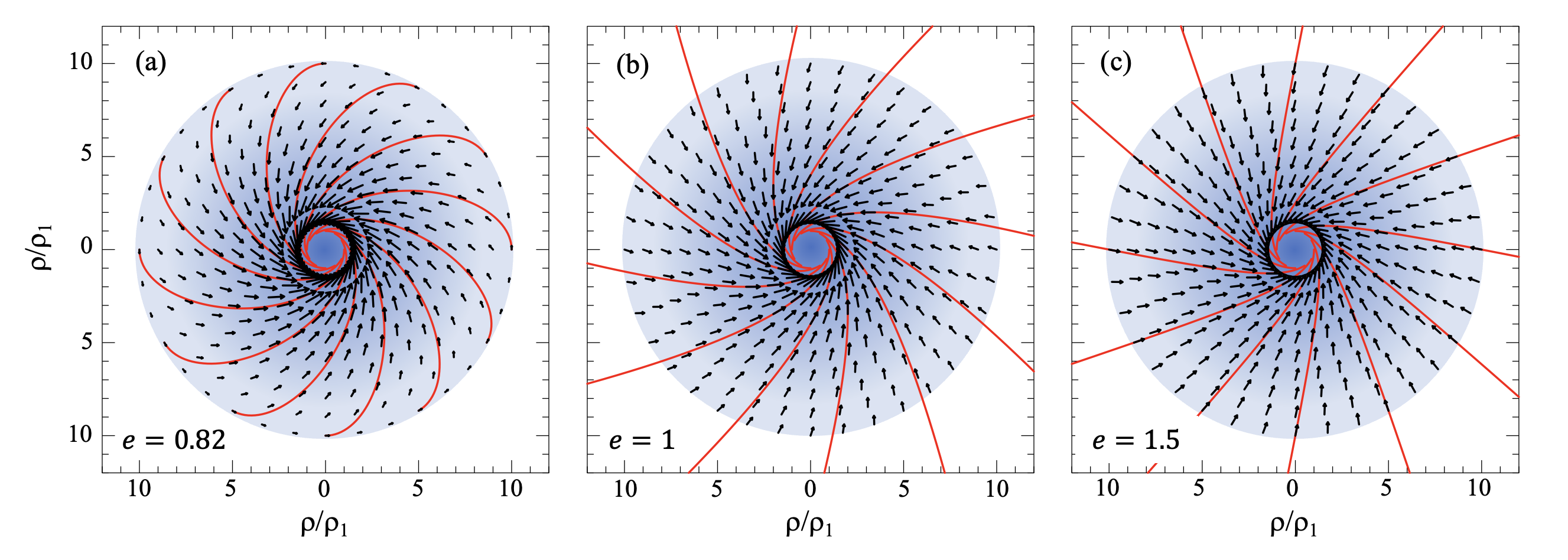}
\caption{\small Planar accretion idealized velocity fields based on Keplerian trajectories in the $z=0$ plane for the geometry $\rho_2/\rho_1 = 10$. Red curves provide the orbital trajectories of the infalling gas, whereas the velocity vector field at $\rho/\rho_1$ is represented by the black arrows (visual aliasing of the orbital paths may occur).  The relative sizes of the arrows are $v/v_1$. (a) Bound spiraling infall following an elliptical orbital path (Eq.~\ref{eq:radialconic} and Eq.~\ref{eq:azimuthalconic} with Eq.~\ref{eq:visviva-elliptical}) for $e=0.82$. (b)  Unbound spiraling infall following a parabolic orbital path (Eq.~\ref{eq:radialconic} and Eq.~\ref{eq:azimuthalconic} with Eq.~\ref{eq:visviva-parabola}) for $e=1.0$. (c) Unbound spiraling infall following a hyperbolic orbital path (Eq.~\ref{eq:radialconic} and Eq.~\ref{eq:azimuthalconic} with Eq.~\ref{eq:visviva-hyperbola}) for $e=1.5$. As these kinematic models enforce the velocity to be strictly azimuthal with magnitude $v_1 = V_c$ at $\rho=\rho_1$, the velocity fields for different orbital eccentricities become less distinguishable as $\rho$ approaches $\rho_1$. The main differences are in the outer regions; for the elliptical trajectories ($e<1$), the velocity is purely azimuthal at $\rho=\rho_2$, whereas the radial component at $\rho=\rho_2$ increases as eccentricity increases for parabolic and hyperbolic trajectories ($e>1$).}
\label{fig:accretion}
\vglue 0.1in
\end{figure*}

The general parametric form of a trajectory on a conic section is 
\begin{equation}
 \rho = \frac{\ell}{1+e\cos \nu }
\label{eq:ellipsecurve}
\end{equation}
where $\rho$ is the distance from the focus (galactic center) as a function of the true anomaly, $\nu$, and where $\ell = h^2/\mu$ is the semi-latus rectum with gravitational parameter $\mu = GM(r)$  and specific angular momentum, $h$.  If we enforce velocity $v_1$ at the periapsis $\rho_1$, we have $h = \rho_1v_1$ and $\mu = \rho_1 v_1^2/(e+1)$.  We then have $\ell = \rho_1(e+1)$. The eccentricity is $0 < e < 1$ for an ellipse, $e=1$ for a parabola, and $e>1$ for a hyperbola.


For an ellipse the trajectories are bound and the specific energy is negative, $\epsilon  = v^2_{\rm e}/\rho - \mu/r = - \mu/2a$, from which we have the {\it vis-viva\/} equation giving the instantaneous orbital speed $v_{\rm e}(\rho)$, 
\begin{equation}
  v_{\rm e} (\rho) =
\sqrt{\mu \left( \frac{2}{\rho} - \frac{1}{a} \right) }  
= v_1
\sqrt{\frac{2}{(e\!+\!1)}\frac{\rho_1}{\rho} - \frac{1-e}{1+e} }   \, ,
\label{eq:visviva-elliptical}
\end{equation}
where $\rho \in [\rho_1,\rho_2]$, and where $a = (\rho_1 + \rho_2)/2$ is the semi-major axis for an ellipse with periapsis $\rho_1$ and apoapsis $\rho_2$. For such an ellipse, the eccentricity is $e=(\rho_2-\rho_1)/2a$.
The {\it vis-viva\/} equation yielding the velocity field of elliptical trajectories is thus fully specified by the choice of $\rho_1$, $\rho_2$, and $v_1$ at $\rho_1$.


For parabolic trajectories, the specific energy is zero and we have $\epsilon = v^2_{\rm p}/\rho - \mu/\rho$ = 0, from which the {\it vis-viva\/} orbital speed equation is 
\begin{equation}
  v_{\rm p}(\rho) = \sqrt{\frac{2\mu}{\rho} } 
  = v_1 \sqrt{\frac{\rho_1}{\rho}} 
 \, ,
\label{eq:visviva-parabola}
\end{equation}
where $\rho \in [\rho_1,\infty)$. For a parabola, $e=1$, so $\ell = 2\rho_1$, where the periapsis is $\rho_1=a$ by definition and there is no apoapsis. The {\it vis-viva\/} velocity field of parabolic trajectories is thus fully specified by the choice of $\rho_1$ and $v_1$ at $\rho_1$.


For hyperbolic trajectories, the specific energy is positive, $\epsilon = v^2_{\rm h}/\rho - \mu/\rho = \mu/(2|a|)$, with
{\it vis-viva\/} equation,
\begin{equation}
\begin{array}{rcl}
  v_{\rm h}(\rho) \!\!&=&\! \displaystyle \sqrt{\mu \left( \frac{2}{\rho} + \frac{1}{|a|} \right) } \, ,\\[15pt]
  \!\! &=&\! \displaystyle v_1
\sqrt{\frac{2}{(e\!+\!1)}\frac{\rho_1}{\rho} + \frac{|1-e|}{1+e} }    \, ,
\end{array}
\label{eq:visviva-hyperbola}
\end{equation}
where $e>1$ and $\rho \in [\rho_1,\infty)$. There is no apoapsis.  Unlike the ellipse, where both $e$ and $a$ are defined in terms of the periapsis and the apoapsis, or the parabola for which $e=1$ and $a$ is the periapsis, there is no geometric constraint on the choice of $e$ for a hyperbola. For a periapsis $\rho_1$, we have $a=\rho_1/(e-1)$, where $e$ is a free parameter. The {\it vis-viva\/} velocity field of hyperbolic trajectories is thus fully specified by the choice of $e$, $\rho_1$, and $v_1$ at $\rho_1$.

Along a any conic section, the axial radial component of the velocity trajectory is $V_\uprho = (\mu/h)e\sin \nu$. We can write this expression as
\begin{equation}
\begin{array}{lcl}
  V_\uprho(\rho)  &=& \displaystyle
   \frac{\mu}{h} \sqrt{ e^2 - \left( 1- \frac{\ell}{\rho}  \right)^2} \, , \\[12pt]
 &=& \displaystyle \frac{v_1}{1\!+\!e}
  \left[ e^2 - 
  \left( 1- (e\!+\!1) \frac{\rho_1}{\rho} \right)^2 
  \right] ^{1/2}\, ,
\end{array}
\label{eq:radialconic}
\end{equation}
which follows from the identity $\sin^2 \nu + \cos^2 \nu = 1$ and the rearrangement of Eq.~\ref{eq:ellipsecurve}, yielding $e\cos \nu =  1- (\ell/\rho)$, where $\ell = \rho_1(e+1)$. It can be shown that the quantity $\mu/h = v_1/(1+e)$. Note that $V_\uprho(\rho_1) = 0$ at the periapsis $\rho_1$, as this velocity is purely azimuthal in the orbital plane. For the ellipse, we also find $V_\uprho(\rho_2)=0$ at the apoapsis, which follows from $\rho_1/\rho_2 = (1-e)/(e+1)$.  

The azimuthal velocity component can readily be obtain from the {\it vis-viva\/} orbital speed and the radial component, which gives
\begin{equation}
  V^2_\upphi(\rho) = v^2_{\rm x}(\rho) - V^2_\uprho(\rho)
\, ,
\label{eq:fullazimuthalconic}
\end{equation}
where $\rho \in [\rho_1,\rho_2]$ for an ellipse and $\rho \in [\rho_1,\infty)$ for a parabola or hyperbola, and where
$v_{\rm x}(\rho)$ is given by Eq.~\ref{eq:visviva-elliptical} for an ellipse,  by Eq.~\ref{eq:visviva-parabola} for a parabola, or by Eq.~\ref{eq:visviva-hyperbola} for a hyperbola. 
Over the spatial domain and range of eccentricity of each of the three Keplerian trajectories, it can be shown that Eq.~\ref{eq:fullazimuthalconic} simplifies to 
\begin{equation}
  V_\upphi(\rho) = \left( \frac{\rho_1}{\rho} \right) v_1 = \frac{h}{\rho}
\, ,
\label{eq:azimuthalconic}
\end{equation}
which is a statement of the conservation of angular momentum. We have
\begin{equation}
 {\bf V}_a(\rho) = \displaystyle 
V_\uprho (\rho) \, {\rhohat}(\rho) +
V_\upphi(\rho)\, {\phihat} (\rho) \, , 
\label{eq:keplervels}
\end{equation}
where $V_\uprho (\rho)$ is given by Eq.~\ref{eq:radialconic} and $V_\upphi(\rho)$ is given by Eq.~\ref{eq:azimuthalconic}.
Applying Eq.~\ref{eq:Vlos-general}, the LOS velocity is 
\begin{equation}
  V^{(a)}\subLOS(t) = \displaystyle V_\uprho(t)  {\cal P}_\uprho(t) + V_\upphi(t) {\cal P}_\upphi(t)  \, . 
\label{eq:VLOS-planaraccretion}
\end{equation}

Since, for all conic sections, $\mu = \rho_1 v_1^2/(e+1)$ and $h = \rho_1 v_1$, we see that we define the specific energy and angular momentum by specifying the orbital eccentricity, $e$, and the velocity at periapsis, $v_1$. To match the disk rotation, we would set $v_1=V_c$ for the \citet{steidel02} disk kinematic model or $v_1=V_\upphi(\rho_1,0)$ for the \citet{bizyaev17} disk kinematic model.  For consistency with the accretion radius of the planar wind structure, one would adopt $\rho_1 =\rho_{a,0}$. 

By employing conic-section infalling trajectories, we are effectively assuming that the accreting material begins its ``observable" journey at $R_{a}$, and, conserving energy and angular momentum, ends its journey accreting at the periapsis.  In Figure~\ref{fig:accretion}, we show representative velocity fields and trajectories in the $z=0$ plane for an ellipse with $e=0.82$, a parabola ($e=1$) and a hyperbola with $e=1.5$.  

\subsubsection{Planar Accretion Kinematics}

The Keplerian model captures many of the most salient attributes of the accretion kinematics as predicted by hydrodynamic cosmological simulations \citep[in particular, see][]{hafen22,trapp22, stern23, kocjan24}.  As the gas spirals radially inward toward the disk, it largely conserves angular momentum until it achieves rotational support, typically {\it just after crossing the gaseous disk edge\/} where it tends to ``pile up." That is, most gas joins the disk slightly interior to the disk edge, which is well beyond the stellar disk. As the accreting gas amalgamates with the disk while transitioning through this narrow ``accretion zone", its radial velocity component diminishes rapidly.  The gas enters the accretion zone at angles ranging from $0^\circ$--$30^\circ$ relative to the vertical of the disk plane; gas accreting from larger angles has larger vertical velocities during the process. Entering the accretion zone, the vertical velocities tend to increase, and once in the disk, the average vertical velocity settles down to small fraction of the circular velocity.  Similarly, the axial average radial velocity settles down to a non-zero finite fraction of the circular velocity.  

\begin{figure*}[tbh]
\centering
\includegraphics[width=1.0\linewidth]{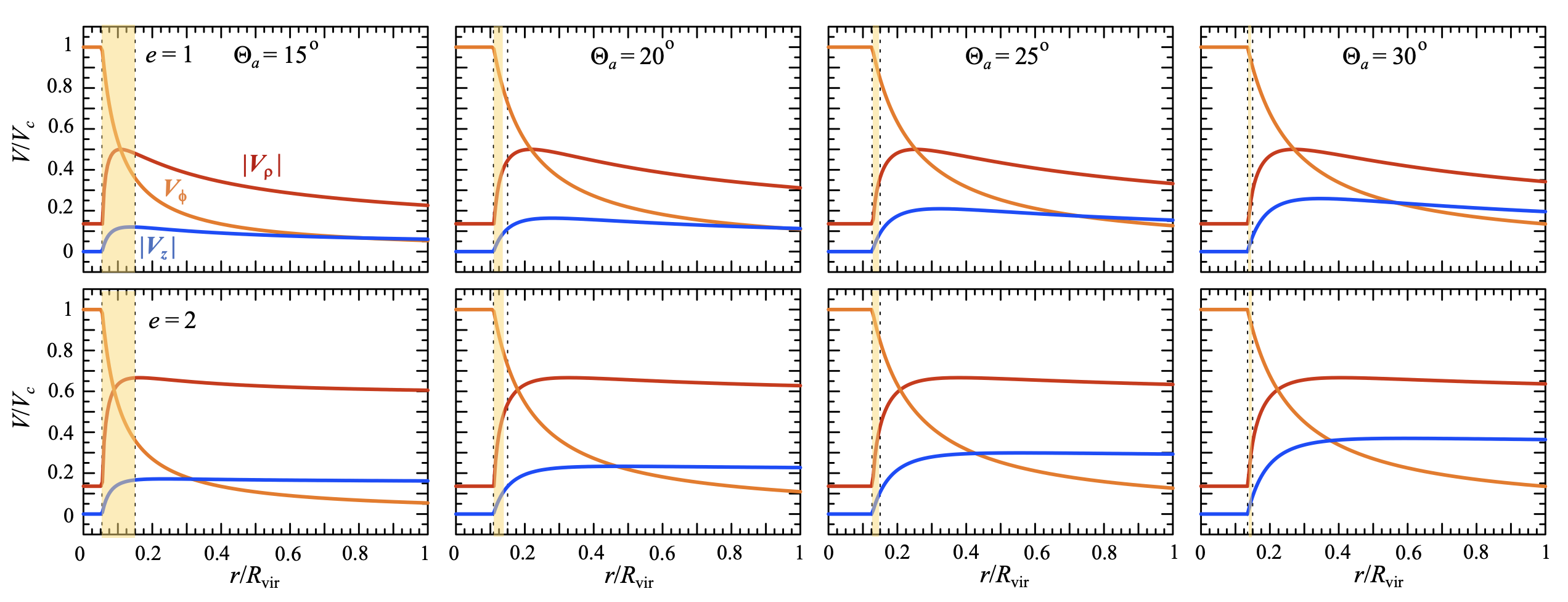}
\vglue -0.05in
\caption{\small The axial velocity profiles normalized to $V_c$ as a function of $r/R_{\rm vir}$ for two Keplerian infall trajectories (upper) the parabolic trajectory with $e=1$, and (lower) a hyperbolic trajectory with $e=2$. The yellow shaded area is the accretion zone as determined by Eq.~\ref{eq:makeAzone} and illustrated in Figure~\ref{fig:Azone}; the inner boundary is at $\rho_{a,0}$ and the outer boundary is at $\rho_d$. The accretion opening angle increases from left to right, ranging from $15^\circ \leq \Theta_a \leq 30^\circ$.  The azimuthal and axial radial velocities, $|V_\uprho(r)|$ and $V_\upphi(r)$,  are obtained from Eq.~\ref{eq:radialconic} and Eq.~\ref{eq:azimuthalconic}, respectively.  The vertical velocity component, $|V_z(r)|$, represents the upper limit as described by Eq.~\ref{eq:zvelmaxmodel}.  Solid blue curves include accretion breaking as described in the text, where as the dashed curve is with no accretion breaking. The model parameters for the galactic disk for this illustration are $\rho_d = 0.15R_{\rm vir}$ and $h_d = 0.2 \rho_d$, with $V_\uprho = -0.14V_c$ and $V_z = - 0.1V_c$ for $r < \rho_{a,0}$.  These curves conserve angular momentum for $\rho_{a,0} \leq r \leq R_{\rm vir}$. These velocity profiles follow theoretical expectations from simulations \citep{hafen22,trapp22,stern23}.  In particular, the parabolic trajectory with $\Theta_a = 20^\circ$ is comparable to Figure~3(c) of \citet{hafen22} and Fig.~2 of \citet{stern23}.}
\label{fig:velprofs}
\vglue 0.1in
\end{figure*}

In Figure~\ref{fig:velprofs}, we show the $V_\upphi(\rho)$, $|V_\uprho(\rho)|$, and $|V_z(\rho)|$ velocity profiles for two Keplerian planar accretion trajectories, a parabola with $e=1$ (upper panels) and a hyperbola with $e=2$ (lower panels).  The small variations in the profiles are highlighted for four accretion opening angles, $\Theta_a = 15^\circ$, $20^\circ$, $25^\circ$, and $30^\circ$. We assumed a disk axial radius of $\rho_d = 0.15R_{\rm vir}$ and height $h_d = 0.2 \rho_d$.  For each $\Theta_a$, we computed the accretion radius, $\rho_{a,0}$, using Eq.~\ref{eq:makeAzone} and defined the accretion zone as illustrated in Figure~\ref{fig:Azone}.  In Figure~\ref{fig:velprofs}, the accretion zones are shaded yellow. Guided by the simulations of \citet{hafen22}, we adopted $V_\uprho = -0.14V_c$. 

The azimuthal velocity conserves specific angular momentum, $h= \rho V_\upphi(\rho) = \rho_{a,0}V_c$, matching the disk rotation at the inner edge of the accretion zone (by design). The axial radial velocity increases slightly as the gas approaches the disk, but this profile is flatter for higher energy trajectories (larger $e$). As seen in simulations, it drops rapidly in the vicinity of the accretion zone.  Note that these velocity profiles are very different than those utilized in the Modified Accretion Disk (MAD) models of \citet{wang23}, which are more similar to the logarithmic spiral model we describe in Appendix~\ref{app:D}.  However, they do have similar features to those of \citet{stern23}.  Note that as the gas approaches the inner edge of the accretion zone, we truncate $V_\uprho$ to have the value adopted for the galactic disk. 

The Keplerian model is defined in a plane.  We assume that the extended accretion comprises a continuum of oblique planes over the range $-\Theta_a \leq \theta \leq +\Theta_a$.  We thus write Eq.~\ref{eq:VLOS-planaraccretion} in terms of radial accretion confined on a plane within the extra planar accretion structure, 
\begin{equation}
  V^{(a)}\subLOS(t) = \displaystyle V_r(t)  {\cal P}_r(t) + V_\upphi(t) {\cal P}_\upphi(t)  \, , 
\label{eq:VLOS-newplanaraccretion}
\end{equation}
where $V_r(t)$ is simply reframed from Eq.~\ref{eq:radialconic}, and, on a given plane, the vertical component of this now radially directed infall velocity is 
\begin{equation}
V_z(\rho,z) = V_\uprho (r)  \tan \theta = \frac{z}{\rho} V_\uprho (r) \, ,
\label{eq:zvelmodel}
\end{equation}
for $\rho > \rho_{a,0}$, where we note that $V_\uprho(r) < 0$ always.
In Figure~\ref{fig:velprofs}, we plot the maximum of $|V_z(\rho,z)|$, which is calculated for the planes of maximum obliquity, for which $z = \pm (\rho^2 - \rho_{a,0}^2)/C_\Theta$.  Thus, the plotted velocity profiles (blue curves) are  \begin{equation}
V_z(r) = \frac{\rho^2 - \rho_{a,0}^2}{C_\Theta \rho} V_\uprho (r)  \, ,
\label{eq:zvelmaxmodel}
\end{equation}
and they should be interpreted as the upper envelope of the distribution of vertical velocity profiles.  

These idealized velocity profiles are in general agreement with the average profiles found in the hydrodynamic cosmological simulations of \cite{hafen22}, \cite{trapp22}, \citet{stern23} and \citet{kocjan24}.  \citet{hafen22} find that this alignment, cooling, and onset of rotational support occurs over a short time period of $\sim 300$--600~Myr. 

\subsection{Summarizing Spatial-Kinematic Models}

Selected idealized spatial-kinematic models are listed in Table~\ref{tab:kinematics-downtocases}. The kinematics of each model is specified by one to four free parameters.  The spatial geometry of each CGM component was described in Section~\ref{sec:geometries} and their LOS locations were summarized in Table~\ref{tab:downtocases}.

For the spherical halo, the simplest kinematics is radial outflow or inflow specified by the constant radial velocity, $V_r$. The stall function, ${\cal S}(r;\eta_s,\Delta\eta_s)$, or the accretion breaking function, ${\cal B}(r;f\subT, \eta_b,\Delta\eta_b)$, can be applied to model deceleration and terminal infall velocities. Alternatively, the radial velocity can be specified to vary with galactocentric radius. 

Disk rotation and extra-planar gas rotational gradients can be modeled with theoretically predicted or empirically motivated velocity fields.  Here, we have considered the models of \citet{steidel02} and \citet{bizyaev17} as constrained by the observations of \citet{bizyaev22-lags}.  Additional kinematics can also be specified.  For example, a gentle axial ``migration'' within the disk can be modeled by adding a $V_\uprho ({\rvec})$ component \citep[see][]{ho17,martin19}.  One could also trivially add a vertical infall or outflow component, $V_z ({\rvec})$, if desired.

\begin{table}[htb]
\centering
\caption{Summary of Spatial-Kinematic Models}
\begin{tabular}{lll}
\hline\\[-8pt]
Spatial & Kinematic & Kinematic \\
Component  & Model  & Parameters       \\[2pt]
\hline\\[-8pt]
halo & radial outflow  &  $V_r$; $V_c$; ${\cal S}(r)$ \\
    & radial infall  &  $V_r$; $V_c$; ${\cal B}(r;f\subT)$ \\
disk/EPG & Steidel & $V_c$; $H_v$  \\
         & Bizyaev & $V_c$; $\rho_{d,0}$; $dV_\upphi/dz$; $dV_\upphi/d\rho$  \\
wind & hyperbolic outflow & $V_w$; $\tilde{V}\subE f\subE(\rho,z)$ \\
     & radial & $V_w$; $\tilde{V}\subE f\subE(\rho,z)$ \\
accretion  & Keplerian infall  &  $V_c$ \\
& logarithmic spiral$^{\rm a}$ &   $V_c$; $k$; $\rho_{a,0}$; $R_a$   \\[2pt]
\hline\\[-8pt]
\multicolumn{3}{l}{(a) see Appendix~\ref{app:E}}
\end{tabular}
\label{tab:kinematics-downtocases}
\end{table}

The bi-polar wind is commonly modeled as a purely radial component. We have slightly modified the radial velocity field to handle purely vertical launch velocities from the disk plane.  We call this a ``hyperbolic outflow."  Variation in the wind speed can be specified as a function of galactocentric distance, or a rapid deceleration can be applied using a stall function. One can trivially add a slight rotational component, ${V}_\upphi ({\rvec})$, if desired.  We have also considered axial variation in wind velocity in the form of a function called the enhancement velocity, $V\subE(\rho,z)$.  

The modeling of extended planar accretion is more challenging \citep[see][]{stern23, wang23}.  Accretion kinematics can be described by a range of velocity fields that exhibit some degree of coupling to the gas disk rotation.  Here, we have presented a simplistic model of Keplerian infall trajectories normalized to the disk circular velocity or rotation velocity at the trajectory periapsides. Note that the energy of the trajectories are higher for larger eccentricity, which provide the highest infall velocities at larger galactocentric distances.


\begin{figure*}[tbh]
\centering
\includegraphics[width=0.99\linewidth]{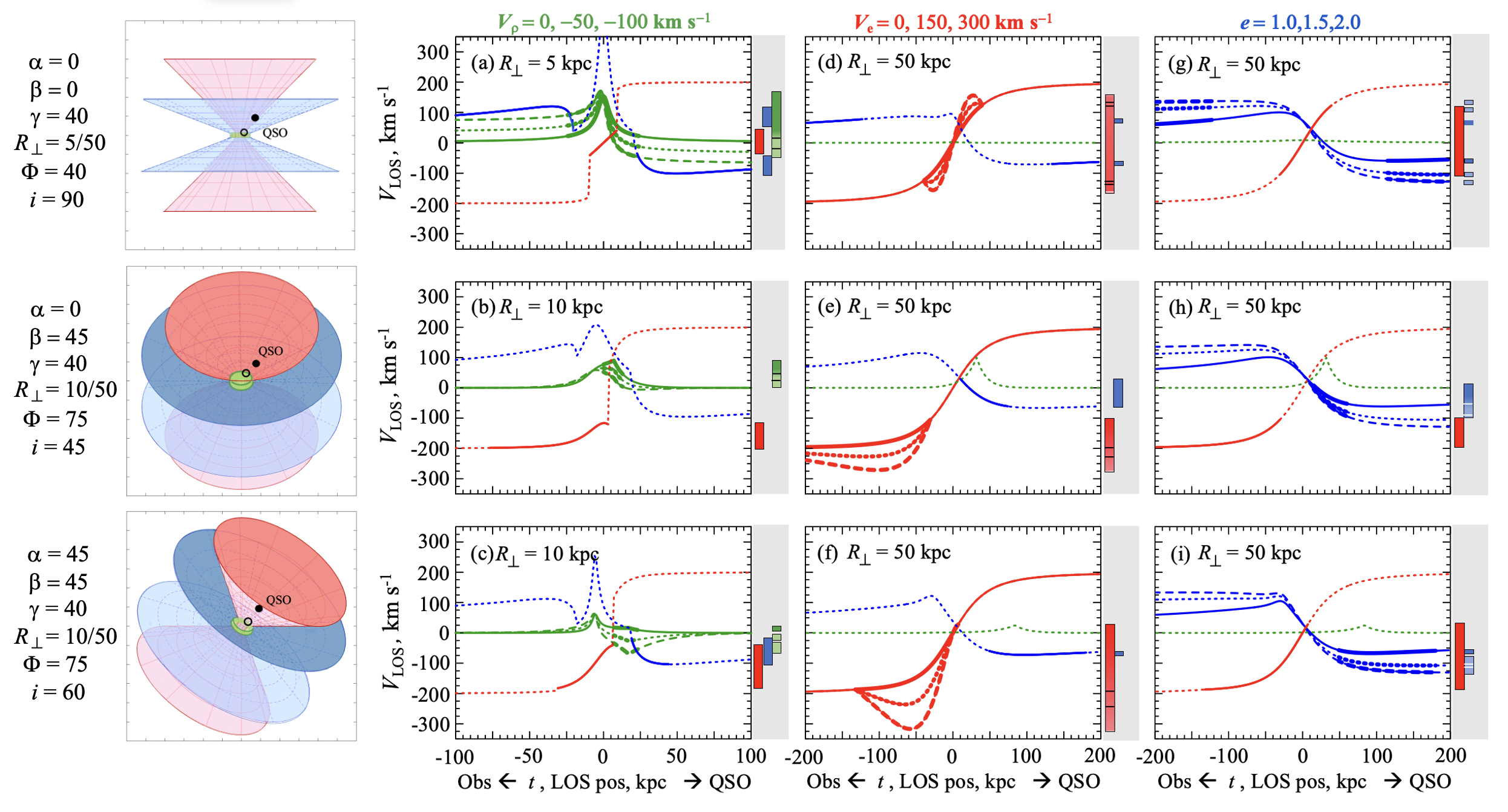}
\caption{\small Example $V\subLOS$--LOS plots showing LOS velocities as a function of LOS position for a spatial-kinematic CGM model of a Milky Way-like galaxy. The model parameters are described in the text. (left panels) The sky plane (observer view) of the galaxy/CGM spatial components showing the gaseous disk (green), bi-conical wind (red), plus extended planar accretion (blue). The observer coordinate rotations and galaxy/LOS orientations are given in the legends. Recall that the outflowing material is confined within the wind hyperboloid structure, whereas and the infalling material is confined within the void formed by the accretion hyperboloid structure.   (right panels) The LOS velocities for the three spatial-kinematic components (colored to match the structures).   Thick curves are LOS locations that probe the material within the structures (see Table~\ref{tab:downtocases}).  Observed LOS velocity ranges for each structure are shown as color bars to the right of each panel. (a,b,c) LOS velocities for the face-on galaxy: (a) Variations in the disk LOS velocities for three values of axial migration, $V_\uprho=0$~{\kms} (solid), $-50$~{\kms} (dotted), and $-100$~{\kms} (dashed), for $R_\perp = 5$~kpc; (b) Variations in the wind enhanced velocity, $V\subE=0$~{\kms} (solid), $150$~{\kms} (dotted), and $300$~{\kms} (dashed), for $R_\perp=50$~kpc; (c) Variation in the accretion eccentricity, $e=1$ (solid), $1.5$ (dotted), and $2.0$ (dashed), for $R_\perp=50$~kpc.  (d,e,f) Same as for (a,b,c) but for the obliquely inclined galaxy and indicated $R_\perp$.  (g,h,i) Same as for (a,b,c) but for the directly inclined galaxy and indicated $R_\perp$.}
\label{fig:vlosmodels}
\vglue 0.1in
\end{figure*}

\section{Example LOS Velocity Plots}
\label{sec:exampleSKAM}

In Figure~\ref{fig:vlosmodels}, we illustrate LOS velocities for selected examples of LOS--galaxy orientations and galaxy/CGM kinematics.  The spatial-kinematic model comprises a Steidel disk (green curves), hyperbolic outflowing wind (red curves), plus Keplerian infall extra planar accretion (blue curves). Each curve represents $V\subLOS(t)$ [{\kms}], where $t$ is the LOS position [kpc] increasing in the direction from the observer to the background source. 

For this Milky-Way like model, the galaxy virial mass and radius is defined by $V_c = 220$~{\kms} and the disk axial radius and height are $\rho_d=25$~kpc and $h_d=5$~kpc, respectively.  The exponential velocity scale height for the Steidel disk model is $H_v = 30$~kpc.  The wind opening angle is $\Theta_w=55^{\circ}$ with base radius $\rho_{w,0}= 10$~kpc and radial extent $R_w=R_{\rm vir}=200$~kpc.  The extra planar accretion has flare angle $\Theta_a=20^{\circ}$ with accretion radius $\rho_{w,0}= 23$~kpc (see Eq.~\ref{eq:makeAzone}) and radial extent $R_a=R_{\rm vir}$. We fix the position angle for the quasar at $\gamma = 40^\circ$.  

In the left hand panels of Figure~\ref{fig:vlosmodels}, we show the sky plane with the observer perspective of the galaxy/CGM structures for three representative orientations.  The first is an edge-on view $(\alpha,\beta = 0^\circ,0^\circ)$, which yields an observer perspective with galaxy-quasar sky-projected azimuthal angle $\Phi\!=\!\gamma\!=\!40^\circ$ and inclination $i=90^\circ$. The second is a purely inclined orientation $(\alpha,\beta=0^\circ,45^\circ)$, which yields $\Phi=40^\circ$ and $i=45^\circ$. The third is an obliquely inclined orientation $(\alpha,\beta=45^\circ,45^\circ)$, which yields $\Phi=75^\circ$ and $i=60^\circ$. For the background quasar, we adopt impact parameters of $R_\perp = 5$ and $10$~kpc for probing the disk kinematics (open points labeled ``QSO"), and $R_\perp = 50$~kpc for the wind and accretion illustrations (solid ``QSO" points).

In Figure~\ref{fig:vlosmodels}(a,b,c), we emphasize variations in disk LOS velocity for three values of the axial migration velocity, $V_\uprho(\rho) = 0$, $-50$, and, $-100$~{\kms}. In panels~\ref{fig:vlosmodels}(d,e,f), we emphasize variation in the wind LOS velocity for three values of the wind enhancement velocity, $V\subE(\rho,z)$, using a truncated cosine enhancement function with ${\cal M} _{V_w}= 0$, 0.75, and 1.5 with $V_w(z) = 200$~{\kms}. In panels~\ref{fig:vlosmodels}(g,h,i), we emphasize variations in the accretion LOS velocities for three values of the eccentricity ($e=1.0$, 1.5, and 2.0), ranging from parabolic to higher eccentricity hyperbolic Keplerian trajectories. 

Consider the effect of different $V_\uprho$ on the disk LOS velocity for the edge-on galaxy, which is shown in Figure~\ref{fig:vlosmodels}(a) at $R_\perp=5$~kpc.  For the disk component (green curves), the thick portions of the LOS velocity curves are where the LOS is probing the disk structure and the thinner portions are what the LOS velocity would be if the structure extended to those LOS positions.  The solid curve represents no axial migration, just pure rotation.  The dotted curve is $V_\uprho = -50$~{\kms}, and the dashed curve is $V_\uprho = -100$~{\kms}.  These are extreme values, but they illustrate the sensitivity of the LOS velocity to inward axial migration in the disk. For the wind (red curves) and accretion (blue curves), the solid portion is where the LOS probes the structure and the dotted portions are where the LOS position is outside the structure.  For $R_\perp=5$~kpc, the LOS probes the accretion structure virtually for the full extent of the CGM, except within the inner $\simeq 25$~kpc near the structure orifice and the galaxy center, but probes the wind near its base over a relatively short path length.  The LOS velocity ranges are presented as color bars along the right axis. Note that the accretion LOS velocities are confined to two very narrow ranges at $V\subLOS \simeq -100$ and $\simeq +100$~{\kms} even the LOS path length is $\simeq 100$~kpc for each.  The wind, on the other hand accounts for a continuous velocity range from $V\subLOS \simeq -50$  to $+50$~{\kms} even though the LOS path length is $\simeq 30$~kpc. 

In Figure~\ref{fig:vlosmodels}(b,c), we show how the disk LOS velocities can vary with $V_\uprho$ for two different inclinations and azimuthal angles. For these examples, we adopt $R_\perp =10$~kpc.  Note that simply inclining the galaxy reduces the LOS velocity projection of the disk so that variations in $V_\uprho$ are diminished. Note also that the wind is probed for a path length of $\sim 100$~kpc on the observer side of the sky plane and that, relative to the edge-on orientation, the LOS velocities are blue shifted ($\simeq -125$ to $-200$~{\kms}) relative to the galaxy rest frame. Finally, inclining the galaxy causes the LOS to fully miss the accretion structure.  This occurs for two reasons.  First, in this case, $\Theta_a < \beta$ and second $R_\perp < \rho_{a,0}$, so that the LOS is passing ``over/under" the flared structure for larger $r(t)$ and is passing ``through" the inner orifice of the structure for smaller $r(t)$. Note however, that when the galaxy is obliquely inclined (panel~\ref{fig:vlosmodels}(c)), the accretion structure is probed along its inner extremity on the far side of the galaxy.

Now consider the wind LOS velocity for different enhanced velocities. For the curves (red) shown in Figure~\ref{fig:vlosmodels}(d,e,f), we adopted a truncated cosine for the enhancement function. The solid curve is no enhancement, i.e., $V\subE\!=\!0$, the dotted curve is $V\subE\!=\!150$~{\kms}, and the dashed curve is $V\subE\!=\!300$~{\kms} for a wind with $V_w=200$~{\kms} probed at $R_\perp \!=\! 50$~kpc.  Thicker portions of the curves are where the wind structure is being probed by the LOS.  In panel~\ref{fig:vlosmodels}(d), we note that enhancing the wind velocity makes a negligible change in the observed LOS velocity, as we are viewing perpendicular to the polar axis (edge on to the galaxy). Furthermore, for $R_\perp = 50$~kpc and $\Phi=40^\circ$, we probe the wind far from the wind axis where the enhancement is minimal.  Note that the LOS probes the wind symmetrically about the sky plane; again, this is because the wind structure is perpendicular to the observer. 

If we incline the galaxy (panel~\ref{fig:vlosmodels}(e)), we see two effects.  First, the wind structure is now tilted with respect to our perspective and we probe the structure that is on the observer side of the sky plane for a path length of $\simeq 170$~kpc.  Second, the effects of different levels of enhancement in the wind velocity are more pronounced, adding an additional LOS velocity spread of $\simeq 75$~{\kms}. However, if we now rotate the polar axis by $\alpha =45^\circ$ (panel~\ref{fig:vlosmodels}(f)) so that it is not parallel to the LOS, then for this obliquely inclined galaxy, we probe closer to the wind axis and we observe enhanced LOS velocities in the range of $V\subLOS \simeq -200$ to $\simeq -300$~{\kms}  (even though we probe a shorter path length through the wind structure).  

Now, consider the accretion LOS velocity for three Keplerian eccentricities, $e=1$ (solid curve), $e=1.5$ (dotted curve), and $e=2.0$ (dashed curve), as illustrated in Figure~\ref{fig:vlosmodels}(g,h,i). We adopted the same LOS/galaxy orientations with $R_\perp=50$~kpc presented in Figure~\ref{fig:vlosmodels}(e,f,g). The thicker portions of the blue curve are where the accretion structure is probed by the LOS.
As the eccentricity is increased, the energy and {\it vis-viva\/} velocity of the accreting material is increased.  As can be seen, changing the eccentricity slightly shifts the LOS velocities.  Note that for the edge-on galaxy (panel~\ref{fig:vlosmodels}(g)), the accretion structure is probed on both sides of the sky plane and, on account of infalling kinematics, both positive and negative LOS velocities are observed. However, once the galaxy is inclined (panel~\ref{fig:vlosmodels}(h)) such that $90-i > \Theta_a$, then the LOS probes the accretion structure only on one side of the sky plane, resulting in either in exclusively positive or exclusively negative accretion LOS velocities. In the examples shown here, the galaxy is inclined using $\beta>0$ and the LOS probes the ``far side" of the accretion structure, yielding negative accretion LOS velocities. For $\beta = -|\beta|$, the same inclination is observed, but the LOS would probe the ``near side" of the accretion structure, yielding positive accretion LOS velocities. For an obliquely inclined galaxy (panel~\ref{fig:vlosmodels}(i)) the path length through the accretion structure is increased but the LOS velocities are relatively unchanged compared to a galaxy with a non-oblique inclination.  However, note that the path length probing the accretion structure through the obliquely inclined galaxy is roughly 150\% longer.



\section{Concluding Remarks}
\label{sec:conclusion}

We have derived the basic mathematical framework for building idealized spatial-kinematic absorption models (SKAMs) comprising multiple geometric galaxy/CGM components each with their own 3D velocity fields. In this Paper~I of a two paper series, we focused on geometric and kinematic relationships. We then presented a few selected models using $V\subLOS$--LOS plots. In \citetalias{churchill25-skamII}, we will develop formalism for generating absorption lines from these SKAMs. This Paper~I can be summarized as follows: 

\begin{enumerate}

\item We defined the central galaxy in terms of its NFW dark matter halo properties, which are characterized by three free parameters, the maximum circular velocity, the overdensity contained within the virial radius, and the concentration parameter. We derived the mathematical formalism describing the galaxy/CGM as a set of geometric structures and their spatial relationships, including the sky-view perspective of an arbitrarily positioned distant observer and the position along the LOS of the background quasar.  

\item We adopted four galaxy/CGM structures to represent the components of a fiducial galaxy/CGM model based on observations and theoretical considerations. The galaxy disk is represented by a cylinder and the ``halo" is represented by a sphere. Hyperboloids of one sheet represents both a bi-polar wind and a flared extended planar accretion zone. The spatial domains of these geometric structures can be easily adjusted using a small set of free parameters.  

\item We derived a set of simple quadratic inequalities for each galaxy/CGM structure that identify the positions along the LOS that reside within the spatial domains of the respective geometric structure. The roots of these quadratic equations yield the position(s) on the LOS that intersect the geometric surfaces. The interpretation is that LOS positions within the spatial domain of a given geometric structure are probing the galaxy/CGM component represented by the structure.

\item We developed the formalism for obtaining the LOS velocity as a function of position along a LOS that probes an arbitrary 3D velocity field from any orientation and impact parameter through the SKAM.  A set of simple LOS projection functions have been tabulated for vertical, radial, azimuthal, axial, and polar velocity components. As a function of position along the LOS, these effectively provide the scalar dot product of the LOS vector and the respective velocity vector component. 

\item We derived expressions for 3D velocity fields for each galaxy/CGM structure that are motivated from observations and theoretical considerations. The velocity field corresponding to a given galaxy/CGM structure is simply nulled outside its spatial domain (or more simply, when the position along the LOS is outside a structure spatial domain). These velocity fields are specified by a small set of adjustable kinematic parameters. Additional physics-based realism can be incorporated in the form of rotational lag and adjustable ``stalling," ``breaking," and ``enhancement" functions. 

\end{enumerate}

Though SKAMs lack the higher degree of realism represented in computational methods such as hydrodynamical simulations, they provide a valuable tool with which we can directly confront the distilled essence of our ideas about the nature of the CGM.  In addition, analytic models provide a useful gateway between observations and simulations.  SKAMs can be immediately understood on an intuitive level. Their intuitive-based predictions are founded on simple physical constructs that help us untangle the complexity of astrophysics and computational constraints bedeviling the predictions of simulations.  

Examination of $V\subLOS$--LOS plots shows that there can be a degeneracy in $V\subLOS$ along the LOS. That is, for a given LOS/galaxy orientation, spatial-kinematic models clearly reveal that similar/overlapping LOS velocities can arise from multiple unique and spatially separated CGM structures (different LOS positions probing different CGM structures can yield a similar/overlapping LOS velocity).  In Figure~\ref{fig:vlosmodels}, the color bars on the right hand sides of the individual panels show the degree of these overlaps for the few simple illustrations in this work.  For example, the LOS may sample the wind paraboloid over a finite LOS velocity, but it might also sample the extended accretion structure over this same LOS velocity range. This indicates that absorption lines that may arise from the different spatial-kinematic components of the CGM would be superposed in the observed spectrum \citep[also see][]{marra22}.  

How is it that we can determine from which structure (galaxy/CGM component) absorption arises?  Or, how can we estimate the relative proportions of the absorption from each galaxy/CGM structure? Forward modeling of absorption systems observed to have multiple ions spanning many ionization states is one method \citep[see][]{sameer21, sameer22, sameer24}. For example, as shown in Fig.~C10 of \citet{sameer24}, what appeared as a single component in {\CIII} at $50$~{\kms} actually comprises two distinct gas phases. Such detailed decomposition is enabled by using profile shapes from all the observed lines.  For such cases, inferring which CGM structures are giving rise to the distinct gas phases from the modeling would require spatial information. Even if the LOS/galaxy orientation is known, there remains ambiguity as to where along the LOS the gas resides, and this is information needed for inferring outflow or infall because it breaks degeneracies.  Incorporating SKAMs that produce realistic absorption profiles into the analysis can provide the requisite spatial-kinematic information.




If there are differences in the average gas phase properties (density, temperature, chemical abundances, and ionization conditions) for the different idealized structures, then the absorption from different galaxy/CGM structures would reveal those signatures.   
An advantage of generating synthetic absorption profiles with SKAMs is that the optical depth from each individual galaxy/CGM structure can be synthesized as a function of LOS velocity.  The magnitudes of the optical depths can be compared and the contribution to the absorption from each structure can be quantified. 

This last consideration motivates us to investigate the predicted observable effects of the gas physical conditions for a given CGM spatial-kinematic structure and what those conditions would need to be in order to yield absorption lines generally consistent with observations.  In \citetalias{churchill25-skamII}, we will extend our formalism beyond models that simply predict LOS velocities as a function LOS positions for idealized spatial-kinematic components of the CGM.  To synthesize absorption profiles for atomic transitions of various ions, we need to include formalism for determining the optical depths of various ion transitions as a function of LOS position and velocity. Since optical depth is proportional to the product of the number density of the absorbing ion and the total absorption cross section, we must specify the densities of all ions and the line-broadening physics of the gas at all locations within each unique CGM structure. The first step will be to populate the CGM structures with gas density-temperature-ionization fields (in a manner similar to how the velocity fields were defined). The second step will be to develop the radiative transfer formalism accounting for degenerate LOS velocities. 

A rudimentary SKAM GUI (with the capability to generate absorption lines) is available as a Fortran 95 code at \href{https://github.com/CGM-World}{github.com/CGM-World}.

\section*{Acknowledgments}

The author thanks Hasti Nateghi for suggesting this project during her extended visit to New Mexico State University. Thanks also to Jane Charlton, Stephanie Ho, Glenn Kacprzak, Nikki Nielsen, Ben Oppenheimer, Sameer, and Evan Schneider for helpful comments. 

\newpage

\appendix

\section{A. The LOS Equations}
\label{app:A}

As shown in Figure~\ref{fig:thecoordinaterotation}, we set up two right-handed 3D coordinate systems with coincident origins. The first is the galaxy frame, {\Gsys}, and the second is the observer frame, {\Osys}. We denote coordinates in {\Gsys} as $x,y,z$ with basis vectors ${\ihat}$, ${\jhat}$, and ${\khat}$ and coordinates in {\Osys} as $x_o,y_o,z_o$ with basis vectors ${\ihato}$, ${\jhato}$, and ${\khato}$.
In {\Gsys}, we place the galaxy at the origin, represented by point $P\subG(0,0,0)$. The galaxy rotational axis is the $z$ axis and the galaxy disk lies in the $xy$ plane. In {\Osys}, the galaxy is also at the origin, represented by point $P\subO(0,0,0)$. 

In {\Osys}, we place the observer at $x_o = +\infty$ and the quasar at $x_o = -\infty$.  Thus, in {\Osys}, the LOS vector direction, ${\shat}$, is fixed and simply given by ${\shat} = - {\ihato}$.  As shown in Figure~\ref{fig:skyplanedefs}, the sky plane is defined as the $y_oz_o$ plane, which is perpendicular to the LOS vector.  The point where the LOS intersects the sky plane is given by the impact parameter, $R_\perp$, and the position angle, $\gamma$. Thus, the LOS vector intersects the sky plane at the observer coordinates $x_o = X\supQ\sky = 0$, $y_o = Y\supQ\sky = R_\perp \cos\gamma$ and, $z_o = Z\supQ\sky = R_\perp \sin\gamma$; this point on the LOS is denoted $P\subO(0,R_\perp \cos\gamma,R_\perp \sin\gamma)$.

We define physical position along the the LOS using the parameter $t$, defined such that $t=0$ is the sky-plane intersection point $P\subO(0,R_\perp \cos\gamma,R_\perp \sin\gamma)$, the observer is at $t=-\infty$ and the quasar is at $t=\infty$.  Thus, the 3D position along the LOS in {\Osys} is simply $x_o(t) = -t$, $y_o(t) = R_\perp \cos\gamma$, and $z_o(t) = R_\perp \sin\gamma$.

To account for arbitrary viewing angles of the galaxy we adopt rotation of {\Osys} with respect to {\Gsys}.  The reason is that it is mathematically expedient to describe the equation of a LOS vector in a rotated 3D space relative stationary structures than it is to describe the spatial distributions of multiple rotated structures.  A minimum of two rotations of {\Osys} is required, as illustrated in Figure~\ref{fig:thecoordinaterotation}. Together, the two rotations account for all possible orientations of the LOS relative to the galaxy. The first rotation is by the angle $\alpha$, which is counterclockwise about the galaxy $z$ axis. The second rotation is by the angle $\beta$ about the {\it rotated\/} $y_o$ axis. 

The rotations are implemented through the operation $[x,y,z]\supT\subG = R_y(\beta)R_z(\alpha) [x_o,y_o,z_o]\supT\subO$, where $R_y(\beta)R_z(\alpha)$ is the rotation matrix for the two rotations $\alpha$ and $\beta$, applied in that order.  We have
\begin{equation}
\begin{bmatrix}
 x \\ y \\ z
 \end{bmatrix} 
 =
\begin{bmatrix}
\cos\beta\cos\alpha & \cos\beta\sin\alpha & -\sin\beta\\
-\sin\alpha & \cos\alpha & 0 \\
\sin\beta\cos\alpha & \sin\beta\sin\alpha & \cos\beta 
\end{bmatrix}
 \begin{bmatrix}
 x_o \\ y_o \\ z_o
 \end{bmatrix} \, .
\label{eq:rotateGaroundO}
\end{equation}
For the reverse rotations, we write $[x_o,y_o,z_o]\supT\subO = [R_y(\beta)R_z(\alpha)]\supT [x,y,z]\supT\subG$, where $[R_y(\beta)R_z(\alpha)]\supT$ is the transpose matrix of $R_y(\beta)R_z(\alpha)$.  We have
\begin{equation}
 \begin{bmatrix}
 x_o \\ y_o \\ z_o
 \end{bmatrix}
 =
\begin{bmatrix}
\cos\beta\cos\alpha & -\sin\alpha  & \sin\beta\cos\alpha \\
\cos\beta\sin\alpha & \cos\alpha & \sin\beta \sin\alpha  \\
-\sin\beta &  0 & \cos\beta 
\end{bmatrix}
  \begin{bmatrix}
 x \\ y \\ z
 \end{bmatrix} \, .
\label{eq:rotateOaroundG}
 \end{equation}
Eq.~\ref{eq:rotateGaroundO} maps the point $P\subO(x_o,y_o,z_o)$ in system {\Osys} to point $P\subG(x,y,z)$ in system {\Gsys}, whereas Eq.~\ref{eq:rotateOaroundG} maps the point $P\subG(x,y,z)$ in {\Gsys} to point $P\subO(x_o,y_o,z_o)$ in {\Osys}.  The latter is useful for viewing projections of the galaxy and CGM structures in the observer frame, especially the view of the sky plane, $P\subO(0,y_o,z_o)$. 

We express the 3D position of the LOS in {\Gsys} as a function of LOS position $t$ using the parametric equations,
\begin{equation}
\begin{array}{rcl}
x(t) \!\!&=&\! X_{0} + \sigma_x t \, , \\[2pt] 
y(t) \!\!&=&\! Y_{0} + \sigma_y t \, , \\[2pt] 
z(t) \!\!&=&\! Z_{0} + \sigma_z t \, , 
\end{array}
\label{eq:xyzlos-fordef}
\end{equation}
where $P\subG(X_{0},Y_{0},Z_{0})$ is the point on the LOS at ${t=0}$ (the intersection with the sky plane) and $\sigma_x$, $\sigma_y$, and $\sigma_z$ are the direction cosines.  As we previously stated, in {\Osys} the LOS intersects the sky plane at point $P\subO(0,R_\perp \cos\gamma,R_\perp \sin\gamma)$ so we see that this point on the LOS is coincident with the point $P\subG(X_{0},Y_{0},Z_{0})$ in {\Gsys}. To determine point $P\subG(X_{0},Y_{0},Z_{0})$, we set $[x,y,z]\supT\subG = [X_0,Y_0,Z_0]\supT\subG$ for the left-hand side of Eq.~\ref{eq:rotateGaroundO}, and set $[x_o,y_o,z_o]\supT\subO = [0,R_\perp \cos\gamma,R_\perp \sin\gamma]\supT\subO$ for the right-hand side of Eq.~\ref{eq:rotateGaroundO}. Carrying out the rotation, we obtain 
\begin{equation}
\begin{array}{rcl}
    X_{0} \!\!&=&\! R_\perp \cos\gamma \cos\beta \sin\alpha - R_\perp \sin\gamma \sin\beta  \, , \\[1pt] 
    Y_{0} \!\!&=&\! R_\perp \cos\gamma \cos\alpha \, ,  
    \\[1pt] 
    Z_{0} \!\!&=&\! R_\perp \cos\gamma \sin\beta \sin\alpha + R_\perp \sin\gamma \cos\beta\, .
\end{array}
\label{eq:skypos-derived}
\end{equation}

Since, in {\Osys} the LOS is anti-parallel to the $x_o$ axis, we immediately have the direction cosines for the LOS in {\Gsys} from the top row of the rotation matrix in Eq.~\ref{eq:rotateOaroundG}. This can be seen by setting $[x,y,z]\supT\subG = [{\ihat},{\ihat},{\ihat}]\supT\subG$, and, given that ${\shat} = - {\ihato}$, setting $[x_o,y_o,z_o]\supT\subO = [-{\shat},0,0]\supT\subO$. The rotation yields ${\shat} = \sigma_x\, {\ihat} + \sigma_y\, {\jhat} + \sigma_z\, {\khat}$, where
\begin{equation}
\begin{array}{rcl}
\sigma_x \!\! &=& \!  -\cos\beta\cos\alpha \, ,\\[1pt]
\sigma_y \!\! &=& \!  + \sin\alpha \, , \\[1pt]
\sigma_z \!\! &=& \!  - \sin \beta \cos \alpha \, .
\end{array}
\label{eq:shat-derived}
\end{equation}
Substituting Eqs.~\ref{eq:skypos-derived} and \ref{eq:shat-derived} into Eq.~\ref{eq:xyzlos-fordef} yields the LOS equation in {\Gsys}, which we presented in Eq.~\ref{eq:lospos}.

\section{B. The Observed Inclination}
\label{app:B}

\begin{figure}[!bt]
\centering
\vglue -0.2in
\includegraphics[width=0.75\linewidth]{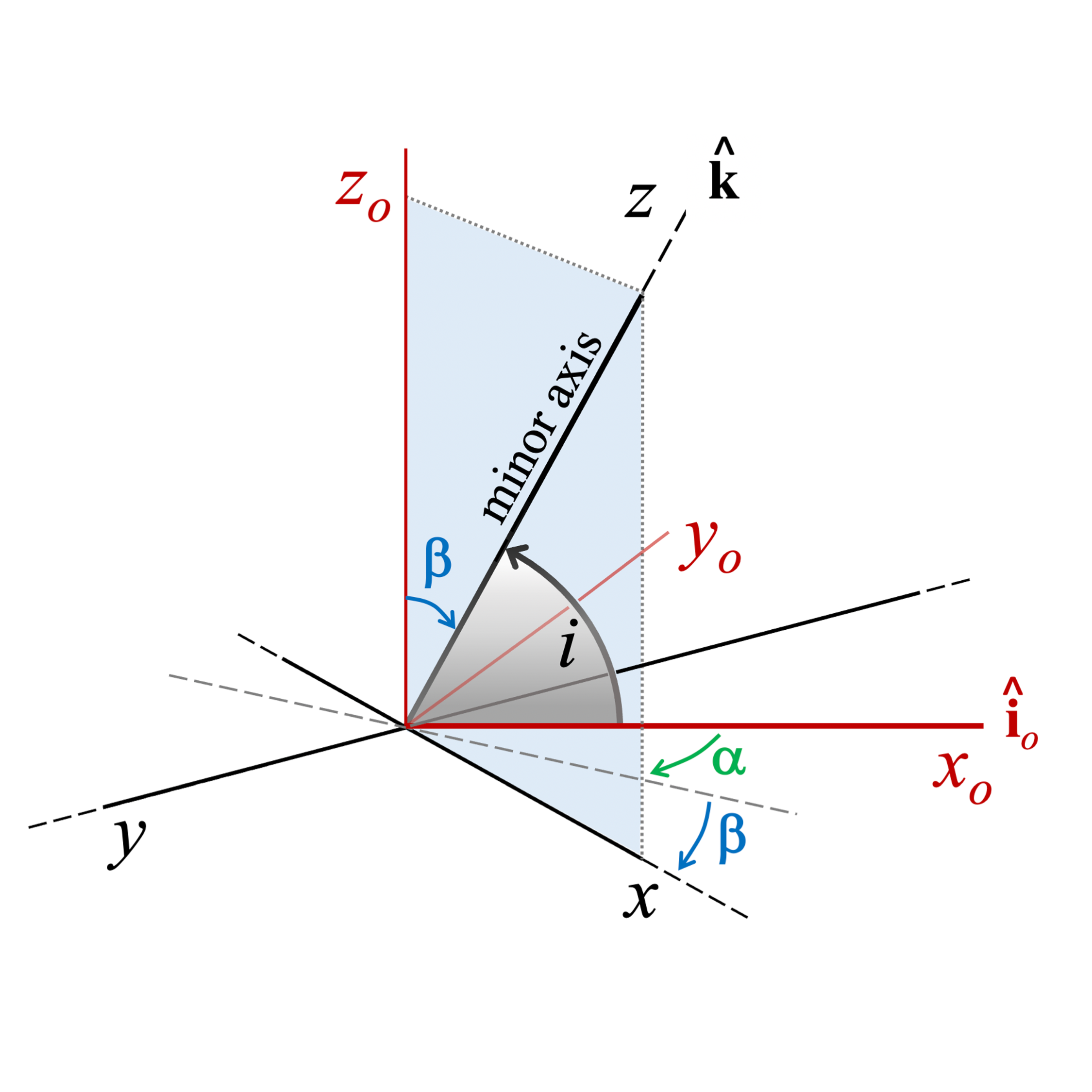}
\vglue -0.2in
\caption{\small The cosmic-eye view emphasizing the $x_oz_o$ plane. The observed galaxy inclination is the angle subtended between the minor axis of the galaxy (the ${\khat} = \hat{\bf k}$ direction) and the LOS toward the observer (the ${\ihato} = \hat{\bf i}_o$ direction).  Note that the galaxy may appear rotated by the angle $\alpha$ with respect to the ${\ihato}$ direction.}
\label{fig:skygeocalcs}
\end{figure}

The observed galaxy inclination is the angle between the galaxy $z$ axis and the LOS. In {\Gsys}, the former is in the ${\khat}$ direction and in {\Osys} the latter is in the ${\ihato}$ direction.  The geometry is illustrated in Figure~\ref{fig:skygeocalcs}. 

We thus have $({\ihato} \! \cdot {\khat}) = \cos i = x_o/z$.  This dot product can be obtained from the operation $[x_0,y_o,z_o]\supT\subO [R_y(\beta)R_z(\alpha)]\supT [0,0,z]\supT\subG$.  For the equation for $x_o$, we obtain $x_o = z (\sin\beta \cos\alpha)$, yielding 
\begin{equation}
    i = \cos^{-1} ( \sin|\beta| \cos\alpha) \, ,
\label{eq:inclination-derived}
\end{equation}
where the absolute value of $\beta$ ensures $i \in [-90^\circ,90^\circ]$.
Inspection of Eq.~\ref{eq:inclination-derived} shows that $i=90^{\circ}$ (edge-on) when $\beta = 0^{\circ}$ for $\alpha \in [0,90^{\circ}]$ and when $\alpha = 90^{\circ}$ for $\beta \in [0,90^{\circ}]$. We also see that $i=0^{\circ}$ (face-on) when $\beta = 90^{\circ}$ and $\alpha = 0^{\circ}$.

\section{C. The Azimuthal Angle}
\label{app:C}

\begin{figure}[h!tb]
\centering
\vglue -0.1in
\includegraphics[width=0.7\linewidth]{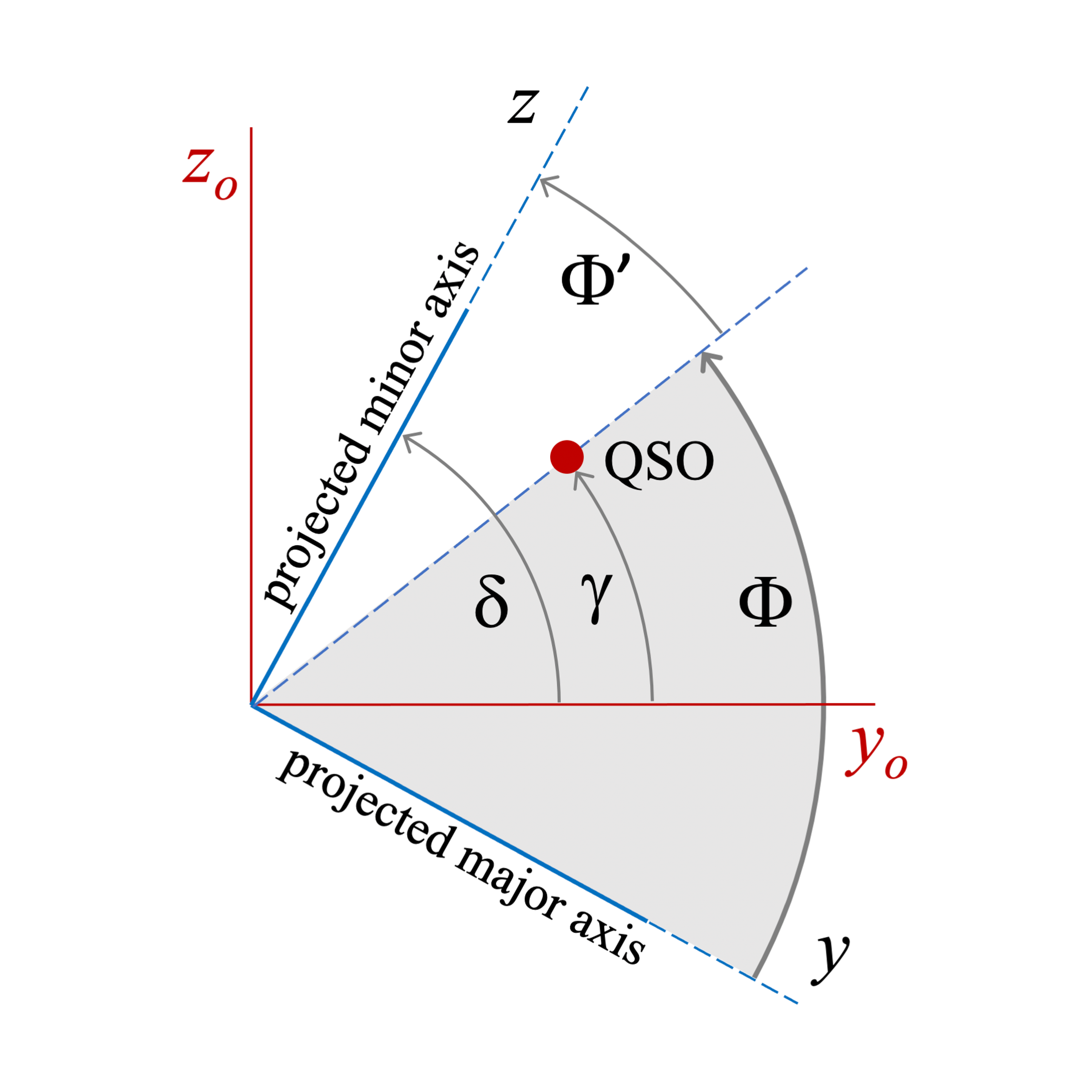}
\vglue -0.05in
\caption{\small  The sky plane ($x_o = 0$) showing the position angle $\gamma$ of the background quasar. By defining the complimentary angles $\Phi + \Phi' = 90^{\circ}$ and the relationship $\Phi' + \gamma = \delta = \tan^{-1} (z_o/y_o)$, we can apply the transpose rotation of the sky projection of the galaxy $z$ axis to obtain an expression for the sky-projected azimuthal angle $\Phi$, which is measured from the projected major axis of the galaxy.}
\label{fig:gettingPhi}
\end{figure}

The sky-projected azimuthal angle, $\Phi$, is defined counterclockwise from the observed sky-projected major axis of the galaxy. For all rotations of {\Osys} about {\Gsys}, the sky-projected minor axis of the galaxy is coincident with the sky-projection of the galaxy $z$ axis in {\Osys}.  Thus, we can write $\Phi = 90^\circ - \Phi'$, where $\Phi'$ is the angular difference between the sky-projected minor axis of the galaxy and the quasar position angle, $\gamma$, on the sky plane.  If we define the angle $\delta$ as as the angle between the sky-projected minor axis of the galaxy and the $y_o$ axis of the sky plane, we have $\delta = \gamma + \Phi'$.  This yields, $\Phi = 90^\circ - (\delta - \gamma)$. The geometric relationships are illustrated in Figure~\ref{fig:gettingPhi}. 

The angle $\delta$ will be a function of the rotation angles $\alpha$ and $\beta$. The components of the projection of the galaxy $z$ axis on the sky plane are $y_o$ and $z_o$ from which we write $\tan \delta = z_o/y_o$.  
These sky-projected quantities can be obtained from the operation $[x_o,y_o,z_o]\supT\subO [R_y(\beta)R_z(\alpha)]\supT [0,0,z]\supT\subG$.  From the equations for $y_o$ and $z_o$, we obtain $ y_o = z (\sin\beta \sin\alpha)$ and $z_o = z \cos\beta$, which yields, $\tan \delta = \cos\beta/(\sin\beta \cos\alpha)$.  We obtain
\begin{equation}
\Phi =  \gamma + \cot ^{-1}
\left( \frac{\cos\beta}{\sin\beta \cos\alpha}  \right) \, ,
\label{eq:phi-derived}
\end{equation}
where we have invoked the identity $\tan(90\!-\!x) = \cot(x)$.

\section{D. Enhancement Functions}
\label{app:D}

Here we will discuss flexible enhancement functions, the application of which was described in Section~\ref{sec:ehanced-vwind} for the wind velocity field.  We will discuss basic attributes and present a few limited examples. 

Spatial enhancement of a some quantity, $Q$, within the wind structure is of the form
\begin{equation}
    Q(\rho,z) = A + B f\subE(\rho,z)
\label{eq:howtoenhance}
\end{equation}
where $Q$ is a quantity such as the velocity, $A$ is the base value and $B$ is the amplitude of the enhancement to $A$, and $f\subE(\rho,z)$ is the enhancement function.  

The enhancement functions are axisymmetric and, because they operate only within the wind structure, at a given height $z$ we have the axial domain $\rho \in [0,\rho_w(z)]$ for $z \in [-R_w,R_w]$, where 
\begin{equation}
\rho_w(z) = (\rho_{w,0}^2 + z^2 T^2_\Theta)^{1/2} \, , 
\label{eq:AppD-rhowz}
\end{equation}
is the axial distance to the surface of the wind at height $z$. The quantity $\rho_{w,0}$ is the base radius of the wind and $T_\Theta = \tan \Theta_w$, where $\Theta_w$ is the opening angle of the wind structure, and $R_w$ is the ``cap" radius of the wind.  Enhancement functions must be designed to have peak value $f\subE(0,z)=1$ when evaluated on the wind axis (the wind core), yielding $Q(0,z) = A + B$, and vanish when evaluated at the wind surface, i.e., $f\subE(\rho_w(z),z) = 0$, yielding  $Q(\rho_w(z),z) = A$.

\begin{figure}[!htb]
\centering
\vglue -0.05in
\includegraphics[width=0.75\linewidth]{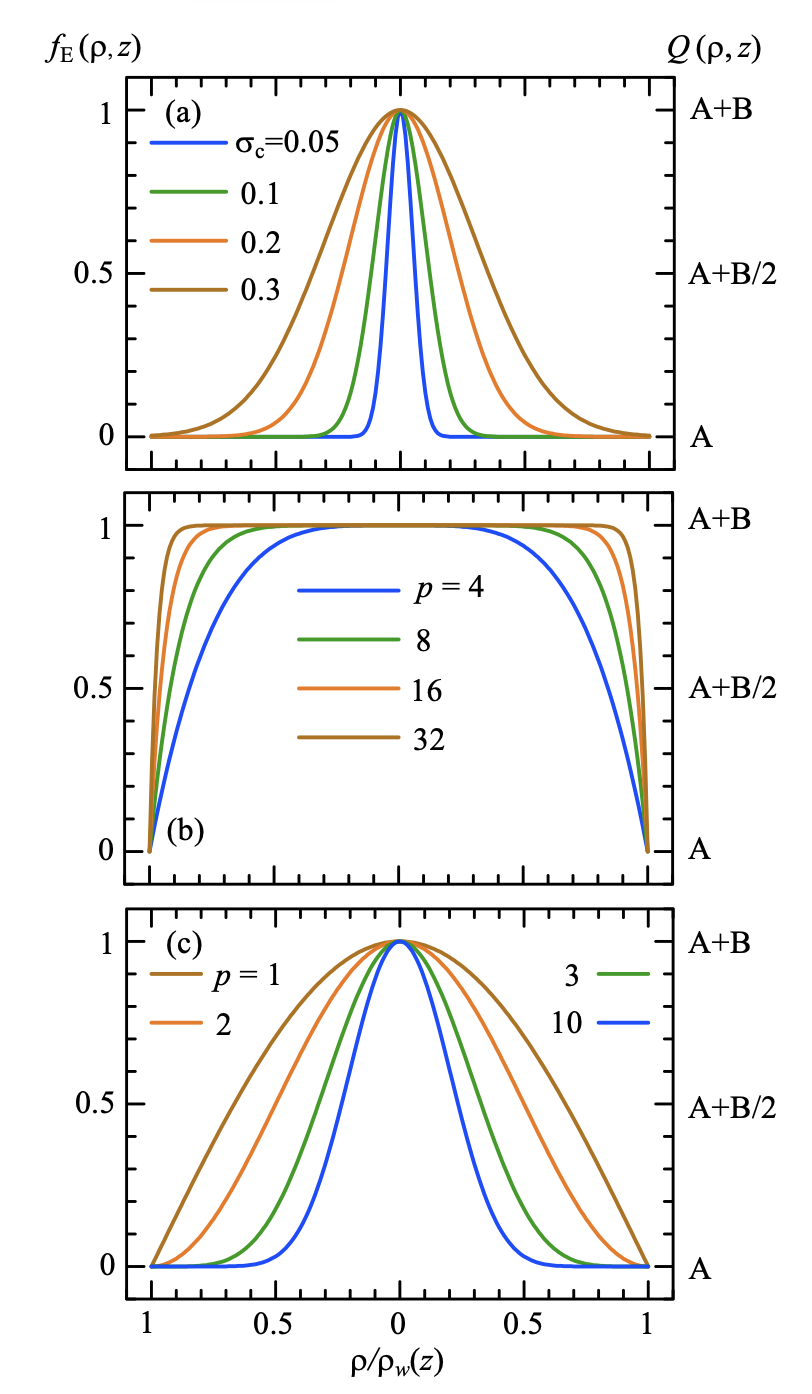}
\caption{\small Three example enhancement functions applied to Eq.~\ref{eq:howtoenhance} as a function or normalized axial distance in the wind structure plotted at arbitrary height $z$ above the galactic plane. When $f\subE(\rho,z)=0$ we have $Q(\rho,z)=A$, and when $f\subE(\rho,z)=1$ we have $Q(\rho,z)=A+B$, where $Q$, $A$, and $B$ could be velocity, density, or temperature. (a) A Gaussian function (Eq.~\ref{eq:fE-gauss}), appropriate for a highly peaked core enhancement.  (b) A polynomial (Eq.~\ref{eq:fE-poly}) of high power, $p$, appropriate for strong enhancement along the wind surface. (c) A truncated cosine (Eq.~\ref{eq:fE-cosine}) also raised to the power $p$, appropriate for enhancement intermediate to the Gaussian and polynomial functions. Note that $B$ can be either positive or negative.  The shown examples are for $B>0$ so that $Q(\rho,z)$ would be larger in the core.  For $B<0$, the curves would be inverted and $Q(\rho,z)$ would be larger at the wind surface.}
\label{fig:fE-efuncs}
\end{figure}

One potential enhancement function is a Gaussian function, which can be invoked to obtain a highly peaked core enhancement,
\begin{equation}
    f\subE(\rho,z) = \exp 
    \left\{ - \frac{1}{2} 
    \left[ \frac{\rho/\rho_w(z)}{\sigma_c} \right]^2
    \right\} \, ,
\label{eq:fE-gauss}
\end{equation}
where $\sigma_c$ is the core ``width" in units of the normalized axial distance, $\rho/\rho_w(z)$.  As illustrated in Figure~\ref{fig:fE-efuncs}(a), for ``narrow" cores ($\sigma_c \leq 0.33$), Eq.~\ref{eq:fE-gauss} will effectively vanish for $\rho/\rho_w(z)\geq 3\sigma_c$.  For broader wind cores ($\sigma_c > 0.33$), the Gaussian function does not vanish at the wind surface (which is to say that the wind surface will also be enhanced, which is a valid condition, if desired).  

Whereas the Gaussian function can provide a highly peaked core enhancement, a high-power polynomial, 
\begin{equation}
    f\subE(\rho,z) = 1 - [\rho/\rho_w(z)]^p \, ,
\label{eq:fE-poly}
\end{equation}
where $p$ is an even integer, can yield a high rapidly changing enhancement along the wind surface.  If $p=2$, then $f\subE(\rho,z)$ is a parabola with apex of unity at $\rho = 0$ (the wind core), and minimum of zero at $\rho=\rho_w(z)$ (the wind surface). If $p=4$, then $f\subE(\rho,z)$ is quartic function, etc.  As $p$ is increased the functional form of Eq.~\ref{eq:fE-poly} behaves as shown in Figure~\ref{fig:fE-efuncs}(b).  A rapid change near the wind surface may be desired, for example, for an extended enhanced core with a diminished  ``thin" circumferential wind ``wall." 

An example of a more gradually changing enhancement profile would be a truncated cosine function,
\begin{equation}
    f\subE(\rho,z) = \cos^p \! \left\{
    \frac{\pi}{2} \left( \frac{\rho}{\rho_w(z)} \right) \right\} \, ,
\label{eq:fE-cosine}
\end{equation}
where $p\geq 1$ is an integer power.
This function, with $p=1$, was introduced in Section~\ref{sec:ehanced-vwind} as applied to the wind velocity field illustrated in Figure~\ref{fig:vwind}(b).  As illustrated in Figure~\ref{fig:fE-efuncs}(c), for $p=1$, Eq.~\ref{eq:fE-poly} peaks at the wind core and vanishes at the wind surface and, as $p$ is increased, the function begins to take on the shape of a Gaussian.
Indeed, 
a Gaussian with $\sigma_c=0.2$ and a truncated cosine with $p=10$ are highly similar.  Similarly, the polynomial of degree $p=2$ (parabola, Eq.~\ref{eq:fE-poly}) is highly similar to the truncated cosine with $p=1$.  Such ``quasi-degeneracy" suggests that functions of different families can yield similar spatial distributions of outflow velocity within the wind structure.  One should adopt functions that are computationally expedient and intuitive in their parameterization.

\begin{figure}[htb]
\centering
\vglue -0.05in
\includegraphics[width=0.75\linewidth]{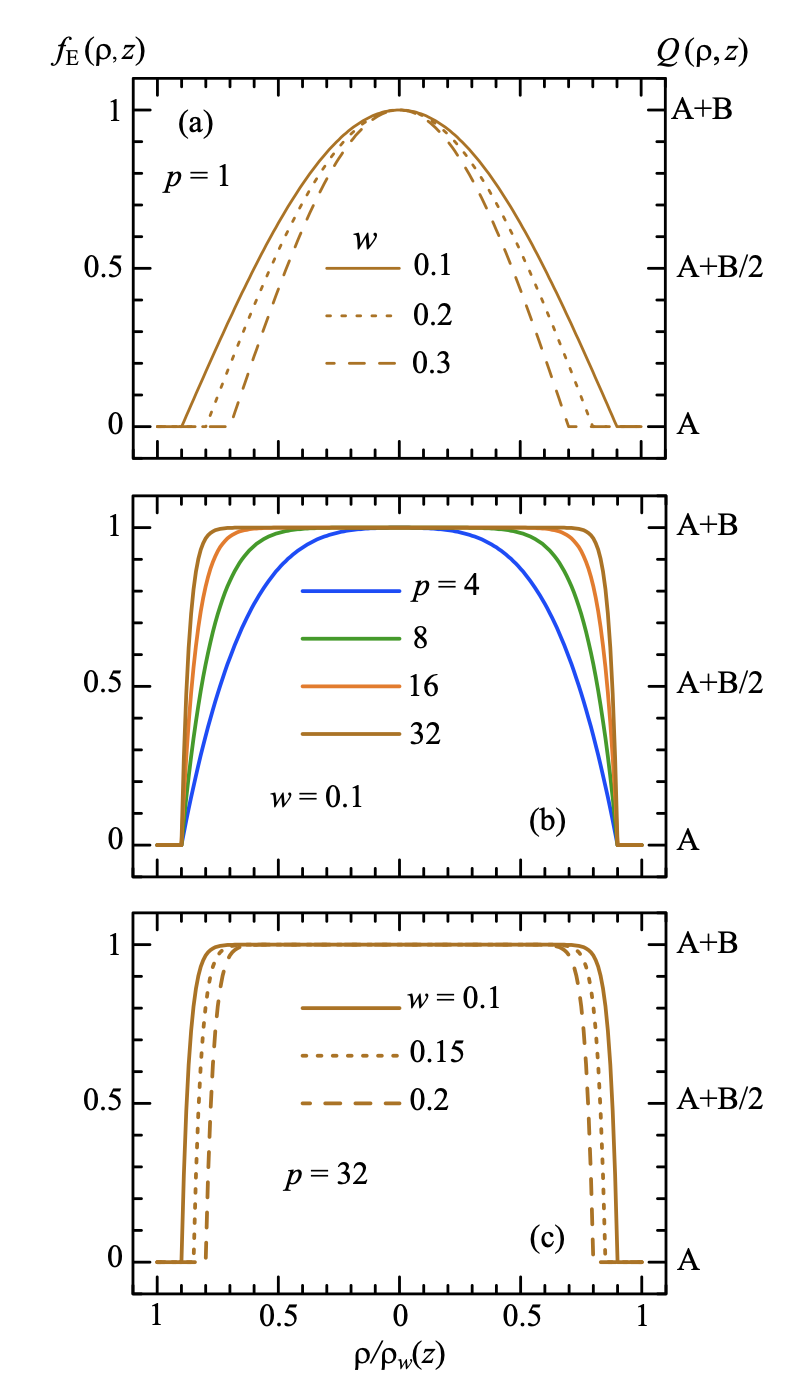}
\caption{\small Wind enhancement functions with small refinements designed to yield  circumferential walls of fractional thickness $w$, defined by $f\subE(\rho,z)=0$ for $(1\!-\!w)\rho_w(z) \leq \rho \leq \rho_w(z)$.  (a) Truncated cosine functions ($p=1$) with $w=0.1$, 0.2, and 0.3, as expressed by Eqs.~\ref{eq:fE-enforce} and \ref{eq:fE-cosine-refined}.   (b)  Four polynomial functions with $p=4$, 8, 16, and 32 with $w=0.1$. (c) The polynomial function with $p=32$ for three different values of $w=0.1$, 0.15, and 0.2.}
\label{fig:fE-refined}
\end{figure}

For galactic winds, a significant mass and momentum of cold gas can be entrained along the wall of the wind structure \citep[e.g.,][]{bustard16, rupke18, zhang18, gronke20, fielding22}. An additional refinement to enhancement might be to model ``thick walls," or a circumferential wind surface with a finite thickness. By applying a ``horizontal dilation," we design an enhancement function that vanishes at an axial distance $\rho < \rho_w(z)$ for all $z$. Such an enhancement function would yield $Q(\rho,z)=A$ within a finite distance from the wind surface (representing a thickness to the wind surface).

Defining $w$ as the fractional thickness of the wind ``walls," we modify a given enhancement function to be null (no enhancement) over the wind circumference with thickness $\Delta \rho(z) = w\rho_{w}(z)$, where $w \in (0,1]$. Note that, with this formalism, the thickness of the wind wall would increase with height above the galactic plane since it is a fixed fraction of $\rho_w(z)$, which increases with $z$ per Eq.~\ref{eq:AppD-rhowz}. Alternatively, one could develop a formalism for a constant thickness or for a thickness that decreases with height.

\begin{figure*}[htb]
\centering
\vglue -0.1in
\includegraphics[width=0.92\linewidth]{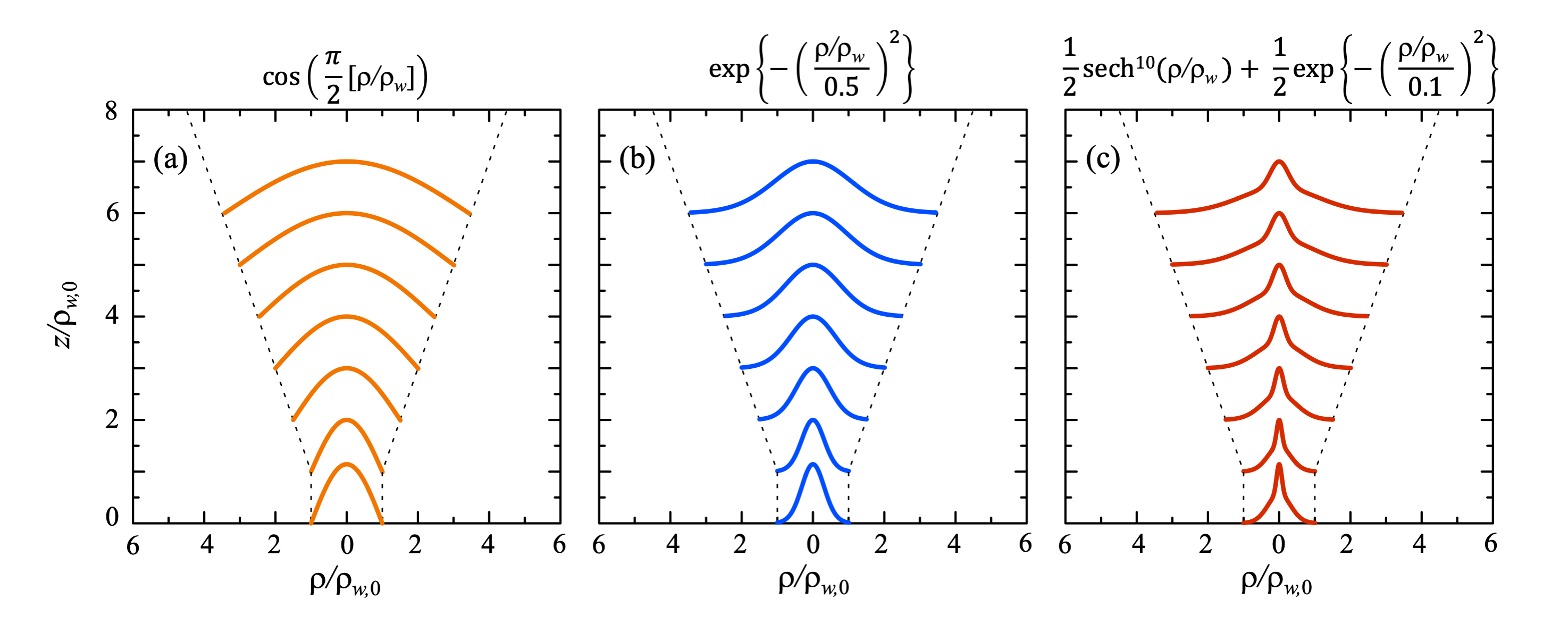}
\vglue -0.1in
\caption{\small Cross sections of example enhancement functions that could be applied to the velocity field or temperature distribution inside the wind structure. These functions are all of the form $Q(\rho,z) = 1 + f\subE(\rho/\rho_w(z))$. (a) A truncated cosine function. (b) A Gaussian function with $\sigma_c = 0.5$. (c) The sum of a hyperbolic secant raised to the power of ten and a Gaussian function with $\sigma_c = 0.1$. Because $\rho_w(z)$ increases with $z$, the functions experience horizontal stretching which manifests as dilation with increasing height.}
\vglue 0.1in
\label{fig:windenhanced}
\end{figure*}

Consider the truncated cosine function with $p=1$. Modifying the periodicity of $\pi/2$ to $\pi/2(1\!-\!w)$, the cosine will cross zero at $\rho/\rho(z) = 1\!-\!w$.  However, for $\rho/\rho(z) > 1\!-\!w$, the function goes negative, a feature we need to suppress. We thus apply the condition
\begin{equation}
f\subE(\rho,z) = {\tt max}(0,g(\rho,z)) \, ,
\label{eq:fE-enforce}
\end{equation}
to enforce $f\subE(\rho,z) \geq 0$, where the horizontally dilated truncated cosine function is then written 
\begin{equation}
    g(\rho,z) = \displaystyle \cos \left\{
    \frac{\pi}{2(1\!-\!w)} \left( \frac{\rho}{\rho_w(z)} \right) \right\}
 \, .
 \label{eq:fE-cosine-refined}
\end{equation}
This function is illustrated in Figure~\ref{fig:fE-refined}(a) for $w=0.1$, 0.2, and 0.3.  At arbitrary height $z$, we have $Q(\rho,z)=A$ for $\rho/\rho_w(z) \geq 1-w$, with an enhancement occurring in the inner axial regions $\rho/\rho_w(z) < 1-w$.  

Similarly, for the polynomial enhancement function, we can write
\begin{equation}
g(\rho,z) = 1 - \{ (\rho/\rho_w(z)) + w \}^p \, , 
\label{eq:fE-poly-refined}
\end{equation}
with the application of Eq.~\ref{eq:fE-enforce}. For  $p \leq 4$, this function will have a cuspy wind core at $\rho=0$, however, the cusp is effectively smoothed out for higher orders. In Figure~\ref{fig:fE-refined}(b), this function is illustrated for $p=4$, 8, 16, and 32 with $w=0.1$.  In Figure~\ref{fig:fE-refined}(c), we focus on $p=32$ for three fractional thicknesses, $w=0.1$, 0.15, and 0.2.  For $p=32$ and $w=0.1$, for example, we see that the wind wall would have thickness $\Delta \rho(z) = 0.1\rho_w(z)$ and exhibit a rapid change in velocity from the value $A$ to the value $B$ across this boundary (the magnitude of the change is given by the parameter $B$ and the fractional change is $(A+B)/A)$. For the Gaussian function, no horizontal dilation needs be built into the function, as a small value of $\sigma_c$ can be chosen such that $f\subE (\rho,z) \simeq 0$ for $\rho/\rho(z) > 3\sigma_c$. For example, for $\sigma_c=0.3$, this would be equivalent to a fractional thickness of $w= 1-3\sigma_c = 0.1$ for the wind wall.

In Figure~\ref{fig:windenhanced}, we illustrate the cross sectional behavior of three example wind enhancement functions. These might serve to description of the velocity field or the temperature distribution inside a wind structure. As these have positive amplitudes, they provide an enhancement in the wind core relative to the wind surface. Note that the form of the function $f\subE(\rho/\rho_w(z))$, which is written above each panel, can be as simple or complex as desired, providing flexibility and allowing an arbitrary model.

\section{E. Alternative Accretion Kinematics}
\label{app:E}

\begin{figure*}[htb]
\centering
\includegraphics[width=0.92\linewidth]{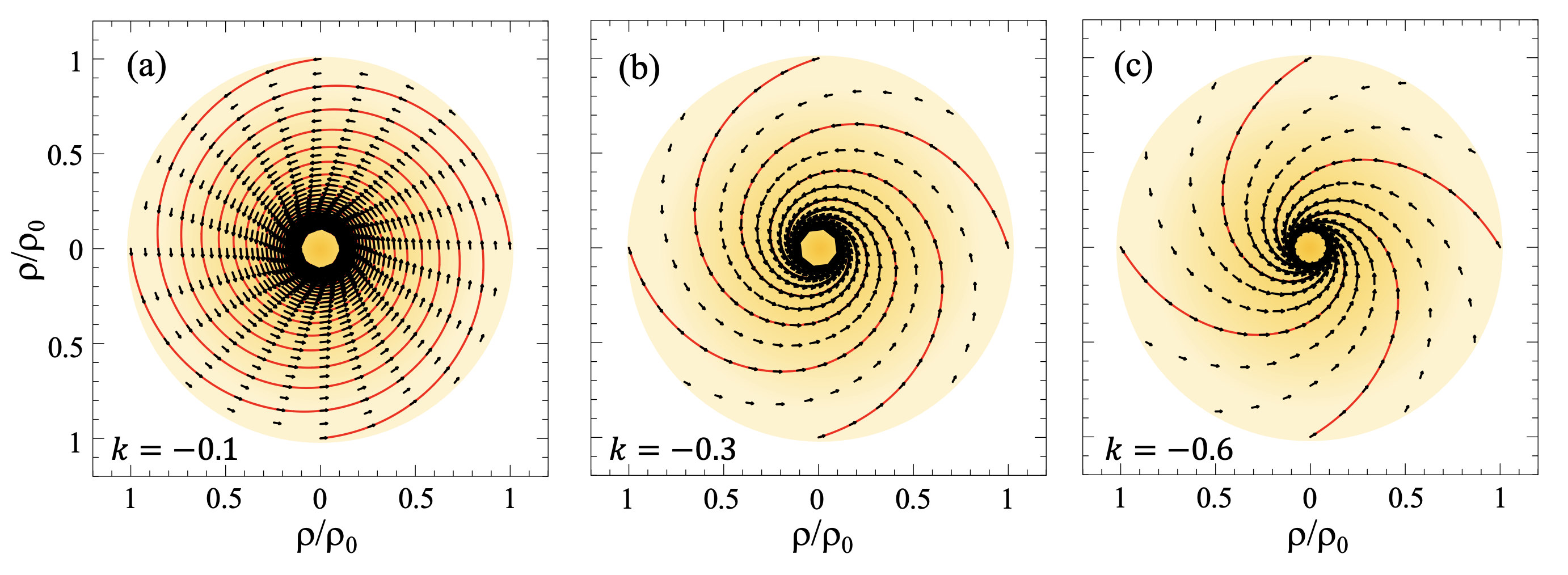}
\caption{\small Idealized velocity fields based on logarithmic spiral infall in the $z=0$ plane for the scenarios in which $\rho/\rho_0 \in (0.1,1)$, or $\rho_0 =  10 \rho_1$. Red curves provide the orbital trajectories of the infalling gas, whereas the velocity vector field at $\rho/\rho_0$ is represented by the black arrows. The relative sizes of the arrows are $v/V_c$. The orbital trajectories and velocity field for (a) $k=-0.1$, which follows for $\phi \simeq -6^{\circ}$, (b) $k=-0.3$ for $\phi \simeq -17^{\circ}$, and (c) $k=-0.6$ for $\phi \simeq -31^{\circ}$. As these kinematic models enforce the velocity to be strictly azimuthal with magnitude $v_1 = V_c$ at $\rho=\rho_1$, the velocity fields for different orbital eccentricities become less distinguishable as $\rho$ approaches $\rho_1$. The main differences are in the outer regions.}
\vglue 0.1in
\label{fig:spiral-accretion}
\end{figure*}
Here we describe an alternative kinematic model for the accretion structure known as  the logarithmic spiral, $\rho = \rho_0 e^{k\phi}$ \citep[see][for details]{wysocki15}. This curve was first presented as the {\it ewige Linie} \citep{durer1525} and later studied as the {\it spira mirabilis\/} by J.\ Bernoulli (circa 1695). It has the dual benefit of providing both a velocity field expressible in closed form that conserves angular momentum and a spiral structure compatible with density wave theory \citep[e.g.,][]{bertin96, pringle19}.  A central force scaling as $r^{-3}$ results in a logarithmic spiral orbit, which is asymptotically consistent with an NFW potential for $r \gg R_s$ \citep[see][]{bassetto22}. 

The constant parameter $k = \tan(\psi)$ is the tangent of the ``pitch angle," measured as the angle between the tangent to the curve and the normal to the axial radial vector at $\rho$. For galactic spiral arms, observations indicate $\psi \sim 15^\circ$--$20^\circ$, or $k=0.27$--0.36 for a range of galaxy mass and spiral morphology \citep[e.g.,][]{savchenko13, lingard21}. 

Defining $V_\uprho = (d\rho/d\phi)(d\phi/dt)$ and $V_\upphi = \rho(d\phi/dt)$, we have
\begin{equation}
 V_\uprho = k\rho_0 e^{k\phi} \frac{d\phi}{dt}\, ,
 \quad 
 V_\upphi = \rho_0 e^{k\phi} \frac{d\phi}{dt} \, .
\label{eq:logspiralvcomps}
\end{equation}
From the solution of the equation of motion including acceleration due to gravity \citep{bassetto22}, 
\begin{equation}
\frac{d\phi}{dt} = 
\left[
\frac{2\mu}{\rho^3} \frac{[1-e^{(k^2+2)\phi/k}]}{k^2+2} 
\right] ^{1/2} \, ,
\label{eq:vlogphodot}
\end{equation}
where $\mu = GM(r)$ is the gravitational parameter for a total mass contained within radius $r$. 
Substituting $\rho = \rho_0 e^{k\phi}$ into Eq.~\ref{eq:vlogphodot} and inserting into Eq.~\ref{eq:logspiralvcomps}, we write
\begin{equation}
 V_\uprho = k \chi(\phi) \, ,
 \quad 
 V_\upphi = \chi(\phi) \, ,
\label{eq:vlogfinal}
\end{equation}
where
\begin{equation}
\chi(\phi) = \displaystyle e^{-(k/2)\phi}
\left[ \frac{2\mu}{\rho_0} \frac{[1-e^{(k^2+2)\phi/k}]}{k^2+2}\right] ^{1/2}  \, . 
\label{eq:vlogchiphi}
\end{equation}
To convert the velocity of the trajectory into  a velocity field that depends only on the axial distance from the $z$ axis, we write 
$\phi = \ln(\rho/\rho_{0})/k$ and substitute into Eq.~\ref{eq:vlogchiphi}.  Adopting $V_c^2 = \mu/\rho_1$, where $\rho_1$ is the periapsis, we have $\mu/\rho_0 = V_c^2 \rho_1/\rho_0$. We obtain,
\begin{equation}
\chi(\rho) = \displaystyle V_c \left[
\frac{2\rho_1}{\rho} \frac{[1-(\rho/\rho_0)^{(k^2+2)/k^2}]}{k^2+2} 
\right] ^{1/2} \, .
\label{eq:vlogchirho}
\end{equation}
For an inward spiral, we define $k<0$, in which case $\rho_0$ is the maximum axial distance, $\rho_1$ is the minimum axial distance, and $\rho \in (\rho_1,\rho_0)$. The resulting velocity field is
\begin{equation}
 {\bf V}_a(\rho) = \displaystyle 
k\chi(\rho) \, {\rhohat}(\rho) +
\chi(\rho) \, {\phihat} (\rho) \, . 
\label{eq:logspiralvels}
\end{equation}
Applying Eq.~\ref{eq:Vlos-general}, the LOS velocity is 
\begin{equation}
  V^{(a)}\subLOS(t) = \displaystyle \left[ k \, {\cal P}_\uprho(t) + {\cal P}_\upphi(t) \right] \chi(t) \, . 
\label{eq:LOSspiralvels}
\end{equation}

In Figure~\ref{fig:spiral-accretion}, we show three logarithmic spiral velocity fields:  $k=-0.1, -0.3$, and $-0.6$.  Whereas $k=-0.3$ is somewhat representative of the velocity field of density wave theory for spiral arms \citep{bertin96, savchenko13, pringle19, lingard21}, the value $k=-0.1$ might more closely model the manner in which accreting streams blend into ``extended rings" similar to those predicted for the highest angular momentum infalling streams \citep[][]{danovich15}.  For $k=-0.3$, the velocity field and gas orbital trajectories are similar to those of \cite[][see their Fig.~2]{wang23} for their ``MAD" model, in which they assume a velocity field of the form $V_\upphi(r) = V_c \tan^{-1}(r/R_t)$, where $R_t$ is a multiple of the scale length of the star formation surface density of the disk. For $k \leq -0.6$, the logarithmic spiral trajectories resemble the trajectories seen in the simulations of \citet[][]{hafen22} and \citet{kocjan24} and begin to approach the ballistic infall kinematics that are more akin to the Keplerian trajectories described in Section~\ref{sec:kepler-vfields}. 

To fully describe a logarithmic spiral model, one needs to specify $k$, $\rho_1$, and $\rho_2$ (assuming $V_c$ for the velocity at $\rho_1$).  Values of $k$ in the range $k=-0.1$ to $k=-0.6$ provide suitable velocity fields that crudely emulate a viable range of accretion trajectories that resemble those seen in multiple simulations \citep[e.g.,][]{keres05, stewart11, stewart13, hafen22, gurvich23, kocjan24, stern23}. The natural choice for the maximum axial distance is $\rho_0 = R_a$, the extent of the accretion hyperboloid structure.  The minimum axial distance, $\rho_1$, could be the accretion radius, $\rho_{a,0}$, the axial radius of the disk $\rho_d$, or the radius of maximum circular velocity, $R_c$.


\bibliography{main}{}
\bibliographystyle{aasjournal}



\end{document}